\documentclass[11pt, twoside]{scrartcl}

\usepackage[tracking=true,expansion=true,stretch=15,shrink=15]{microtype}
\DeclareMicrotypeSet*[tracking]{my}
{ font = */*/*/sc/* } 
\SetTracking{encoding = *, shape = sc }{ 45 }
\SetProtrusion{encoding={*},family={bch},series={*},size={6,7}}
{1={ ,750},2={ ,500},3={ ,500},4={ ,500},5={ ,500},
	6={ ,500},7={ ,600},8={ ,500},9={ ,500},0={ ,500}}
\SetExtraKerning[unit=space]
{encoding={*}, family={qhv}, series={b}, size={large,Large}}
{1={-200,-200}, 
	\textendash={400,400}}

\usepackage{setspace}
\usepackage[
	left=1in,
	right=1in,
	top=0.7in,
	bottom=0.7in,
	includeheadfoot
]{geometry}
\usepackage[automark, headsepline]{scrlayer-scrpage}
\usepackage[final]{pdfpages}

\usepackage{amssymb}
\usepackage{amsthm}
\usepackage{yhmath}
\usepackage{dsfont}
\usepackage{bm}
\usepackage{enumitem}
\usepackage{subfig}
\usepackage{graphicx}
\usepackage{float} 
\usepackage{tikz}
\usetikzlibrary{shapes,arrows,decorations}
\usepackage[round,comma]{natbib}
\usepackage{notoccite}

\usepackage{chngcntr}
\usepackage{comment}

\usepackage{hyperref}
\definecolor{LUblue}{cmyk}{1, .75, .0, .35}
\hypersetup{
	colorlinks   = true, %Colours links instead of ugly boxes
	urlcolor     = LUblue, %Colour for external hyperlinks
	linkcolor    = LUblue, %Colour of internal links
	citecolor   = LUblue %Colour of citations
}
\usepackage[noabbrev,nameinlink]{cleveref}
\usepackage{algorithm}
\usepackage[noend]{algpseudocode}

\usepackage[mathlines]{lineno} %% <- with [mathlines] to number lines in equations
\usepackage{amsmath}           %% <- after lineno
\usepackage{etoolbox}          %% <- for \cspreto, \csappto and \patchcmd

\newcommand*\linenomathpatch[1]{%
  \cspreto{#1}{\linenomath}%
  \cspreto{#1*}{\linenomath}%
  \csappto{end#1}{\endlinenomath}%
  \csappto{end#1*}{\endlinenomath}%
}
\newcommand*\linenomathpatchAMS[1]{%
  \cspreto{#1}{\linenomathAMS}%
  \cspreto{#1*}{\linenomathAMS}%
  \csappto{end#1}{\endlinenomath}%
  \csappto{end#1*}{\endlinenomath}%
}

\expandafter\ifx\linenomath\linenomathWithnumbers
  \let\linenomathAMS\linenomathWithnumbers
  \patchcmd\linenomathAMS{\advance\postdisplaypenalty\linenopenalty}{}{}{}
\else
  \let\linenomathAMS\linenomathNonumbers
\fi

\linenomathpatch{equation}
\linenomathpatchAMS{gather}
\linenomathpatchAMS{multline}
\linenomathpatchAMS{align}
\linenomathpatchAMS{alignat}
\linenomathpatchAMS{flalign}

\makeatletter
\patchcmd{\mmeasure@}{\measuring@true}{
  \measuring@true
  \ifnum-\linenopenaltypar>\interdisplaylinepenalty
    \advance\interdisplaylinepenalty-\linenopenalty
  \fi
  }{}{}
\makeatother

\newtheoremstyle{new}{12pt}{12pt}{\itshape}{}{\bfseries}{.}{1em}{}
\theoremstyle{new}
\newtheorem{theorem}{Theorem}
\newtheorem{corollary}{Corollary}

\newtheorem{lemma}{Lemma}
\newtheorem{definition}{Definition}
\newtheorem{example}{Example}
\newtheorem{remark}{Remark}

\newcommand{\ind}{\mathds{1}}
\newcommand{\E}{\mathds E}
\newcommand{\R}{\mathds{R}}
\renewcommand{\Pr}{\mathrm{P}}
\newcommand{\N}{\mathds{N}}
\newcommand{\var}{{\mathrm{Var}}}
\newcommand{\cov}{{\mathrm{Cov}}}

\newcommand{\wh}{\widehat}
\newcommand{\wt}{\widetilde}
\newcommand{\wb}[1]{\widehat{\bm{#1}}}

\newcommand{\PT}{\mathds P_{_T}}

\DeclareFontFamily{U}{mathx}{\hyphenchar\font45}
\DeclareFontShape{U}{mathx}{m}{n}{
      <5> <6> <7> <8> <9> <10>
      <10.95> <12> <14.4> <17.28> <20.74> <24.88>
      mathx10
      }{}
\DeclareSymbolFont{mathx}{U}{mathx}{m}{n}
\DeclareMathSymbol{\bigtimes}{1}{mathx}{"91}

\allowdisplaybreaks
\newcommand{\change}[1]{#1}

\title{Stationary vine copula models for multivariate time series}

\author{Thomas Nagler\footnote{Corresponding author, Delft Institute of Applied Mathematics, Delft University of Technology, Mekelweg 4, 2628 CD Delft, Netherlands (email: mail@tnagler.com)}, Daniel Kr{\"u}ger\footnote{Technical University of Munich}, and Aleksey Min\footnote{Technical University of Munich}}

\date{\hspace{3pt} \normalsize\today}

\begin{document}

\maketitle

%---------------------------------------------------------%

\begin{abstract} 
\noindent {\bfseries \sffamily Abstract}\\
Multivariate time series exhibit two types of dependence: across variables and across time points. Vine copulas are graphical models for the dependence and can conveniently capture both types of dependence in the same model. We derive the maximal class of graph structures that guarantee stationarity under a natural and verifiable condition called translation invariance. We propose computationally efficient methods for estimation, simulation, prediction, and uncertainty quantification and show their validity by asymptotic results and simulations. The theoretical results allow for misspecified models and, even when specialized to the \emph{iid} case, go beyond what is available in the literature.  The new model class is illustrated by an application to forecasting returns of a portfolio of 20 stocks, where they show excellent forecast performance. The paper is accompanied by an open source software implementation.  \\[12pt]
\end{abstract}

%---------------------------------------------------------%

\pagestyle{scrheadings}
\clearscrheadings
\lohead{Stationary vine copula models for multivariate time series}
\rohead{\pagemark}
\lehead{T.~Nagler, D.~Kr{\"u}ger, A.~Min}
\rehead{\pagemark}

\section{Introduction} 

In multivariate time series there are two types of dependence: \emph{cross-sectional} and \emph{serial}. The first is dependence between variables at a fixed point in time. The second is dependence of two random vectors at different points in time. Copulas are general dependence models and have been used for both types. One line of research considers copula models for serial dependence in  univariate Markov processes \citep[including][]{darsow:et:al:1992,chen2006estimation,chen2009, ibragimov:2009, beare2010copulas, nasri2019b}.  An orthogonal, but equally popular approach is to filter serial dependence by univariate time series models, like the ARMA-GARCH family, and model the cross-sectional dependence by a copula for the residuals \citep{patton2006modelling, hu:2006, CHEN2006125, oh2017modeling, nasri2019a, CHEN2021484}.  See also \citet{Patton2009, patton2012, aas2016pair} for surveys in the context of financial and economic time series. 

Copulas can be used to capture both types of dependence in a single model \citep[e.g.,][]{ remillard2012copula, simard2015}. 
In this context, vine copulas \citep{bedford2002,AasKjersti2009Pcom} have been proven particularly useful. Vine copulas  are graphical models that build a $d$-dimensional dependence structure from two-dimensional building blocks, called \emph{pair-copulas}. The underlying graph structure consists of a nested sequence of trees, called \emph{vine}. Each edge is associated with a pair-copula and each pair-copula encodes the (conditional) dependence between a pair of variables.  \citet{brechmann2015copar}, \citet{smith2015copula}, and \citet{BeareSeo2015} proposed different vine structures suitable for time series models. 
The three models are quite similar. The vine graphs start with copies of a cross-sectional tree that connect variables observed at the same point in time. These trees are constrained to be either stars \citep{brechmann2015copar} or paths \citep{smith2015copula, BeareSeo2015}.  Cross-sectional trees are then linked by a specific building plan.

Inspired by the three latter works, we propose more flexible vine models for stationary time series (\Cref{sec:svines}). But we approach the problem from the opposite direction. Previous models aim to guarantee stationarity of the model through a condition called \emph{translation invariance} \citep{BeareSeo2015}: pair-copulas stay the same when corresponding random variables are shifted in time.  Translation invariance is a necessary condition for stationarity  and  the only practicable condition to check. We derive a characterization of the class of vines for which translation invariance is also sufficient for stationarity (\Cref{thm:svines,thm:explicit}). The class allows for general vine structures for cross-sectional dependence and leaves flexibility for linking them across time. The class includes the D-vine and M-vine models of \citet{smith2015copula} and \citet{BeareSeo2015} as special cases, but not the COPAR model of \citet{brechmann2015copar}. Hence, the COPAR model is not stationary in general (see \Cref{ex:COPAR}).

For practical purposes, it is convenient to restrict to Markovian models, which are easily obtained by placing independence copulas on most edges in the vine (\Cref{thm:markov}).
Parameters of such models can be estimated by adapting the popular sequential maximum-likelihood method to take time invariances into account. In \Cref{sec:estsel}, we show consistency and normality of parametric and semiparametric versions of this method (\Crefrange{thm:P_consistency}{thm:SP_an}). 
%For the semiparametric estimator, the results of \citet{hobaekhaff2013} in  the \emph{iid} case and \citet{chen2006estimation} for univariate time series are recovered as special cases, under weaker conditions \am{Lassen wir das so? Vielleicht ohne 'under weaker conditions'}. The results for the parametric estimator are new, even in the \emph{iid} case. It is noteworthy that parametric and semiparametric results are proven simultaneously by more high-level theorems (\Cref{thm:aux_consistency,thm:aux_an}). These theorems apply to general semiparametric method-of-moment estimators with non-negligible nuisance parameter and may prove useful beyond the current scope.
By simulating from an estimated model, conditionally  on the past, we can easily compute predictions. In \Cref{sec:prediction}, we translate this into a  theoretical framework and prove consistency and asymptotic normality. The results also cover Monte Carlo integrals from estimated \emph{iid} models, which are widely used but have not been formalized yet.
We further propose a computationally  efficient bootstrap method for both parameter estimates and predictions in \Cref{sec:uncertainty}.
In \Cref{sec:application}, we apply the methodology to forecast portfolio returns, showing that our generalized models improve both performance and interpretability. \Cref{sec:conclusion} offers concluding remarks.  Abstract mathematical results on general method-of-moment estimators, which empower all our main theorems, are stated in \Cref{sec:gen}.  Additional illustrations, simulation results, and all proofs are provided in the supplementary materials.  
All methodology is implemented in the open source R package \texttt{svines} (available at \url{https://github.com/tnagler/svines}), which is built on top the C++ library \texttt{rvinecopulib} \citep{rvinecopulib}.

%The article is structured as follows. \Cref{sec:background} introduces some necessary concepts and notation for vine copula models and reviews previously proposed vine models for stationary time series. The class of stationary vine models and some properties are derived  in \Cref{sec:svines}. \Cref{sec:estsel} discusses estimation and model selection, \Cref{sec:prediciton} simulation and simulation-based prediction. The bootstrap method is explained in \Cref{sec:uncertainty}.  The models are illustrated with simulations in \Cref{sec:sim} and an application to financial time series in \Cref{sec:application}. \Cref{sec:conclusion} offers concluding remarks. Our abstract results, empowering all our main theorems, are stated in \Cref{sec:gen}.  Additional illustrations, simulation results, and all proofs are provided in the supplementary materials.  

%\section{Background on vine copula models} \label{sec:background}

\section{Multivariate time series based on vine copulas} \label{sec:background}

\subsection{Copulas}

Copulas are models for the dependence in a random vector. 
By Sklar's theorem \citep{Sklar1959}, any multivariate distribution $F$ of a $d-$dimensional random vector $\bm X=(X_1, \ldots, X_d)^\prime$ with marginal distributions $F_1, \dots, F_d$ can be expressed as
\begin{linenomath}
\begin{align*} 
	F(x_1, \dots, x_d) = C\bigl\{F_1(x_1), \dots, F_d(x_d)\bigr\} \qquad \mbox{for all } \bm x \in \R^d,
\end{align*}
\end{linenomath}
for some function $C\colon [0, 1]^d \to [0, 1]$ called \emph{copula}. It characterizes the dependence in $F$ because it determines how margins interact. If $F$ is continuous, then $C$ is the unique joint distribution function of the random vector $\bm U = (F_1(X_1), \dots, F_d(X_d))^\prime$. A  similar formula can be stated for the density provided $F$ is absolutely continuous:
\begin{align*}
	f(x_1, \dots, x_d) = c\bigl\{F_1(x_1), \dots, F_d(x_d)\bigr\} \times \prod_{k = 1}^d f_k(x_k) \qquad  \mbox{for all }  \bm x \in \R^d,
\end{align*}
where $c$ is the density of $C$ and called the \emph{copula density}, and $f_1,\dots,f_d$ are the marginal densities. 

\subsection{Regular vines} \label{sec:vines}

Vine copulas are a particularly flexible class of copula models. They are based on an idea of \citet{joe1996, joe:1997} to decompose the copula into a cascade of bivariate copulas. This decomposition is not unique, but all possible decomposition can be organized as a graphical model, called \emph{regular vine (R-vine)} \citep[see,][]{bedford2001,bedford2002}.
We shall briefly outline the basics of R-vines; for more details on R-vines, we refer to \cite{Dissmann2012, joe2014dependence, czado2019analyzing}. Additional illustrations of the following definitions can be found in Section S1 of the supplementary materials.

A regular vine is a sequence of nested trees. A tree $(V, E)$ is a connected acyclic graph consisting of vertices $V$ and edges $E$.
\begin{definition}\label{def:R-Vine} 
	A collection of trees $\mathcal{V}=(V_k, E_k)_{k = 1}^{d - 1}$ on a set $V_1$ with $d$ elements is called R-vine if
	\begin{enumerate}[label=(\roman*)]
		\item $T_1$ is a tree with vertices $V_1$ and edges $E_1$,
		\item for $k=2,\dots,d-1$, $T_k$ is a tree with vertices $V_k = E_{k-1}$,
		\item (\emph{proximity condition}) for $k=2,\dots,d-1$: if vertices $a, b \in V_k$  are connected by an edge $e \in E_k$, then the corresponding edges  $a=\{a_1,a_2\}$, $b=\{b_1,b_2\} \in E_{k - 1}$, must share a common vertex: $|a \cap b| = 1$.
	\end{enumerate}
\end{definition}

\begin{figure}
	\centering
	\begin{tikzpicture}
		\tikzstyle{every node}=[ellipse, minimum width=80pt,align=center,scale=0.68]
		\node[shape=rectangle,draw=black] (1) at (0,5.5) {$1$};
		\node[shape=rectangle,draw=black] (2) at (3,5.5) {$2$};
		\node[shape=rectangle,draw=black] (3) at (6,5.5) {$3$};
		\node[shape=rectangle,draw=black] (4) at (9,5.5) {$4$};
		\node[shape=rectangle,draw=black] (5) at (12,5.5) {$5$};

		\node[shape=rectangle,draw=black] (6) at (1.5,4) {$1,2$};
		\node[shape=rectangle,draw=black] (7) at (4.5,4) {$2,3$};
		\node[shape=rectangle,draw=black] (8) at (7.5,4) {$3,4$};
		\node[shape=rectangle,draw=black] (9) at (10.5,4) {$4,5$};

		\node[shape=rectangle,draw=black] (10) at (3,2.5) 	{$1,3|2$};
		\node[shape=rectangle,draw=black] (11) at (6,2.5) 	{$2,4|3$};
		\node[shape=rectangle,draw=black] (12) at (9,2.5) 	{$3,5|4$};

		\node[shape=rectangle,draw=black] (13) at (4,1) 	{$1,4|2,3$};
		\node[shape=rectangle,draw=black] (14) at (8,1) 	{$2,5|3,4$};

		\path [-] (1)  edge node[above] {$1,2$} (2);
		\path [-] (2)  edge node[above] {$2,3$} (3);
		\path [-] (3)  edge node[above] {$3,4$} (4);
		\path [-] (4)  edge node[above] {$4,5$} (5);

		\path [-] (6)  edge node[above] {$1,3|2$} (7);
		\path [-] (7)  edge node[above] {$2,4|3$} (8);
		\path [-] (8)  edge node[above] {$3,5|4$} (9);

		\path [-] (10)  edge node[above] {$1,4|2,3$} (11);
		\path [-] (11)  edge node[above] {$2,5|3,4$} (12);

		\path [-] (13)  edge node[above] {$1,5|2,3,4$} (14);
	\end{tikzpicture}
		\caption{A five-dimensional D-vine.}\label{fig-d-vine}
\end{figure}
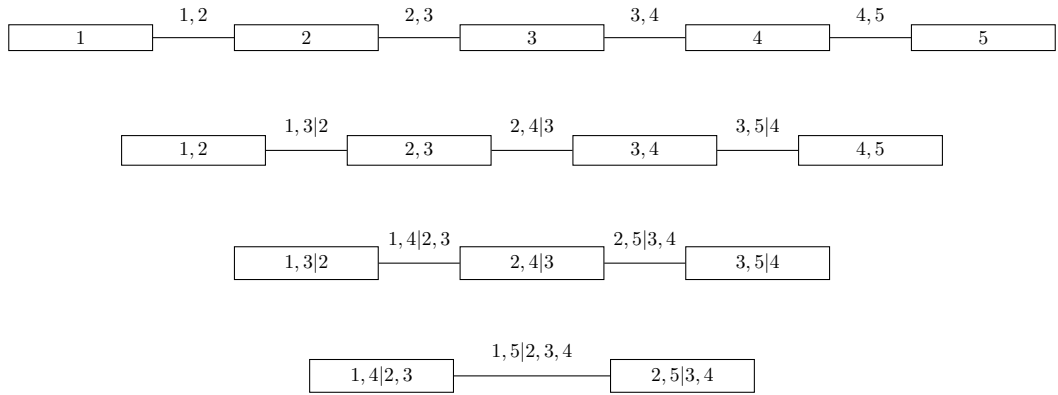

\Cref{fig-d-vine} shows a graphical example of special regular vine, called D-vine. We call a vine a D-vine if each tree is a path. A tree is a path if each vertex is connected to at most two other vertices. Such a structure is most natural when there is a natural ordering (e.g., in time or space) of the variables. When one tree of the vine is a path, all trees at higher levels are fixed uniquely by the proximity condition. Another prominent sub-class are C-vines, where  each tree is a star (see also Figure S2 in the supplementary material). A tree is a star if there is one vertex that is  connected to all remaining vertices. This structure is most natural when there is a single variable driving the others (e.g., a market factor driving individual stocks). In that case, the proximity condition poses no restrictions on the next tree. 

The connection of regular vines to a decomposition of the dependence becomes apparent through a specific labeling of the edges. Each edge corresponds to a pair of random variables conditioned on some others.
This is encoded in the \emph{conditioned} and \emph{conditioning sets} of an edge. We first need the definition of a \emph{complete union}.
\begin{definition} \label{def:cu}
	The complete union of an edge $e \in E_k$ is given by
	$$\mathcal U_{e}=
		\{i \in V_1| \ i \in e_1 \in e_2\in \dots \in e \text{ for some } (e_1,\dots, e_{k-1}) \in E_1\times \dots \times E_{k-1}\}$$
	and for a singleton $i \in V_1$ it is given by the singleton, i.e. $\mathcal U_i=\{i\}$.
\end{definition}
In other words, the complete union of an edge $e \in E_k$ is just the set of  all vertices from the first tree $T_1$, which are involved in the iterative construction of the edge $e$.
\begin{definition}\quad \\[-12pt] \label{def:labels}
	\begin{enumerate}[label=(\roman*)]
		\item The \emph{conditioning set} of an edge  $e$ connecting $v_1$ with $v_2$  is  $D_{e}=\mathcal U_{v_1}\cap \mathcal U_{v_2}$.
		\item The \emph{conditioned set} of an edge $e$ connecting $v_1$ with $v_2$ is defined as $(a_e,b_e)$, where
		      $a_e=\mathcal U_{v_1}\setminus D_{e}$ and $b_k=\mathcal U_{v_2}\setminus D_{e}$.
	\end{enumerate}
	We will then label an edge by $e = (a_e, b_e | D_{e})$.
\end{definition}
\Cref{def:labels} complements  \Cref{def:R-Vine} by relating each edge in the R-vine to a pair-wise conditional distribution, as we shall see in the following section. Note that the conditioning set $D_{e}$ is  empty for edges of the first tree level.

\subsection{Vine copulas} \label{sec:vine_copulas}

By fixing the  marginal   distributions $F_1, \ldots, F_d$, it suffices to consider a random vector  $\bm U = (F_1(X_1), \dots, F_d(X_d))^\prime$ with standard uniform margins to describe the dependence structure of  $\bm X$. A vine copula model for the  $\bm U$ identifies each edge of an R-vine with a bivariate copula. We shall write the model as $(\mathcal V, \mathcal C(\mathcal V))$, where $\mathcal V = (V_k, E_k)_{k = 1}^{d - 1}$ is the vine structure, $d$ the number of variables, and $\mathcal C(\mathcal V) = \{c_e\colon e \in E_k, k = 1, \dots, d - 1\}$ the set of associated bivariate copulas. As an example, consider the regular vine shown in \Cref{fig-d-vine}. The vertices in the first tree represent the random variables $U_1, \dots, U_5$. All edges connecting them are identified with a bivariate copula (or \emph{pair-copula}). The edge $(a_e, b_e)$ then encodes the dependence between $U_{a_e}$ and $U_{b_e}$. In the second tree, the edges have labels $(a_e, b_e| D_e)$ and encode the dependence between $U_{a_e}$ and $U_{b_e}$ conditional on $U_{D_e}$. In the following trees, the number of conditioning variables increases.

\citet{bedford2001} showed that the density of a such a copula model for the vector $\bm U$ has a product form:
\begin{align*}
	c(\bm u) = \prod_{k=1}^{d-1} \prod_{e \in E_k} c_{a_e,b_e|D_e} \left(u_{a_e|D_e},u_{b_e|D_e} \mid \bm{u}_{D_e}\right),
\end{align*}
where $u_{a_e|D_e} := C_{a_e|D_e}(u_{a_e}\mid  \bm{u}_{D_e})$, $\bm{u}_{D_e}:= (u_l)_{l\in D_e}$ is a subvector of $\bm{u} = (u_1, \dots, u_d) \in [0,1]^d$ and $C_{a_e|D_e}$ is the conditional distribution of $U_{a_e}$ given $\bm U_{D_e}$. The functions $c_{a_e,b_e|D_e}$ are copula densities describing the dependence of $U_{a_e}$ and $U_{b_e}$ conditional on $\bm U_{D_e} = \bm u_e$. For every $e \in E_k$, the conditional distributions $C_{a_e|D_e}$ can be expressed recursively as
\begin{align*}
	u_{a_e|D_e} = \frac{\partial C_{a_{e'}, b_{e'}|D_{e'}}(u_{a_{e'} \mid  D_{e'}}, u_{b_{e'} | D_{e'}} \mid  \bm{u}_{D_{e'}})}{\partial  u_{b_{e'} | D_{e'}}},
\end{align*}
where $e' \in E_{k - 1}$, $a_e = a_{e'}$, $b_{e'} \in D_e$ and $D_{e'} = D_e \setminus b_{e'}$. At the end of the recursion, the right hand side involves an edge $e' \in E_1$, for which
$u_{a_{e'} | D_{e'}} = u_{a_{e'}}$ and $u_{b_{e'} | D_{e'}} = u_{b_{e'}}$.

To make the model tractable, one commonly ignores the influence of $\bm{u}_{D_e}$ on the pair-copula density $c_{a_e,b_e|D_e}$. Under this assumption, the density simplifies to
\begin{align*}
	c(\bm u) = \prod_{k=1}^{d-1} \prod_{e \in E_k} c_{a_e,b_e|D_e} \left(u_{a_e|D_e},u_{b_e|D_e}\right).
\end{align*}
Since each pair-copula can be modeled separately, simplified vine copulas remain quite flexible. Conditional independence properties can be imposed by setting appropriate pair-copulas to the independence copula. This shall prove convenient when we construct Markovian time series models in \Cref{sec:markov}.
We further note that a similar factorization holds when some variables are discrete \citep[see,][Section 2.1]{stober2013regular}.  Although both continuity and the simplifying assumption are immaterial for our theoretical results, we will stick to the simplified, continuous case for convenience. For a more extensive introduction to vine copulas, we refer to \cite{AasKjersti2009Pcom} and \cite{czado2019analyzing}.

\subsection{Vine copula models for multivariate time series} \label{sec:literature}
 
Now suppose  $(\bm{X}_t)_{t = 1, \dots, n} = (X_{t, 1}, \dots, X_{t, d})_{t = 1, \dots, T}$ is a  stationary time series,  whose cross-sectional and temporal dependence structure we model by a vine copula. Throughout the paper, stationarity refers to strict stationarity. By fixing the stationary marginal distributions $F_1, \ldots, F_d$, it suffices to consider time series   $(\bm{U}_t)_{t = 1, \dots, T} = (U_{t, 1}, \dots, U_{t, d})^\prime_{t = 1, \dots, T}$ of marginally standard uniform variables,  where $U_{t,1} =F_1(X_{t,1}), \ldots,  U_{t,d} = F_d(X_{t,d})$. 
Note that the random variables $U_{t, j}$ in our model have two sub-indices. The first sub-index $t$ indicates the time point and the second sub-index $j$ determines the marginal variable.  In the time series context, each vertex of a vine's first tree is identified with a   tuple $(t, i)$, where $t$ is the time index and $i$ is the variable index. The vertex $(t, i)$   corresponds to the random variable $U_{t,i}$. In particular, the components of edge labels $e=(a_e,b_e|D_{e})$ (i.e., $a_e, b_e$ as well as elements of $D_e$)  are tuples. 
%\am{e.g., $(2, 4),(3, 4)| (3, 1),(3, 2),(3, 3)$}.

\begin{figure}[p]
	\centering
	\resizebox{\textwidth}{!}{
		\begin{tikzpicture}
			\tikzstyle{every node}=[ellipse, minimum width=20pt,align=center,scale=0.68]
			\node[shape=rectangle,draw=black] (1)  at (0,5)  {$U_{1, 1}$};
			\node[shape=rectangle,draw=black] (2)  at (0,4)  {$U_{1, 2}$};
			\node[shape=rectangle,draw=black] (3)  at (0,3)  {$U_{1, 3}$};
			\node[shape=rectangle,draw=black] (4)  at (0,2)  {$U_{1, 4}$};
			\node[shape=rectangle,draw=black] (6)  at (6,5)  {$U_{2, 1}$};
			\node[shape=rectangle,draw=black] (7)  at (6,4)  {$U_{2, 2}$};
			\node[shape=rectangle,draw=black] (8)  at (6,3)  {$U_{2, 3}$};
			\node[shape=rectangle,draw=black] (9)  at (6,2)  {$U_{2, 4}$};
			\node[shape=rectangle,draw=black] (11) at (12,5) {$U_{3, 1}$};
			\node[shape=rectangle,draw=black] (12) at (12,4) {$U_{3, 2}$};
			\node[shape=rectangle,draw=black] (13) at (12,3) {$U_{3, 3}$};
			\node[shape=rectangle,draw=black] (14) at (12,2) {$U_{3, 4}$};

			\begin{scope}
				\path [-] (1)  edge node[right] {} (2);
				\path [-] (2)  edge node[right] {} (3);
				\path [-] (3)  edge node[right] {} (4);
				\path [-] (1)  edge node[right] {} (6);
				\path [-] (6)  edge node[right] {} (7);
				\path [-] (7)  edge node[right] {} (8);
				\path [-] (8)  edge node[right] {} (9);
				\path [-] (6)  edge node[right] {} (11);
				\path [-] (11) edge node[left] {} (12);
				\path [-] (12) edge node[left] {} (13);
				\path [-] (13) edge node[left] {} (14);
			\end{scope}
		\end{tikzpicture}}
	\caption{Example for the first tree level of a four-dimensional M-vine on three time points.}
	\label{fig:mvine}
\end{figure}
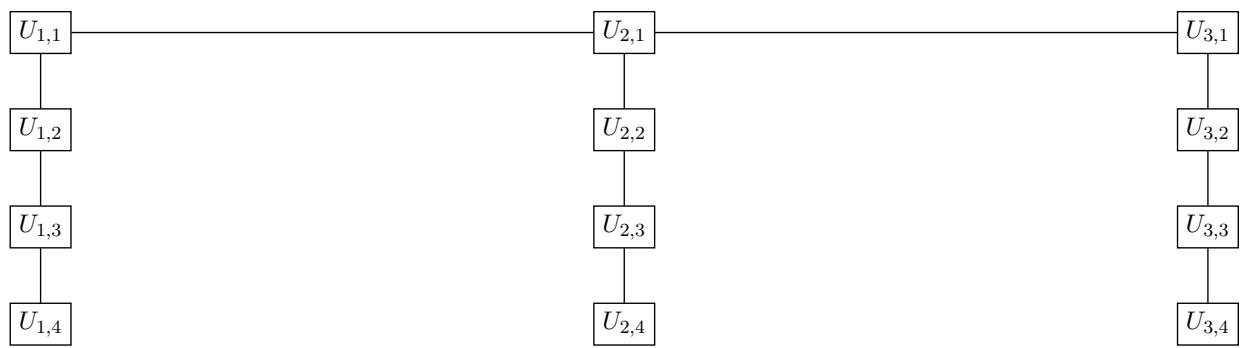

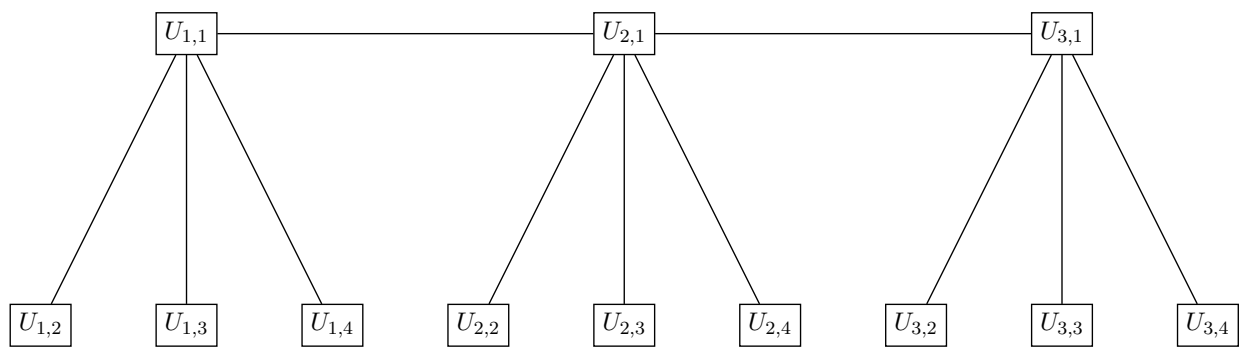
\begin{figure}[p]
	\centering
	\resizebox{\textwidth}{!}{
		\begin{tikzpicture}
			\tikzstyle{every node}=[ellipse, minimum width=20pt,align=center,scale=0.68]
			\node[shape=rectangle,draw=black] (1)  at (1.5,5)  {$U_{1, 1}$};
			\node[shape=rectangle,draw=black] (2)  at (0,2)  {$U_{1, 2}$};
			\node[shape=rectangle,draw=black] (3)  at (1.5,2)  {$U_{1, 3}$};
			\node[shape=rectangle,draw=black] (4)  at (3,2)  {$U_{1, 4}$};
			\node[shape=rectangle,draw=black] (6)  at (6,5)  {$U_{2, 1}$};
			\node[shape=rectangle,draw=black] (7)  at (4.5,2)  {$U_{2, 2}$};
			\node[shape=rectangle,draw=black] (8)  at (6,2)  {$U_{2, 3}$};
			\node[shape=rectangle,draw=black] (9)  at (7.5,2)  {$U_{2, 4}$};
			\node[shape=rectangle,draw=black] (11) at (10.5,5) {$U_{3, 1}$};
			\node[shape=rectangle,draw=black] (12) at (9,2) {$U_{3, 2}$};
			\node[shape=rectangle,draw=black] (13) at (10.5,2) {$U_{3, 3}$};
			\node[shape=rectangle,draw=black] (14) at (12,2) {$U_{3, 4}$};

			\begin{scope}
				\path [-] (1)  edge node[right] {} (2);
				\path [-] (1)  edge node[right] {} (3);
				\path [-] (1)  edge node[right] {} (4);
				\path [-] (1)  edge node[right] {} (6);
				\path [-] (6)  edge node[right] {} (7);
				\path [-] (6)  edge node[right] {} (8);
				\path [-] (6)  edge node[right] {} (9);
				\path [-] (6)  edge node[right] {} (11);
				\path [-] (11) edge node[left] {} (12);
				\path [-] (11) edge node[left] {} (13);
				\path [-] (11) edge node[left] {} (14);
			\end{scope}
		\end{tikzpicture}}
	\caption{Example for the first tree level of a four-dimensional COPAR on three time points.}
	\label{fig:copar}
\end{figure}

All existing regular vine models for multivariate time series follow the same idea \citep{BeareSeo2015, smith2015copula, brechmann2015copar}. There is a vine capturing cross-sectional dependence of $\bm U_t \in \R^d$ for all time points $t = 1, \dots, T$. The first trees of the cross-sectional structures at time $t$ and $t+1$ are then linked by one edge connecting a vertex from the structure at $t$ to one vertex from the one at $t+1$. Because the time series is stationary, it is reasonable to assume that the cross-sectional structure and the linking vertices are time invariant.
The existing models make specific choices for the cross-sectional structure and connecting edge. Their first tree for a four-dimensional model on three time points is illustrated in \Crefrange{fig:dvine}{fig:copar}.  Graphs of the first five trees $T_1, \ldots, T_5$ and additional details can be found in Section S2 of the supplementary materials. In short, there are three different models: 

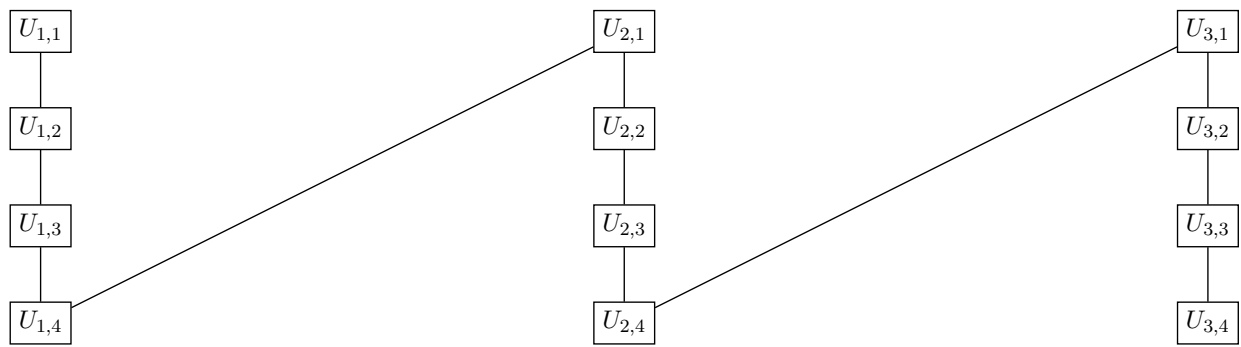
\begin{figure}[p]
	\centering
	\resizebox{\textwidth}{!}{
		\begin{tikzpicture}
			\tikzstyle{every node}=[ellipse, minimum width=20pt,align=center,scale=0.68]
			\node[shape=rectangle,draw=black] (1)  at (0,5)  {$U_{1, 1}$};
			\node[shape=rectangle,draw=black] (2)  at (0,4)  {$U_{1, 2}$};
			\node[shape=rectangle,draw=black] (3)  at (0,3)  {$U_{1, 3}$};
			\node[shape=rectangle,draw=black] (4)  at (0,2)  {$U_{1, 4}$};
			\node[shape=rectangle,draw=black] (6)  at (6,5)  {$U_{2, 1}$};
			\node[shape=rectangle,draw=black] (7)  at (6,4)  {$U_{2, 2}$};
			\node[shape=rectangle,draw=black] (8)  at (6,3)  {$U_{2, 3}$};
			\node[shape=rectangle,draw=black] (9)  at (6,2)  {$U_{2, 4}$};
			\node[shape=rectangle,draw=black] (11) at (12,5) {$U_{3, 1}$};
			\node[shape=rectangle,draw=black] (12) at (12,4) {$U_{3, 2}$};
			\node[shape=rectangle,draw=black] (13) at (12,3) {$U_{3, 3}$};
			\node[shape=rectangle,draw=black] (14) at (12,2) {$U_{3, 4}$};

			\begin{scope}
				\path [-] (1)  edge node[right] {} (2);
				\path [-] (2)  edge node[right] {} (3);
				\path [-] (3)  edge node[right] {} (4);
				\path [-] (4)  edge node[right] {} (6);
				\path [-] (6)  edge node[right] {} (7);
				\path [-] (7)  edge node[right] {} (8);
				\path [-] (8)  edge node[right] {} (9);
				\path [-] (9)  edge node[right] {} (11);
				\path [-] (11) edge node[left] {} (12);
				\path [-] (12) edge node[left] {} (13);
				\path [-] (13) edge node[left] {} (14);
			\end{scope}
		\end{tikzpicture}}
	\caption{Example for the first tree level of a four-dimensional D-vine on three time points. }
	\label{fig:dvine}
\end{figure}

\begin{itemize}
	\item \emph{D-vine} of \cite{smith2015copula}: (i) the cross-sectional structure is a D-vine, (ii) two cross-sectional D-vines  at time points $t$ and $t+1$ are connected at the two distinct vertices that lie at opposite borders of the D-vine trees. For the first tree, illustrated in \Cref{fig:dvine}, we can assume without loss of generality that these vertices are $(t,d)$ and $(t+1,1)$.  With these choices, there is only one global vine model satisfying the proximity condition, which is a long D-vine spanning all variables at all time points. 

	\item \emph{M-vine} of \cite{BeareSeo2015}: (i) the cross-sectional structure is a D-vine, (ii) two cross-sectional D-vines  at time points $t$ and $t+1$ are connected at one vertex that lies at the same border of the D-vine trees. For the first tree, illustrated in \Cref{fig:mvine}, we may assume without loss of generality that the vertices $(t,1)$ and $(t+1,1)$ are connected.  
	With the additional restriction that vertices of adjacent time points are connected first, this also uniquely fixes all further trees of the vine. 

	\item \emph{COPAR} of \cite{brechmann2015copar}: (i) the cross-sectional structure is a C-vine,  (ii) the first trees of two C-vines at time points $t$ and $t + 1$ are connected at the root vertex of the C-vine. For the first tree, illustrated in \Cref{fig:copar}, we may assume without loss of generality that vertices $(t,1)$ and $(t+1,1)$ are connected. This leaves a lot of flexibility for higher trees and the authors settled on a specific set of rules. In particular, the model contains all edges of a D-vine on the variables $U_{1, 1}, U_{2, 1}, \dots, U_{T, 1}$. 
\end{itemize}

There is obvious potential for generalization. First, we would like to allow for arbitrary R-vines in the cross-sectional structure. Second, we would like to connect two cross-sectional trees at arbitrary variables. Specific versions of such models were constructed in preliminary work by \citet{krueger2018} (called \emph{temporal vine}) and in unpublished work by Harry Joe. But where should we stop? In principle, we could take any $(T \times d)$-dimensional vine as a model for the vector $(\bm U_1, \dots, \bm U_T)$. We address this question comprehensively in the following section.

\section{Stationary vine copula models} \label{sec:svines}

The time series context is special. To facilitate inference, it is common to assume that the series is stationary, i.e., the distribution is invariant in time. When a time series is stationary, also its copula must satisfy certain invariances. This is a blessing and a curse: invariances reduce the complexity of the model, but not all vine structures guarantee stationarity under practicable conditions on the pair-copulas. We shall derive a generalization of the existing models that is maximally convenient in this sense. All proofs are collected in Section S5 of the supplementary material. 

\subsection{Stationary time series}
As explained in \Cref{sec:literature}, any stationary time series can be transformed into one with uniform marginal distributions by the probability integral transform.
To ease our exposition, we shall therefore assume uniform marginal distributions in what follows.
Let $(\mathcal{V}, \mathcal{C}(\mathcal{V}))$ be a vine copula model for the random vector $(\bm U_1^\prime, \dots, \bm U_T^\prime)^\prime \in \mathds{R}^{T \times d}$. The time series $\bm U_1, \dots, \bm U_T \in \mathds{R}^d$ is strictly stationary if and only if
$\bm U_{t_1}, \dots, \bm U_{t_m}$ and $\bm U_{t_1+ \tau}, \dots, \bm U_{t_m+ \tau}$ have the same joint distribution for all $ 1\leq t_1 < t_2 < ... < t_m \leq T$, $1 \le \tau \le T - \max_{j = 1}^m t_j$, and $1 \le m \le T$.

For vine copulas, this condition can involve intractable functional equations.
The reason is that only some pairwise (conditional) dependencies are explicit in the model. Explicit pairs are those that correspond to edges in the vine $\mathcal{V}$. All other dependencies are only implicit, i.e., they are characterized by the interplay of multiple pair-copulas. By only focusing on the explicit pairs, we see that \emph{translation invariance}, defined below, is a necessary condition for stationarity. Recall that, in the time series context,  the elements of the set $V_1$ of a vine are  tuples $(t, j)$ with time index $t=1, \ldots, T$ and variable index $j=1, \ldots, d$.
\begin{definition}[Translation invariance] \label{def:invar}
	A vine copula model  $(\mathcal{V}, \mathcal{C}(\mathcal{V}))$  on the set $V_ 1 = \{1, \dots, T\} \times \{1, \dots, d\} $ is called \emph{translation invariant} if $c_{a_e, b_e | D_e} = c_{a_{e'}, b_{e'} | D_{e'}}$ holds	for all edges $e, e' \in \bigcup_{k = 1}^{Td - 1} E_{k}$ for which there is $\tau \in \mathds{Z}$ such that
	\begin{align} \label{eq:timeshift}
		a_e = a_{e'} + ( \tau, 0), \quad b_e = b_{e'} + (\tau, 0), \quad D_e = D_{e'} + ( \tau, 0),
	\end{align}
	where the last equality is short for $D_e = \{v + (\tau, 0)\colon v \in D_{e'}\}$.
\end{definition}
\begin{remark}
 The notation $e = e' + (\tau, 0)$  will be used short for \eqref{eq:timeshift} and indicates a shift in time by $\tau$ steps. For example, if $a_{e'}=(t,j)$ then $a_{e}=(t+\tau, j)$ and similarly for $b_e$ and $D_e$. 
\end{remark}
Translation invariance was formally defined by \citet{BeareSeo2015}, and also used as an implicit motivation for the models of  \cite{brechmann2015copar} and \cite{smith2015copula}. For example, in the M-vine of \citet{BeareSeo2015} shown in \Cref{fig:mvine}, the copulas associated with the edges $(1, 1)-(1, 2)$ and $(2, 1)-(2, 2)$ must be the same.
As mentioned above, this is only a necessary condition for stationarity.
For all non-explicit pairs, stationarity requires more complex integral equations to hold.
Provided with sufficient computing power, they could be checked numerically for any given model. But even if it holds for a specific model, a slight change in the parameter of a single pair-copula may break it. This is problematic in practice.  Hence, the practically relevant vine structures are those for which translation invariance is also a sufficient condition for stationarity. These structures are characterized in what follows.

\subsection{Preliminaries}

First we need some graph theoretic definitions. The first is a version of Definition 6 of \citet{BeareSeo2015}.
\begin{definition}[Restriction of vines]
	Let $\mathcal{V} = (V_{k}, E_{k})_{k = 1}^{Td - 1}$ be a vine on $\{1 \dots, T\} \times \{1, \dots, d\} $ and 
	$V_1' = \{t, \dots, t + m\} \times \{1, \dots, d\}$ for some $t, m$ with $1 \le t \le T$, $0 \le m \le T - t$. For all $k \ge 1$, define $E_k' = E_k \cap \binom{V_{k}'}{2}$ and $V_{k + 1}' = E_{k}'$. Then the sequence of graphs $\mathcal{V}_{t, t + m} = (V_k', E_k')_{k = 1}^{(m + 1)d - 1}$ is called \emph{restriction of} $\mathcal V$ on the time points $t, \dots, t + m$.
\end{definition}
Simply put: to restrict a vine on time points $t$ to $t + m$, we delete all edges and vertices where time indices outside the range $[t, t+m]$ appear in the labels. Note that the restriction $\mathcal{V}_{t, t + m} = (V_k', E_k')_{k = 1}^{(m + 1)d - 1}$ is not necessarily a vine;  the graphs $(V_k', E_k')$ can be disconnected (hence, no trees). For example, if the first tree of the vine $\mathcal{V}$ contains a path $(1, i) - (3, i) - (2, i)$, the vertices $(1, i)$ and $(2, i)$ will be disconnected in $\mathcal{V}_{1, 2}$. 

The \emph{translation} of a vine $\mathcal V$ is obtained by shifting all vertices and edges in time by the same amount.
\begin{definition}[Translation of vines] \label{def:translation}
	Let  $m \ge 0$, and $\mathcal{V}= (V_{k}, E_{k})_{k = 1}^{(m + 1)d - 1}$ be vine on $\{t, \dots, t + m\} \times \{1, \dots, d\} $ and $\mathcal{V}' =(V_k', E_k')_{k = 1}^{(m + 1)d - 1}$ be a vine on $ \{s, \dots, s +m\} \times \{1, \dots, d\}$. We say that $\mathcal{V}$ is a \emph{translation} of $\mathcal{V}'$ (denoted by $\mathcal{V} \sim \mathcal{V}'$) if for all $k = 1, \dots, d - 1$ and edges $e \in E_{k}$, there is an edge $e' \in E_{k}'$ such that $e = e' + (t - s, 0)$ (and vice versa).
\end{definition}
\begin{remark}
	We shall call two edges $e, e'$  satisfying $e = e' + (\tau, 0)$ translations of another and write $e \sim e'$. This defines an equivalence relationship between edges.
\end{remark}
Additional illustrations of these concepts are provided in Section S1 of the supplementary material.

\subsection{Stationary vines}
The last definitions are key to ensure stationarity in vine copula models. If for all time points $s, t$ and gap $m$, the restriction of a vine on $t, \dots, t + m$ is a translation of the restriction on $s, \dots, s + m$, translation invariance will guarantee stationarity. 
\begin{theorem} \label{thm:svines}

	Let $\mathcal{V}$ be a vine on the set $V_ 1 = \{1, \dots, T\} \times  \{1, \dots, d\}$. Then the following statements are equivalent:
	\begin{enumerate}[label = (\roman*)]
		\item \label{cond:stat}  The vine copula model $(\mathcal{V}, \mathcal{C}(\mathcal{V}))$  is stationary for all translation invariant choices of $\mathcal{C}(\mathcal{V})$.
		\item \label{cond:equi_graph} There are vines $\mathcal{V}^{(m)}, m = 0, \dots, T - 1$, defined on $\{1, \dots,  m + 1\} \times \{1, \dots, d\}$, such that for all $m = 0, \dots, T - 1$,  $1 \le t \le T - m$,
		      \begin{align} \label{eq:vstat}
			      \mathcal{V}_{t, t + m} \sim \mathcal{V}^{(m)}.
		      \end{align}
	\end{enumerate}
\end{theorem} 
An important word in condition \ref{cond:stat} is \emph{all}. There are vines violating \ref{cond:equi_graph} that are stationary for a specific choice of $\mathcal C(\mathcal V)$. For example, $c_e \equiv 1$ for all edges always leads to a stationary model. But these structures are impractical, because they limit the choices of $\mathcal C(\mathcal V)$ to a restrictive and unknown set.
Condition \ref{cond:equi_graph} can be seen as a graph theoretic notion of stationarity for vine structures.
\begin{definition}[Stationary vines or S-vines]
	A vine $\mathcal{V}$ on the set $V_ 1 =  \{1, \dots, T\} \times \{1, \dots, d\}$ is called \emph{stationary} or if it satisfies condition \ref{cond:equi_graph} of \Cref{thm:svines}.
\end{definition}

S-vines have a distinctive structure. There is a $d$-dimensional vine $\mathcal{V}^{(0)}$ that contains only pairs for cross-sectional dependence. We will therefore call  $\mathcal{V}^{(0)}$ the \emph{cross-sectional structure} of $\mathcal{V}$. Next, there is a $2d$-dimensional vine  $\mathcal{V}^{(1)}$ that nests two duplicates of $\mathcal{V}^{(0)}$. Besides these cross-sectional parts, the vine contains $d^2$ pairs for dependence across two subsequent time points that are not yet constrained by translation invariance. A similar principal applies for vines $\mathcal V^{(m)}, m \ge 2$, with $d^2$ unconstrained edges entering in every step (i.e., going from  $\mathcal V^{(m-1)}$ to $\mathcal V^{(m)}$). 
An illustrative example for a five-dimensional S-vine on three time points is given in Section S2.5 of the  supplementary materials.

It is easy to check that M-vines and D-vines of \citet{BeareSeo2015} and \citet{smith2015copula} are stationary. As the following example shows, the structure of the COPAR model of \citet{brechmann2015copar} is not stationary, however. In particular, the graph $\mathcal{V}_{t, t+1}$ is not a vine for $t \ge 2$ because the second level of the restricted graph is disconnected. This poses an additional constraint on the choice of pair copulas that went seemingly unnoticed.

\begin{example} \label{ex:COPAR}
	Let us illustrate the tricky part of the proof  \Cref{thm:svines} with the COPAR model for  $d = 2$, $T = 3$. The second tree of the model is given in \Cref{fig:copar2}, see also Figure S11 in the supplement for the remaining trees. For simplicity, we assume that all pair-copulas in trees $k = 1$ and $k \ge 3$ are independence copulas. The restriction $\mathcal V_{2, 3}$ of the model is obtained by 
	deleting all vertices and edges where a time index 1 occurs. Clearly, $\mathcal V_{2, 3}$  is not a vine, because the vertex $(U_{2, 1}, U_{2, 2})$ is disconnected from the others. Now let us see why this is problematic. The joint copula density of vertices
	$(U_{2, 1}, U_{2, 2})$ and $(U_{2, 1}, U_{3, 1})$ equals the product of copulas associated with the edges along the path joining them, integrating over all intermediate vertices. That is,
	\begin{align*}
		c_{(2, 2), (3, 1) \mid (2, 1)}(u, v) = \int_0^1 c_{(1, 1), (2, 2) \mid (2, 1)}(w, u) c_{(1, 1), (3, 1) \mid (2, 1)}(w, v)  d w.
	\end{align*}
	By translation invariance, it must further hold 
	\begin{align*}
		c_{(1, 2), (2, 1) \mid (1, 1)}(u, v) = \int_0^1 c_{(1, 1), (2, 2) \mid (2, 1)}(w, u) c_{(1, 1), (3, 1) \mid (2, 1)}(w, v)  d w.
	\end{align*}
	The copula on the left hand side is an explicit dependence in the model, because it is associated with an edge in the graph (the leftmost one). Thus the equation contains three pair-copulas of the model that are not constrained by translation invariance. For most combinations of pair-copulas, the equality does not hold and the model is not stationary.
\end{example}

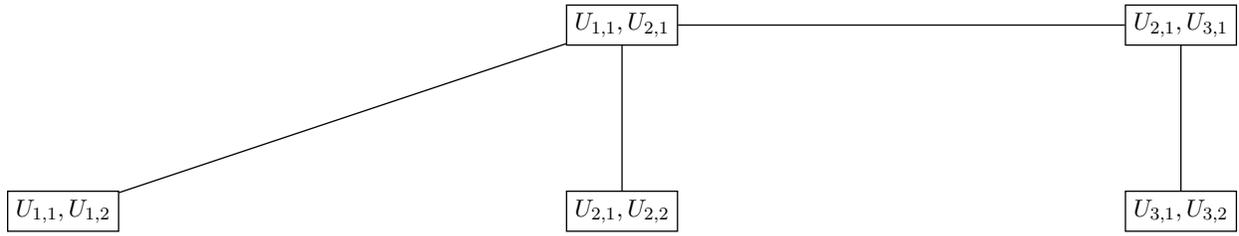
\begin{figure}[t]
	\centering
	\resizebox{\textwidth}{!}{
		\begin{tikzpicture}
			\tikzstyle{every node}=[ellipse, minimum width=20pt,align=center,scale=0.68]
			\node[shape=rectangle,draw=black] (1)  at (0, 0)  {$U_{1, 1}, U_{1, 2}$};
			\node[shape=rectangle,draw=black] (2)  at (6,2)  {$U_{1, 1}, U_{2, 1}$};
			\node[shape=rectangle,draw=black] (3)  at (6,0)  {$U_{2, 1}, U_{2, 2}$};
			\node[shape=rectangle,draw=black] (4)  at (12,2)  {$U_{2, 1}, U_{3, 1}$};
			\node[shape=rectangle,draw=black] (5) at (12,0) {$U_{3, 1}, U_{3, 2}$};

			\begin{scope}
				\path [-] (1)  edge node[right] {} (2);
				\path [-] (2)  edge node[right] {} (3);
				\path [-] (2)  edge node[right] {} (4);
				\path [-] (4)  edge node[right] {} (5);
			\end{scope}
		\end{tikzpicture}}
	\caption{Example for the second tree level of a COPAR model with $d = 2, T = 3$.}
	\label{fig:copar2}
\end{figure}

\subsection{An explicit characterization of stationary vines} \label{sec:explicit}

Stationary vines can also be characterized more explicitly. Somewhat surprisingly, it suffices to pick a cross-sectional structure $\mathcal{V}^{(0)}$ and two permutations of $(1, \dots, d)$.   The permutations determine how the first $d$ trees of the cross-sectional structures are connected across two adjacent time points. The first permutation, called \emph{in-vertices}, determines how trees of one cross-sectional vine are connected to trees of the preceding time point; the second permutation, called \emph{out-vertices}, regulates the connections to the succeeding time point. The permutations are constrained by the choice of cross-sectional structure.
For simplicity, we omit the time index $t$ in the following definition.
\begin{definition}[Compatible permutations] \label{def:compatible}
	We call a permutation $(i_1, \dots, i_d)$ of $(1, \dots, d)$ \emph{compatible} with a vine $\mathcal{V}$ on $\{1, \dots, d\}$  if for all $k = 2, \dots, d$, there is an edge $e \in E_{k - 1}$ with conditioned set $\{i_k, i_r\}$ and conditioning set $\{i_{1}, \dots, i_{k - 1} \} \setminus i_r$ for some $r \in \{1, \dots, k -1\}$.
\end{definition}
The first index of the permutation ($i_1$) is not constrained by compatibility, but the remaining ones are. A permutation is only compatible if the vine contains the edge $\{i_2, i_1\}$ in the first tree. Further, the vine must contain an edge with either a) conditioned set $\{i_3, i_1\}$ and conditioning set $\{i_2\}$, or b) an edge with conditioned set $\{i_3, i_2\}$ and conditioning  set $\{i_1\}$, etc. Because $i_1$ is unconstrained, this also implies that any $d$-dimensional vine has at least $d$ compatible permutations (see \Cref{lem:compatible} below).
The following theorem shows that S-vines are characterized by the cross-sectional structure and two compatible permutations.

\begin{theorem} \label{thm:explicit}
	A vine $\mathcal{V}$ on  $\{1, \dots, T\} \times \{1, \dots, d\}$ is stationary if and only if
	\begin{enumerate}[label = (\roman*)]
		\item \label{cond:cs} there is a vine $\mathcal{V}^{(0)}$ on $\{0\} \times \{1, \dots, d\}$ such that $\mathcal{V}_{t, t} \sim \mathcal{V}^{(0)}$, for all $1 \le t \le T$, 
		\item \label{cond:con} there are two permutations $(i_1, \dots, i_d)$ and $(j_1, \dots, j_d)$ compatible with $\mathcal{V}^{(0)}$, such that
		      \begin{align*}
			      E_k = & \bigcup_{t = 1}^{T} E_k^{(0)} + (t, 0) \;\; \cup                                                                                                                                                                    \\
			            & \bigcup_{t = 1}^{T - 1} \bigcup_{r = 1}^{k} \biggl\{e \colon a_e = (t, i_{k +1 - r}), b_e = (t + 1, j_{r}), D_e = \bigcup_{s = 1}^{k - r} \{(t, i_{s })\} \cup \bigcup_{s = 1}^{r - 1} \{(t + 1, j_{s })\} \biggr\}
		      \end{align*}
		      for $k = 1, \dots, d$.
	\end{enumerate}
\end{theorem}

 As mentioned earlier, S-vines generalize previous models: 
	\begin{enumerate}[label = (\roman*)]
		\item If $\mathcal V^{(0)}$ is a D-vine and $(i_1, \dots, i_d) = (j_1, \dots, j_d)$, we obtain the M-vine of \citet{BeareSeo2015}. 
		\item If $\mathcal V^{(0)}$ is a D-vine and $(i_1, \dots, i_d) = (j_d, \dots, j_1)$, we obtain the long D-vine of \citet{smith2015copula}. 
		\item If we choose $i_s, j_s$, iteratively for $s \ge 2$ as the smallest compatible indices, we obtain the T-vine model of \citet{krueger2018}.
	\end{enumerate}
Compared to the models of \citet{BeareSeo2015} and \citet{smith2015copula}, S-vines do not require the cross-sectional structure to be D-vines. Further, we have some degree of freedom in how we connect variables across different time points. This relaxation can improve both interpretability and performance of associated copula models, as illustrated in \Cref{sec:application} and Section S5 of the supplementary materials.
The explicit characterization of \Cref{thm:explicit} also makes it easy to establish conditions for existence and uniqueness of a stationary vine. The first step is to show that a compatible permutation always exists.
\begin{lemma} \label{lem:compatible}
	For $d$-dimensional vine $\mathcal{V}$ and any $i_1 \in \{1, \dots, d\}$, there exists at least one permutation $(i_1, \dots, i_d)$ compatible with $\mathcal V$.
\end{lemma}

Now the following result is an immediate consequence of \Cref{thm:explicit} and \Cref{lem:compatible}.
\begin{corollary} \quad \\[-12pt]
	\begin{enumerate}[label = (\roman*)]
		\item \label{thm:exist} (Existence)  For any vine $\mathcal{V}^*$, there exists a stationary vine with cross-sectional structure $\mathcal V^{(0)} = \mathcal{V}^*$.
		\item \label{thm:unique} (Uniqueness) Given a cross-sectional structure $\mathcal V^{(0)}$ and two sequences of compatible in- and out-vertices, the stationary vine is unique.
	\end{enumerate}
\end{corollary}

\subsection{Markovian models} \label{sec:markov}

Stationarity is a convenient property because it limits model complexity.
An arbitrary vine copula model for $\bm U_1, \dots, \bm U_T \in [0,1]^d$ requires to specify (or estimate) $Td(Td-1)/2 = O(T^2d^2)$ pair-copulas. In a stationary vine copula model, cross-sectional dependencies are associated with the same pair copulas for each time point. Similarly, serial dependencies are modeled with identical copulas for each lag. This significantly reduces the number of free pair-copulas in the model. We only need to specify $(T - 1)d^2+d(d-1)/2 = O(Td^2)$ of them, all other pair-copulas are constrained by translation invariance.
When the time series contains more than a few  dozen time points, this is still  too much.
Most popular time series models also satisfy the Markov property:
\begin{definition}
	A time series $\bm U_1, \dots, \bm U_T \in [0,1]^d$ is called Markov (process) of order $p$ if for all $\bm u \in \R^d$,
	\begin{align*}
		\Pr\bigl(\bm U_{t} \le \bm u \mid \bm U_{t - 1}, \dots, \bm U_{1} \bigr) = \Pr\bigl(\bm U_{t} \le \bm u \mid \bm U_{t - 1}, \dots, \bm U_{t - p}).
	\end{align*}
\end{definition}
The Markov property limits complexity further. For the $M$-vine model, \citet[Theorem 4]{BeareSeo2015} showed that it is equivalent to an independence constraint on the pair-copulas, used similarly by the Markovian models of \cite{brechmann2015copar} and \cite{smith2015copula}. The same arguments apply for the general class of stationary vines. 
\begin{theorem} \label{thm:markov}
	A vine copula model $(\mathcal V, \mathcal C(\mathcal V))$ on a stationary vine $\mathcal V$ is Markov of order $p$ if and only if  $c_e \equiv 1$ for all $e \notin \mathcal V_{t, t + p}$, $t = 1, \dots, T - p$.
\end{theorem}

In a stationary Markov model of order $p$, the independence copula is assigned to all edges reflecting serial dependence of lags larger than $p$. This reduces the number of distinct pair copulas further to $pd^2+d(d-1)/2 = O(pd^2)$.
\Cref{tab-number-of-copulas} shows the number of distinct copulas in an unrestricted model for the full time series, a stationary vine model, and a stationary vine model with Markov order $p=1,2$. We can see a significant reduction when imposing stationarity and the Markov property. 
\begin{table} 
		\begin{tabular}{l|r|r|r}
		                 & \textbf{$T=100, d=5$}                                           & \textbf{$T=100$, $d=20$} & \textbf{$T=1\,000$, $d=20$} \\
		\hline general model              & 124\,750                                                        & 1\,999\,000              & 199\,990\,000               \\
		\hline stationary model           & 2\,485                                                          & 39\,790                  & 399\,790                    \\

		\hline stationary Markov(2) model & 60                                                              & 990                      & 990                         \\
		\hline stationary Markov(1) model & 35                                                              & 590                      & 590
	\end{tabular}
	\caption{Number of distinct pair-copulas to specify for different vine models.}
	\label{tab-number-of-copulas}
\end{table}

\section{Parameter estimation} \label{sec:estsel}

%Estimation of copula-based Markov chains was discussed earlier by \citet{chen2006estimation} and \citet{chen2009} for $d =1$, $p = 1$. \citet{remillard2012copula} extended their results for $d \ge 1$ for the joint maximum-likelihood estimator (MLE) in a semiparametric model. 

Joint maximum-likelihood is unpopular for vine copula models, because they have many parameters even in moderate dimension. \citet{BeareSeo2015} discussed a version of the popular step-wise maximum likelihood estimator \citep{AasKjersti2009Pcom} for M-vine copula models, but without theoretical guarantees. We shall introduce such a method for the more general class of stationary vines and prove its validity, allowing for either parametric and nonparametric marginal models. The step-wise method is fast also for large models, but incurs a small loss in efficiency according to \citet{hobaekhaff2013}. 

\subsection{Estimation of marginal models} \label{sec:est_margins}

We follow the common practice to estimate marginal models first. Given estimates $\wh F_1, \dots, \wh F_d$ of the marginal distributions, the copula parameters can then be estimated based on `pseudo-observations' $\wh U_{t, j} = \wh F_{j}(X_{t, j})$, $t = 1, \dots, T$, $j = 1, \dots d$. 

Suppose we are given parametric models $f_j(\cdot; \bm \eta_j)$, $j = 1, \dots, d$, for the marginal densities. Then the parameters can be estimated by the maximum-likelihood-type estimator 
\begin{align} \label{eq:ll_marg}
	\wb\eta_j & = \arg\max_{\bm \eta_j} \sum_{t = 1}^T \ln f_j(X_{t, j}; \bm \eta_j), \quad j = 1, \dots, d.
\end{align}
Given estimates of the marginal parameters, we then generate pseudo-observations from the copulas model via $\wh U_{t, j}^{(P)} = F_{j}(X_{t, j}; \wb \eta_j)$, $t = 1, \dots, T$, $j = 1, \dots, d.$

We can also consider semiparametric copula models by estimating the marginal distributions by   empirical distribution functions $\wh F_{j}(x) = \sum_{t = 1}^T \ind(X_{t, j} \le x) / (T + 1)$. This leads to the pseudo-observations $\wh U_{t, j}^{(SP)} = \wh F_{j}(X_{t, j})$, $t = 1, \dots, T$, $j = 1, \dots, d$. In what follows, pseudo-observations $\wh U_{t, j}$ are used generically in place of $\wh U_{t, j}^{(P)}$ or $\wh U_{t, j}^{(SP)}$.

\subsection{Estimation of copula parameters} \label{sec:est_copula}

For all edges $e$ in the vine, let $c_{[e]}(\cdot; \bm \theta_{[e]})$ be a parametric model with parameter $\bm \theta_{[e]}$. Because of translation invariance, many of the edges must have the same families and parameters. This is reflected by the notation $[e]$ which assigns a family $c_{[e]}(\cdot; \bm \theta_{[e]})$ and parameter $\bm \theta_{[e]}$ for the entire equivalence class $[e] = \{e'\colon e' \sim e\}$.
Recall from \Cref{sec:vine_copulas} that the joint density of the model involves conditional distributions of the form $C_{a_e|D_e}$ which can be expressed recursively. We again write $C_{a_{[e]} | D_{[e]}}$ to highlight the invariance of the function with respect to shifts in time.
For an edge $e \in E_k$, denote by $S_a(e)$ the set of edges $e' \in \{E_1, \dots, E_{k - 1}\}$ involved in this recursion and $\bm \theta_{S_a([e])} = (\bm \theta_{[e']})_{e' \in S_a(e)}$. Finally, write $[E_k] = \{[e]\colon e \in E_k\}$, $\bm \theta_{[E_k]} = (\bm \theta_{[e]})_{[e] \in [E_k]}$ and $\bm \theta = (\bm \theta_{[E_k]})_{k = 1}^{(p + 1)d - 1}$ as the stacked parameter vector.

The joint (pseudo-)log-likelihood of a stationary vine copula model for $(\bm X_1, \dots, \bm X_T)$ is
\begin{align*}
	\ell(\bm \theta) & = \sum_{k = 1}^{d(p + 1) - 1} \sum_{e \in  E_k} \ln c_{[e]}\bigl\{C_{a_{[e]}\mid D_{[e]}}(\wh U_{a_e} \mid \wb U_{D_e}; \bm \theta_{S_a([e])}),C_{b_{[e]}\mid D_{[e]}}(\wh U_{b_e} \mid \wb U_{D_e}; \bm \theta_{S_b([e])|D_{[e]})}) ; \bm \theta_{[e]}\bigr\},
\end{align*}
where $(\wb U_1, \dots, \wb U_T)$ can be either the parametric or nonparametric pseudo-observations.
The joint MLE, $\arg\max_{\bm \theta}\ell(\bm \theta)$, is often too demanding. The step-wise MLE of \citet{AasKjersti2009Pcom} estimates the parameters of each pair-copula separately, starting from the first tree. We can adapt it to the setting of a Markov process of order $p$:
for $k = 1, \dots, d(p + 1) - 1$ and every $e' \in E_k$
\begin{align}  \label{eq:ll_step}
	\wb \theta_{[e']} = \arg\max_{\bm \theta_{[e']}} \sum_{e \sim e'} \ln c_{[e]}\bigl\{C_{a_{[e]}\mid D_{[e]}}(\wh U_{a_e} \mid \wb U_{D_e}; \wb \theta_{S_a([e])},C_{b_{[e]}\mid D_{[e]}}(\wh U_{b_e} \mid \wb U_{D_e}; \wb \theta_{S_b([e])}) ; \bm \theta_{[e']}\bigr\}. 
\end{align}
The vectors $\wb \theta_{S(a_{[e]})}$, $\wb \theta_{S(b_{[e]})}$ in \eqref{eq:ll_step} only contain parameter estimates from previous trees, i.e., ones that were already found in earlier iterations. 

\subsection{Asymptotic results}

In what follows, we establish consistency and asymptotic normality of the parametric and semiparametric parameter estimates.
Their proofs are given in Section S6 of the supplementary material. All results are derived as consequences of the more general  \Cref{thm:aux_consistency} and \Cref{thm:aux_an} given in \Cref{sec:gen}. A discussion of the results is given at the end of this section.

We shall assume in the following that the series $(\bm X_t)_{t \in \mathds Z}$ is strictly stationary and absolutely regular. For $p = 1, d = 1$, it is sufficient that the copula density is strictly positive on a set of measure 1 \citep[Proposition 2]{longla2012some}. The proof can be easily extended to $p, d \ge 1$, which leads to the mild sufficient condition that all pair-copula densities are strictly positive on $(0, 1)^2$. \change{In what follows, $\| \cdot \|$ denotes the Euclidean norm.}

\subsubsection{Parametric estimator}  \label{sec:asym-p}

To state the asymptotic results, it is convenient to introduce some more notation. For S-vines of Markov order $p$, one can check that any edge $e \in E_k, 1 \le k \le d(p + 1) - 1,$ the set $\{a_e, b_e, D_e\}$ only contains variables at most $p$ time points apart.  More precisely, if  $a_e = (t_1, j_1)$, $b_e = (t_2, j_2)$ and $t = \min\{t_1, t_2\}$, then $(x_{a_e}, x_{b_e}, \bm x_{D_e})$ is a sub-vector of  $(\bm x_t, \dots, \bm x_{t + p})$. 
Hence, we denote
\begin{align*}
F_{[e], 1, \bm \eta, \bm \theta}(\bm x_1, \dots, \bm x_{1 + p})	
	&= F_{a_{[e]}\mid D_{[e]}}(x_{a_{[e]}} \mid \bm x_{D_{[e]}}; \bm \eta, \bm \theta_{S_a([e])})  , \\
 F_{[e], 2, \bm \eta, \bm \theta}(\bm x_1, \dots, \bm x_{1 + p})
	&= 	F_{b_{[e]}\mid D_{[e]}}(x_{b_{[e]}} \mid \bm x_{D_{[e]}}; \bm \eta, \bm \theta_{S_b([e])}),
\end{align*}
and
\begin{align*}
	\bm s_{j, \bm \eta_j}(\bm x_t, \dots, \bm x_{t + p}) &= \nabla_{\bm \eta_j} \ln f_j(x_{t, j}), \\
	\bm s_{[e], \bm \eta, \bm \theta}(\bm x_t, \dots, \bm x_{t + p}) &=  \nabla_{\bm \theta_{[e]}} \ln c_{[e]}\{F_{[e], 1, \bm \eta, \bm \theta}(\bm x_t, \dots, \bm x_{t + p}), F_{[e], 2, \bm \eta, \bm \theta}(\bm x_t, \dots, \bm x_{t + p}); \bm \theta_{[e]}\},
\end{align*}
and define
\begin{align*}
	\bm \phi_{\bm \eta, \bm \theta}^{(P)} &= \begin{pmatrix}
		\bigl(\bm s_{j, \bm \eta_j} \bigr)_{j = 1, \dots, d} \\
		\bigl(\bm s_{[e], \bm \eta, \bm \theta}\bigr)_{[e] \in [E_k], k = 1,\dots, (p+1)d}
	\end{pmatrix}.
\end{align*}
Up to a finite number of terms, the parametric step-wise MLE $(\wb \eta^{(P)}, \wb \theta^{(P)})$ is then defined as the solution of the estimating equation 
\begin{align*}
	\frac 1 {T} \sum_{t = 1}^{T} \bm \phi_{\bm \eta, \bm \theta}^{(P)}(\bm X_t, \dots, \bm X_{t + p}) = 0.
\end{align*}
To allow for misspecified parametric models, we further define pseudo-true values $(\bm \eta^*, \bm \theta^*)$ via 
$$\E\{\bm \phi_{\bm \eta^*, \bm \theta^*}^{(P)}(\bm X_1, \dots, \bm X_{1 + p})\} = 0,$$
and note that they agree with the true parameters if the model is correctly specified. Here and in the sequel,  all expectations are taken with respect to unknown true distribution of $\bm X_1, \dots, \bm X_{1 + p}$.

We impose the following regularity conditions:

\begin{enumerate}[label = (P\arabic*)]
	\item \label{cond:P_identify}
	The pseudo-true values $(\bm \eta^*, \bm \theta^*)$ lie in the interior of $\mathcal H \times \Theta$ and for every $\epsilon > 0$, 
	$$\inf_{\|\bm \eta - \bm \eta^*\| > \epsilon} \inf_{\|\bm \theta - \bm \theta^*\| > \epsilon} \bigl\| \E\bigl\{\bm \phi_{\bm \eta, \bm \theta}^{(P)}(\bm X_1, \dots, \bm X_{1 + p})\bigr\} \bigr\| > 0.$$

	\item  \label{cond:P_derivs}
	The function $\bm \phi_{\bm \eta, \bm \theta}^{(P)}$ is continuously differentiable with respect to $(\bm \eta, \bm \theta)$ and satisfies
	\begin{align*}
		& \E\left[\sup_{\bm \eta \in \tilde{\mathcal H}} \sup_{ \bm \theta \in \tilde{\Theta}} \biggl\{ \bigl\|\bm  \phi_{\bm \eta, \bm \theta}^{(P)}(\bm X_1, \dots, \bm X_{1 + p})\bigr\| + \bigl\| \nabla_{(\bm \eta, \bm \theta) }^\prime \bm \phi_{\bm \eta, \bm \theta}^{(P)}(\bm X_1, \dots, \bm X_{1 + p})\bigr\| \biggr\}\right] < \infty
	\end{align*} 
	for any compact $\tilde{\mathcal H} \times \tilde \Theta \subseteq {\mathcal H} \times  \Theta$.

	\item \label{cond:P_hessinv} The matrix  $\bm J_{\bm \eta^*, \bm \theta^*} = \E\bigl\{ \nabla_{(\bm \eta^*, \bm \theta^*)}^\prime \bm  \phi_{\bm \eta^*, \bm \theta^*}^{(P)}(\bm X_1, \dots, \bm X_{1 + p}) \bigr\} $	is invertible.

	\item \label{cond:P_mixing2}
	The $\beta$-mixing coefficients of  $(\bm X_t)_{t \in \mathds Z}$ satisfy 
	$\sum_{t = 0}^\infty \int_0^{\beta(t)} Q^2(u) du < \infty,$
	where $Q$ is the inverse survival function of $\|\bm \phi_{\bm \eta^*, \bm \theta^*}^{(P)}(\bm X_1, \dots, \bm X_{t + p})\|$.
\end{enumerate}

Condition \ref{cond:P_identify} ensures identifiability of the model parameters, \ref{cond:P_derivs} and \ref{cond:P_hessinv} are standard regularity condition for maximum-likelihood methods. Condition \ref{cond:P_mixing2} quantifies a trade-off between moments of the `score' function $\bm \phi_{\bm \eta^*, \bm \theta^*}^{(P)}$ and the mixing rate, see the discussion in \Cref{sec:asym-discussion}.

\begin{theorem}	\label{thm:P_consistency}
	Under \ref{cond:P_identify}--\ref{cond:P_derivs}, it holds $(\wb \eta^{(P)}, \wb \theta^{(P)}) \to_p (\bm \eta^*, \bm \theta^*)$.
\end{theorem}

\begin{theorem} \label{thm:P_an}
	Under \ref{cond:P_identify}--\ref{cond:P_mixing2}, it holds
	$\|(\wb \eta^{(P)}, \wb \theta^{(P)}) - (\bm \eta^*, \bm \theta^*)\| = O_p(T^{-1/2})$ and
	\begin{align*}
		\sqrt{T}\begin{pmatrix}
			\wb \eta^{(P)} - \bm \eta^* \\
			\wb \theta^{(P)} - \bm \theta^*
		\end{pmatrix}\stackrel{d}{\to} \mathcal N\bigl\{\bm 0, \bm J^{-1}_{\bm \eta^*, \bm \theta^*} \bm I_{\bm \eta^*, \bm \theta^*} (\bm J^{-1}_{\bm \eta^*, \bm \theta^*})^\prime\bigr\},
	\end{align*}
	where $\bm I_{\bm \eta^*, \bm \theta^*} = \sum_{t = 1}^\infty \{1 + \ind(t \ge 2)\}\E\bigl\{\bm  \phi_{\bm \eta^*, \bm \theta^*}^{(P)}(\bm X_1, \dots, \bm X_{1 + p}) \bm  \phi_{\bm \eta^*, \bm \theta^*}^{(P)}(\bm X_t, \dots, \bm X_{t + p})^\prime \bigr\}$.
\end{theorem}
Note that  \emph{iid} models are included as a special case:
\begin{corollary} \label{cor:P}
    If $(\bm X_{t})_{t \in \N}$ is \emph{iid} and $p = 0$, then \Cref{thm:P_consistency} and \Cref{thm:P_an} hold with  $\bm I_{\bm \eta^*, \bm \theta^*} = \E\bigl\{\bm  \phi_{\bm \eta^*, \bm \theta^*}^{(P)}(\bm X_1) \bm  \phi_{\bm \eta^*, \bm \theta^*}^{(P)}(\bm X_1)^\prime \bigr\}$. If the latter expectation exists, condition \ref{cond:P_mixing2} can be dropped.
\end{corollary}

\subsubsection{Semiparametric estimator} \label{sec:asym-sp}

Similarly to the parametric case, we define
\begin{align*}
	C_{[e], 1, \bm \theta}(\bm u_t, \dots, \bm u_{t + p}) 
	&= C_{a_{[e]}\mid D_{[e]}}(u_{a_e} \mid \bm u_{D_e}; \bm \theta_{S_a([e])}) , \\
	C_{[e], 2, \bm \theta}(\bm u_t, \dots, \bm u_{t + p}) &=  C_{b_{[e]}\mid D_{[e]}}(u_{a_e} \mid \bm u_{D_e};  \bm \theta_{S_b([e])}),
\end{align*}
and
\begin{align*}
	\bm s_{[e], \bm \theta}(\bm u_1, \dots, \bm u_{1 + p}) &=  \nabla_{\bm \theta_{[e]}} \ln c_{[e]}\bigl\{C_{[e], 1, \bm \theta}(\bm u_t, \dots, \bm u_{t + p}), C_{[e], 2, \bm \theta}(\bm u_t, \dots, \bm u_{t + p}) ; \bm \theta_{[e]}\bigr\}.
	% \partial_{(t', j)} \bm s_{[e], \bm \theta}(\bm u_1, \dots, \bm u_{1 + p}) &=  \partial \bm s_{[e], \bm \theta}(\bm u_1, \dots, \bm u_{1 + p}) / \partial u_{t', j}.
\end{align*}
For a \change{generic} vector of functions $\bm G = (G_1, \dots, G_d)$ write $\bm G(\bm x) = \bigl(G_1(x_1), \dots, G_d(x_d)\bigr)$.
Let $\wb F = (\wh F_1, \dots, \wh F_d)$  be the vector of empirical distribution functions and  $\bm F = (F_1, \dots, F_d)$ be the true distributions.
Setting $ \bm \phi^{(SP)}_{\bm \theta} = (\bm s_{[e], \bm \theta})_{[e] \in [E_k], k = 1, \dots, (p + 1)d}$, the semiparametric step-wise MLE $\wb \theta^{(SP)}$ is then defined as the solution of the estimating equation 
\begin{align*}
	\frac 1 {T} \sum_{t = 1}^{T} \bm \phi_{\bm \theta}^{(SP)}\{\wb F(\bm X_t), \dots, \wb F(\bm X_{t + p})\} = 0.
\end{align*}
Further, we define the pseudo-true value $\bm \theta^*$ via $$\E\bigl[\bm \phi_{\bm \theta^*}^{(SP)}\{\bm F(\bm X_1), \dots, \bm F(\bm X_{1 + p})\} \bigr] = 0.$$

For $u \in (0,1)$ and some $\gamma \in [0, 1)$, define the weight function $w(u) = u^\gamma(1 - u)^\gamma$ and set $\mathcal F_\delta = \bigtimes_{j = 1}^d \{G\colon \R \to [0, 1], \sup_x |F_j(x) - G(x)|/w\{F_j(x)\} \le \delta\}$.

\begin{enumerate}[label = (SP\arabic*)]
	% \item \label{cond:SP_margin}
	% The estimator $\wb F$ satisfies $\max_{1 \le j \le d}\sup_{x}| \wh F_j(x) - F_j(x)|/w(x) = o_p(1)$.

	\item \label{cond:SP_identify}
	The pseudo-true value $\bm \theta^*$ lies in the interior of $\Theta$ and for every $\epsilon > 0$, 
	$$ \inf_{\|\bm \theta - \bm \theta^*\| > \epsilon} \bigl\| \E\bigl[\bm\phi_{\bm \theta}^{(SP)}\{\bm F(\bm X_1), \dots, \bm F(\bm X_{1 + p})\}\bigr] \bigr\| > 0.$$

	\item \label{cond:SP_derivs}
	The functions $\bm \phi_{\bm \theta}^{(SP)}$ are continuously differentiable with respect to $\bm \theta$ and its arguments and  there is $\delta > 0$ such that  
	\begin{align*}
		&\E\left[\sup_{ \bm \theta \in \tilde{\Theta}} \sup_{ \bm G \in \mathcal F_\delta } \biggl\{ \bigl\|\bm  \phi_{\bm \theta}^{(SP)}\{\bm F(\bm X_1), \dots, \bm F(\bm X_{1 + p})\}\bigr\| 
		+ \bigl\| \nabla_{\bm \theta }\bm \phi_{\bm \theta}^{(SP)}\{\bm G(\bm X_1), \dots, \bm G(\bm X_{1 + p})\} \bigr\| \biggr\} \right] < \infty, \\
		& \E\left[\sup_{ \bm \theta \in \tilde{\Theta}}  \sup_{\bm G \in \mathcal F_\delta } \biggl\| \frac{\partial}{\partial \{G_j(X_{t, j})\}} \bm \phi^{(SP)}_{\bm \theta}\{\bm G(\bm X_1), \dots, \bm G(\bm X_{1 + p})\}    w\{F_j(X_{t, j})\}  \biggr\|\right] < \infty
	\end{align*} 
	for any compact $\tilde \Theta \subseteq  \Theta$, $t = 1, \dots, 1 + p$, $j = 1, \dots, d$.

	\item \label{cond:SP_mixed}
	The mixed derivatives $ \nabla_{\bm \theta}\nabla_{\bm u}^\prime \bm \phi_{\bm \theta}^{(SP)}(\bm u)$ are continuous in $\bm u \in (0, 1)^{(p + 1)d}$ and $\bm \theta$ in a neighborhood of $\bm \theta^*$ and there is $\delta > 0$ such that for all $t = 1, \dots, 1 + p$, $j = 1, \dots, d$, 
	\begin{align*}
		& \E\left[\sup_{\|\bm \theta - \bm \theta^*\| \le \delta} \sup_{\bm G \in \mathcal F_\delta}  \biggl\| \frac{\partial}{\partial \{G_j(X_{t, j})\} }\nabla_{\bm \theta} \bm \phi_{\bm \theta}^{(SP)}\{\bm G(\bm X_1), \dots, \bm G(\bm X_{1 + p})\}  w\{F_j(X_{t, j})\} \biggr\|\right] < \infty.
	\end{align*}

	\item \label{cond:SP_hessinv} The matrix  $\bm J_{\bm \theta^*} = \E\bigl[ \nabla_{\bm \theta^*}^\prime \bm  \phi_{\bm \theta^*}^{(SP)}\{\bm F(\bm X_1), \dots, \bm F(\bm X_{1 + p})\} \bigr] $	is invertible.

	\item \label{cond:SP_mixing2}
	\change{For $\gamma \in [0, 1/2)$,} the $\beta$-mixing coefficients of $(\bm X_t)_{t \in \mathds Z}$ satisfy   
	 $\beta(t) = O(t^{-a})$ with $ a > 1 / (1 - 2\gamma)$ 
	  \change{and it holds} $\sum_{t = 0}^\infty \int_0^{\beta(t)} Q^2(u) du < \infty,$
	where $Q$ is the inverse survival function of $\|\bm \phi_{ \bm \theta^*}^{(SP)}\{\bm F(\bm X_1), \dots, \bm F(\bm X_{1 + p})\}\|$.
\end{enumerate}

Similar to the parametric case,  \ref{cond:SP_identify} ensures identifiability of the model parameters and \ref{cond:SP_hessinv} is a standard regularity condition. 
Conditions \ref{cond:SP_derivs}--\ref{cond:SP_mixed} are more involved due to the suprema over function classes $\mathcal F_\delta$. This is typical for semiparametric copulas models \citep[see,][]{genest1995,tsukahara2005semiparametric,chen2006estimation,hobaekhaff2013,chen2020}. Derivatives of copula functions tend to blow up in the corners of the unit hypercube. This can be offset by exploiting stronger convergence properties of the empirical margins in these corners. The function $w$ is used to strengthen the metric accordingly.  \ref{cond:SP_mixing2} quantifies the trade-off between moments of the `score' function $\bm \phi_{\bm \theta^*}^{(SP)}$ and the mixing rate, see also our discussion below.  

\begin{theorem} \label{thm:SP_consistency}
	Under \ref{cond:SP_identify}--\ref{cond:SP_derivs}, it holds  $\wb \theta^{(SP)}\to_p  \bm \theta^*$.
\end{theorem}

\begin{theorem} \label{thm:SP_an}
	Under \ref{cond:SP_identify}--\ref{cond:SP_mixing2}, it holds
	$\|\wb \theta^{(SP)} - \bm \theta^*\| = O_p(T^{-1/2})$ and
	\begin{align*}
		T^{1/2}(\wb \theta^{(SP)} - \bm \theta^*)
		\stackrel{d}{\to} \mathcal N\bigl\{\bm 0, \bm J^{-1}_{\bm \theta^*} \bm I_{\bm \theta^*} (\bm J^{-1}_{\bm \theta^*})^\prime\bigr\},
	\end{align*}
	where $\bm I_{\bm \theta^*} = \E(\bm Z_1 \bm Z_1^\prime ) + 2\sum_{t = 2}^\infty \E(\bm Z_1 \bm Z_t^\prime )$ and  
	\begin{align*}
		\bm Z_t &= \bm  \phi_{\bm \theta^*}^{(SP)}\{\bm F(\bm X_t), \dots, \bm F(\bm X_{t + p})\} +  \change{\bm  D(\bm X_{t})}, \\
		\change{\bm D(x_1, \dots, x_d)} &=  \sum_{j = 1}^d \sum_{t = 1}^{1 + p} \change{\E\biggl[\{\ind (x_j \le  X_{t, j}) - F_j(X_{t, j}) \} \frac{\partial \bm \phi_{\bm \theta^*}^{(SP)}\{\bm F(\bm X_1), \dots, \bm F(\bm X_{1 + p})\}}{\partial \{F_j(X_{t, j})\}}   \biggr].}
	\end{align*}
\end{theorem}
\begin{corollary} \label{cor:SP}
    If $(\bm X_{t})_{t \in \N}$ is \emph{iid} and $p = 0$, then \Cref{thm:SP_consistency} and \Cref{thm:SP_an} hold with  
    \begin{align*}
        \bm I_{\bm \theta^*} = \E\bigl([\bm  \phi_{\bm \theta^*}^{(SP)}\{\bm F(\bm X_1)\} + \bm D(\bm X_1)][\bm  \phi_{\bm \theta^*}^{(SP)}\{\bm F(\bm X_1)\} + \bm D(\bm X_1)]^\prime \bigr) .
    \end{align*}
    If $\E[\|\bm \phi_{\bm \theta^*}\{\bm F(\bm X_t)\}\|^2] < \infty$, condition \ref{cond:SP_mixing2} can be dropped.
\end{corollary}

\subsubsection{Discussion} \label{sec:asym-discussion}

The $\sqrt{T}$-convergence predicted by our theorems is confirmed in numerical experiments in Section S4.1 of the supplementary materials.
The results extend the existing literature in various ways. 
The results on the fully parametric sequential MLE (\Cref{thm:P_consistency,thm:P_an}) appear to be new --- even in the \emph{iid} case. \citet{joe2005asymptotic} provided a similar result when there are only two steps: one for the marginal parameters and one for the copula parameters. For semiparametric models  (\Cref{thm:SP_consistency,thm:SP_an}), a similar result was obtained by \citet{remillard2012copula} for $p = 1$, and a joint MLE for the copula parameters. It does not apply to the step-wise MLE commonly used in vine copula models, however. The only known results for the semiparametric step-wise MLE were provided by \citet{hobaekhaff2013} in the \emph{iid} ($p = 0$) case.
These results assume a D-vine structure and correctly specified copula model. The latter assumption is especially questionable in view of the common simplifying assumption (see \Cref{sec:vine_copulas}). 

Also the results in \citet{chen2006estimation} are obtained as a special case with $d = 1$, $p = 1$. The regularity conditions here are slightly weaker than theirs, and also than those of  \citet{tsukahara2005semiparametric} and \citet{hobaekhaff2013}  in the \emph{iid} case. 
Specifically, conditions \ref{cond:SP_derivs}--\ref{cond:SP_mixed} require a first moment uniformly in $\bar \Theta \times \mathcal F_\delta$, whereas previous results require  a second moment. A higher order moment constraint is only imposed on $\bm \phi_{\bm \theta^*}^{(SP)}\{\bm F(\bm X_1), \dots, \bm F(\bm X_{1 + p})\}$ and only at the single point $(\bm \theta^*, \bm F)$ via \ref{cond:SP_mixing2}.  %This  is made possible by a more careful argument based on modern empirical process techniques. 
Conditions \ref{cond:P_mixing2}  and \ref{cond:SP_mixing2}  ensure existence of the asymptotic covariance and are, to the best of our knowledge, the weakest known for $\beta$-mixing time series. Writing generically $\bm \phi$ for either  $\bm \phi_{ \bm \eta^*,  \bm \theta^*}^{(P)}$ in \ref{cond:P_mixing2} or $\bm \phi_{ \bm \theta^*}^{(SP)}$ in \ref{cond:SP_mixing2}, the conditions are satisfied in each of  the following cases \citep[see,][Section 1.4]{rio2013}:
\begin{enumerate}[label = (\roman*)]
 \item $(\bm X_t)_{t \in \mathds Z}$ is \emph{iid} and $\E\bigl\{ \|\bm\phi(\bm X_1, \dots, \bm X_{1 + p})\|^{2} \bigr\} < \infty$.
 \item There is $b \in (0, 1)$ such that $\beta(t) = O(b^t)$ and 
 $$\E \{\|\bm\phi(\bm X_1, \dots, \bm X_{1 + p})\|^{2} \ln(1 + \|\bm\phi(\bm X_1, \dots, \bm X_{1 + p})\|)\bigr\} < \infty.
 $$
 \item  $\beta(t) = O(t^{-a})$ with $a > 1/(1 - 2\gamma)$ and  there is $q  > \max\{2, 2/(a - 1)\}$ such that
 $$
 \E\bigl\{ \|\bm\phi(\bm X_1, \dots, \bm X_{1 + p})\|^{q} \bigr\} < \infty.
 $$
\end{enumerate} 
If the tail decay is fast, we can therefore use weaker moment conditions.
The latter two conditions already appeared in similar form in \citet{chen2006estimation}.
%, but the general formulation in \ref{cond:P_mixing2} and \ref{cond:SP_mixing2} allows to cover the \emph{iid} and similar edge cases without sacrificing sharpness.
%, but a simple consequence of the fact that (uniform) strong laws of large numbers are equivalent in the absolutely regular and \emph{iid} case \citep{nobel1993}. To allow for this generality in \Cref{thm:SP_consistency,thm:SP_an}, we had to improve Lemma 4.1 of \citet{chen2006estimation} about  weighted convergence of the empirical distributions accordingly. A formal result, \Cref{lem:weighted}, is given in the appendix.
For $p = 1, d = 1$, a sizeable literature  \citep[including][]{chen2006estimation, chen2009, beare2010copulas, beare2012archimedean, longla2012some} suggests that all popular parametric models exhibit exponentially decaying mixing coefficients, which is stronger than necessary. However, extending these results to the multivariate case is nontrivial and poses an important open problem. 

%\Crefrange{thm:P_consistency}{thm:SP_an} are specific to step-wise estimation in vine copula models. However, results for other likelihood or method-of-moment based estimators  can be obtained as easy corollaries of \Cref{thm:aux_consistency} and \Cref{thm:aux_an}, under similarly weak conditions. 

\section{Prediction} \label{sec:prediction}

Vine copula models are quite complex and rarely allow closed-form expressions of conditional means, quantiles, or the predictive distribution. One may instead simulate (conditionally) from the estimated model and approximate such quantities by Monte Carlo methods. The standard simulation algorithm \citep[e.g.,][Chapter 6]{czado2019analyzing} poses unnecessary computational demands, however.  An efficient algorithm exploiting the Markov property is given in Section S3.2 of the supplementary material.

With the ability to simulate conditionally on the past, it is easy to compute predictions for all sorts of quantities, like conditional means or quantiles. 
More specifically, suppose we are interested in a functional $\mu = \psi(F_{k, p})$ of the conditional distribution  $F_{k, p}(\cdot \mid \bm x_{t - 1}, \dots, \bm x_{t - p}) = F_{\bm X_t, \dots,  \bm X_{t+ k}| \bm X_{t - 1} , \dots, \bm X_{t - p}}(\cdot \mid \bm x_{t - 1}, \dots, \bm x_{t - p})$ of the next $k$ time points given the past.
We construct an estimator of this functional as follows:
\begin{enumerate}
	\item Simulate $N$ \emph{iid} replicates $(\bm X_{t}^{(i)}, \dots, \bm X_{t + k}^{(i)})_{i = 1}^{N}$ from the (estimated) conditional distribution of $(\bm X_{t}, \dots, \bm X_{t + k})$ given $\bm X_{t - 1} = \bm x_{t - 1}, \dots, \bm X_{t - p} = \bm x_{t - p}$ using the estimated model (either parametric or semiparametric; see  \Cref{sec:estsel}).
	\item Compute $\psi(\wh F_{k, p})$ where 
	\begin{align*}
		\wh F_{k, p}(\bm x_t, \dots, \bm x_{t + k}) = \frac 1 n \sum_{i = 1}^N \ind(\bm X_t^{(i)} \le \bm x_t, \dots,  \bm X_{t+ k}^{(i)} \le \bm x_{t + k}).
	\end{align*}
\end{enumerate}
When simulating from an estimated parametric model $ F_{k, p}(\cdot \mid \bm x_{t - 1}, \dots, \bm x_{t - p}; \wb \eta^{(P)}, \wb \theta^{(P)})$, we call the resulting estimator $\wh \mu^{(P)}$; when simulating from a semiparametric model $ F_{k, p}(\cdot \mid \bm x_{t - 1}, \dots, \bm x_{t - p};  \wb F, \wb \theta^{(SP)})$, we call the resulting estimator $\wh \mu^{(SP)}$. The corresponding pseudo-true value  $\mu^*$ is defined as $\psi\{F_{k, p}(\cdot \mid \bm x_{t - 1}, \dots, \bm x_{t - p}; \bm \eta^*, \bm \theta^*)\}$ or $\psi\{F_{k, p}(\cdot \mid \bm x_{t - 1}, \dots, \bm x_{t - p};   \bm F, \bm \theta^*)\}$ for the parametric and semiparametric cases respectively.

The following results account for the fact that we simulate from an estimated model. They are an immediate consequence of \Cref{thm:aux_mu} in \Cref{sec:gen}. In general, we assume that the map $F \mapsto \psi(F)$ is Frechet differentiable.

\begin{theorem} \label{thm:P_mu}
	Suppose the map $(\bm \theta, \bm \eta) \mapsto \psi\{F_{k, p}(\cdot \mid \bm x_{t - 1},  \dots, \bm x_{t - p}; \bm \eta, \bm \theta)\}$ is continuously differentiable at $(\bm \theta^*, \bm \eta^*)$ with gradient $\bm \Psi_{\bm \theta^*, \bm \eta^*}$.
	\begin{enumerate}
		\item If $N \to \infty$ and conditions  \ref{cond:P_identify}--\ref{cond:P_derivs} hold, then  $\wh \mu^{(P)} \to_p \mu^*$.
		\item If additionally $T = o(N)$ and conditions \ref{cond:P_mixing2}--\ref{cond:P_hessinv} hold, then 
		$\wh \mu^{(P)} - \mu^* = O_p(T^{-1/2})$ and 
		\begin{align*}
			\sqrt{T}(\wh \mu^{(P)} - \mu^*) \to_d \mathcal{N}\bigl( 0, \bm \Psi_{\bm \theta^*, \bm \eta^*}^\prime \bm J^{-1}_{\bm \eta^*, \bm \theta^*} \bm I_{\bm \eta^*, \bm \theta^*} (\bm J^{-1}_{\bm \eta^*, \bm \theta^*})^\prime \bm \Psi_{\bm \theta^*, \bm \eta^*}\bigr),
		\end{align*}
		where $\bm I_{\bm \eta^*, \bm \theta^*}, \bm J_{\bm \eta^*, \bm \theta^*}$ are defined in \Cref{sec:asym-p}.
\end{enumerate}
\end{theorem}

For any function $G$, let $W(G) = G(\cdot) / w\{G(\cdot)\}$ and denote $W(\bm G) = (W(G_1), \dots, W(G_d))$.
\begin{theorem} \label{thm:SP_mu}
	Suppose the map $(W(\bm G), \bm \theta) \mapsto \psi\{F_{k, p}(\cdot \mid \bm x_{t - 1},  \dots, \bm x_{t - p}; \bm G, \bm \theta)\}$ is Frechet differentiable at $(W(\bm F), \bm \theta^*)$ with derivative $(W(\bm h), \bm \theta) \mapsto \sum_{j = 1}^d \Psi_j(h_j)  + \bm \Psi_{\bm \theta}^\prime \bm \theta$.
	\begin{enumerate}
		\item If $N \to \infty$ and conditions \ref{cond:SP_identify}--\ref{cond:SP_derivs} hold, then  $\wh \mu^{(SP)} \to_p \mu^*$.
		\item If additionally $T = o(N)$ and conditions \ref{cond:SP_mixed}--\ref{cond:SP_hessinv} hold, then	$\wh \mu^{(SP)} - \mu^* = O_p(T^{-1/2})$ and  $\sqrt{T}(\wh \mu^{(SP)} - \mu^*) \to_d \mathcal{N}( 0, \sigma^2),$
		where $\sigma^2 = \var(Z_1) + 2\sum_{t = 2}^{\infty} \cov(Z_1, Z_t)$ with 
		\begin{align*}
			Z_t = \sum_{k = 0}^{p} \sum_{j = 1}^d \Psi_j\{\ind (X_{t + k, j}   \le \cdot ) - F_j(\cdot)  \} + \bm \Psi_{\bm \theta}^\prime [\bm  \phi_{\bm \theta^*}^{(SP)}\{\bm F(\bm X_t), \dots, \bm F(\bm X_{t + p})\} +  \bm  D(\bm X_{t}) ]
		\end{align*}
		and $\bm D(\bm X_{t}) $ given in \Cref{thm:SP_an}.
\end{enumerate}
\end{theorem}

Simulation-based prediction from vine copula models has been used widely in the last decade, despite a lack of theoretical justification. A consistency result for extreme quantile estimation in semiparametric \emph{iid} models was previously established by \citep[Theorem 1]{gong2015}. In contrast, the results above allow for both parametric and semiparametric models, time series data, and  a generic prediction target. In addition, \Cref{thm:P_mu} and \Cref{thm:SP_mu} characterize a distributional limit for such predictions. This is of practical importance because it allows to properly assess estimation/prediction uncertainty. Of course, the results specialize to the \emph{iid} case similarly to \Cref{cor:P,cor:SP}. The asymptotic covariances are generally unknown and must be estimated. We propose computationally efficient methods in the following section.

\section{Uncertainty quantification} \label{sec:uncertainty}

In principle, the asymptotic covariances in the preceding theorems can be estimated by HAC methods \citep[e.g.,][]{andrews1991}. In the prediction context such methods become numerically demanding, especially for the semiparametric estimator. (Semi-)parametric or block bootstrap methods \citep{kunsch1989,chen2006estimation, genest2008} are general alternatives, but  similarly demanding because the model has to be fit many times. We propose a more efficient bootstrap method based on an asymptotic approximation of the parameter estimates. In essence, we avoid refitting the entire model by performing only a single Newton-Raphson update on the bootstrapped likelihood.

We employ a depedendent multiplier bootstrap scheme similar to \citet{buecher2016}. Its idea is as follows.
Let $\ell_T$ be a sequence with $\ell_T \to \infty$. We simulate a stationary time series of bootstrap weights $\xi_1, \dots, \xi_T$ that is $\ell_T$-dependent, independent of the data, and satisfies $\E(\xi_1) = \var(\xi_1) = 1$ and $\cov(\xi_1, \xi_{1 + t}) = 1 - o(t/\ell_T)$. Given a sufficiently regular stationary time series $Z_1, \dots, Z_T$, one can then show that $T^{1/2}\{ T^{-1}\sum_{t = 1}^T Z_t - \E(Z_1)\}$ and $T^{-1/2}\sum_{t = 1}^T(\xi_t - 1)Z_t$ converge to independent copies of the same random variable  \citep[see,][]{buhlmann1993blockwise, bucher2019note}.  We shall apply this principle to bootstrap the step-wise log-likelihood and (if necessary) empirical marginal distributions.

Consider the bootstrapped estimating equation 
\begin{align*}
	\frac 1 T \sum_{t = 1}^T \xi_t \bm \phi^{(P)}_{\bm \eta, \bm \theta}(\bm X_{t}, \dots, \bm X_{t + p}) = 0.
\end{align*}
We define our bootstrap replicates as an approximate one-step Newton-Raphson update from $(\wb \eta^{(P)}, \wb \theta^{(P)})$, i.e.,
\begin{align*} 
 	\begin{pmatrix}
		\wt{\bm \eta}^{(P)} \\ \wt{\bm \theta}^{(P)} 
\end{pmatrix}
&= \begin{pmatrix}
	\wb \eta^{(P)} \\ \wb \theta^{(P)}
 \end{pmatrix}  	-  \biggl(  \sum_{t = 1}^T  \nabla_{(\bm \eta, \bm \theta)} \bm \phi^{(P)}_{(\wb \eta^{(P)}, \wb \theta^{(P)})}(\bm X_{t}, \dots, \bm X_{t + p})\biggr)^{-1}  \sum_{t = 1}^T  \xi_t \bm \phi^{(P)}_{(\wb \eta^{(P)}, \wb \theta^{(P)})}(\bm X_{t}, \dots, \bm X_{t + p}). \nonumber
\end{align*}
Because $\nabla_{(\bm \eta, \bm \theta)} \bm \phi^{(P)}_{(\wb \eta^{(P)}, \wb \theta^{(P)})}$ and $\bm \phi^{(P)}_{(\wb \eta^{(P)}, \wb \theta^{(P)})}$
have already been evaluated when computing  $(\wb \eta^{(P)}, \wb \theta^{(P)})$, this update has negligible computational cost.
One may then show that 
\begin{align*}
	 &\frac 1 T  \sum_{t = 1}^T  \nabla_{(\bm \eta, \bm \theta)} \bm \phi^{(P)}_{\wb \eta^{(P)}, \wb \theta^{(P)}}(\bm X_{t}, \dots, \bm X_{t + p})\to_p \bm J_{\bm \eta^*, \bm \theta^*}, \\
	&\frac 1 T \sum_{t = 1}^T \xi_t  \bm \phi^{(P)}_{(\wb \eta^{(P)}, \wb \theta^{(P)})}(\bm X_{t}, \dots, \bm X_{t + p}) \to_d \mathcal N(0, \bm I_{\bm \eta^*, \bm \theta^*}),
\end{align*}
with limiting variable independent of $(\wb \eta^{(P)}, \wb \theta^{(P)})$.

In semiparametric models, we also have to bootstrap the empirical marginal distribution:
\begin{align*}
	\wt F_j(x) = \frac{1}{T} \sum_{t = 1}^T \xi_t \ind(X_t \le x), \quad j = 1, \dots, d.
\end{align*}
Now consider the bootstrapped estimating equation 
\begin{align*}
	\frac 1 T \sum_{t = 1}^T \xi_t \bm \phi^{(SP)}_{\bm \theta}\{\wt{\bm F}(\bm X_{t}), \dots,\wt{\bm F}(\bm X_{t + p})\} = 0
\end{align*}
and the approximate one-step update 
\begin{align*}
 \wt{\bm \theta}^{(SP)} 
=  \wb \theta^{(SP)}
	-  \biggl( \frac 1 T  \sum_{t = 1}^T \nabla_{\theta} \bm \phi^{(SP)}_{\bm \theta}\{\wh{\bm F}(\bm X_{t}), \dots,\wh{\bm F}(\bm X_{t + p})\} \biggr)^{-1} \frac 1 T \sum_{t = 1}^T \xi_t \bm \phi^{(SP)}_{\wb \theta^{(SP)}}\{\wt{\bm F}(\bm X_{t}), \dots,\wt{\bm F}(\bm X_{t + p})\}.
\end{align*}
Note that the function $\bm \phi^{(SP)}_{\wb \theta^{(SP)}}$ on the far right is evaluated at the bootstrapped margins. 
This is necessary to account for the estimation uncertainty in the margins.
It also makes the update slightly more demanding, since we have to evaluate the function $\bm \phi^{(SP)}_{\wb \theta^{(SP)}}$ for every bootstrap replication.
This cost is manageable, however.
One may again show that $\wt{\bm \theta}^{(SP)}$ and $\wb \theta^{(SP)}$ converge in distribution to two \emph{iid} variables.

To get bootstrap replicates for a prediction $\wh \mu$, we simply simulate from bootstrapped models: for $r = 1, \dots R$, 
\begin{enumerate}[label = \arabic*.]
	\item Simulate multipliers $\xi_1, \dots, \xi_T$ independently from previous steps.
	\item Compute a  bootstrapped model (parametric or semiparametric) as outlined above.
	\item Compute $\wt \mu_r$ as in \Cref{sec:prediction}, where $\bm X_{t}^{(i)}, \dots, \bm X_{t + k}^{(i)}, i = 1, \dots, N$ are simulated conditionally on the past from the bootstrapped model.
\end{enumerate}

Validity of the above procedure can be established along the lines of \Crefrange{thm:P_consistency}{thm:SP_mu}  and arguments similar to \citet[Chapter 3]{buhlmann1993blockwise}. A formal proof is beyond the scope of this paper, but our simulation experiments in Section S4.2 of the supplementary materials indicate approximately correct coverage in a range of scenarios.

\section{Application} \label{sec:application}

Vine copula models are widely used in finance, in particular for modeling cross-sectional dependence in time series of financial returns \citep{aas2016pair}. The most common approach is to model marginal series with ARMA/GARCH-models and the cross-sectional dependence of their residuals with a vine copula. Stationary vine copula models are different; they incorporate both serial and cross-sectional dependence in a single vine copula model. 

We consider daily stock returns of 20 companies retrieved from Yahoo Finance\footnote{https://de.finance.yahoo.com/}. These companies belong to several industry branches and can be found in \Cref{tab-data}. The data covers the time slot from 1st January 2015 until 31st December 2019, containing in total 1296 trading days.

\subsection{In-sample analysis} \label{sec:insample}

\begin{table}[!t]
	\centering
		{\small
			\begin{tabular}{l|l|l||l|l|l}
			 Coding &	Company      & Industry Branch  & Coding &	Company      & Industry Branch  \\ \hline
			 1   &	Allianz        & Insurance   & 11  &	Microsoft      & IT                \\
			 2   &	AXA            & Insurance   &			 12  &	Apple          & IT                \\
			 3   &	Generali       & Insurance   &			 13  &	Amazon         & IT/Consumer goods \\
			 4   &	MetLife        & Insurance   &			 14  &	Alphabet       & IT                \\
			 5   &	Prudential     & Insurance   &			 15  &	Alibaba        & IT/Consumer goods \\
			 6   &	Ping An        & Insurance   &			 16  &	Exxon          & Oil and gas       \\
			 7   &	BMW            & Automotive  &			 17  &	Shell          & Oil and gas       \\
			 8   &	General Motors & Automotive  &			 18  &	PetroChina     & Oil and gas       \\
			 9   &	Toyota         & Automotive  &			 19  &	Airbus         & Aerospace         \\
			 10  &	Hyundai        & Automotive  &			 20  &	Boeing         & Aerospace      
			\end{tabular}}
			\caption{Companies, their coding, and industry branches.}
			\label{tab-data}
\end{table}

We start with an in-sample illustration of models fit to the whole data set. We first fit skew-$t$ distributions to the individual time series of each company. We then apply the probability integral transform to obtain \emph{pseudo-observations} of the copula model.
We consider the M-vine, D-vine, and a general stationary (S-)vine models from the previous section, each with Markov orders $p=1$ (higher order models were fit in preliminary experiments, but did not improve fit/performance). Vine structures are selected by a modification of the algorithm by \citet{Dissmann2012}, the pair-copula families by the AIC criterion; we refer to Section S3 of the supplementary materials for details. We allow for all parametric families implemented in the \texttt{rvinecopulib} R package \citep{rvinecopulib}. This includes families without tail dependence (e.g., Gaussian copula), and with tail dependence in either one (e.g., Clayton copula), two (e.g., BB7 copula), or all four tails (e.g., the two-parameter $t$ copula).

\begin{figure}
	\centering
	\includegraphics[width=\textwidth]{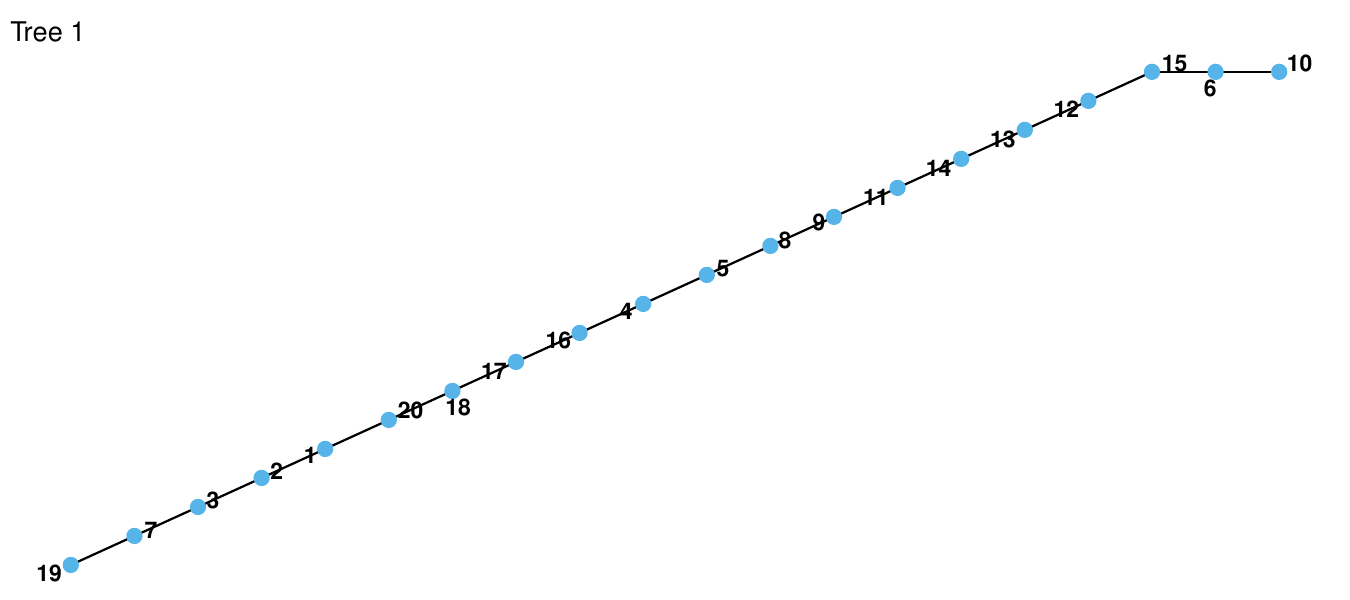}
	\caption{First tree of M- and D-vines fitted on the whole data set. Trees across time-steps are connected at $(i_1, j_1) = (8, 8)$ (with $\wh \tau \approx 0.02)$ for the M-vine, and $(i_1, j_1) = (19, 10)$  (with $\wh \tau \approx 0.05)$ for the D-vine.}
	\label{figMDVine}
\end{figure}

In \Cref{figMDVine} we illustrate the first trees of the M- and D-vine obtained via the previously described approach.  We observe that the cross-sectional D-vine for both approaches is described by a path $19 - \dots - 10$. The M-vine makes the serial connection by an edge linking the same stock from time $t$ to time $t + 1$. In this case, the connection is $(i_1, j_1) =  (19, 19)$ (Airbus$\rightarrow$Airbus) which has an empirical Kendall's $\tau$ of 0.02. The only other viable choice would have been $(10, 10)$ (Hyundai$\rightarrow$Hyundai), but it had a lower Kendall's $\tau$ of around 0.01. The D-vine connects two opposites ends of the path, here from Hyundai $(10)$  to Airbus $(19)$  ($\wh \tau = 0.05$).

\begin{figure}
	\centering
	\includegraphics[width=\textwidth]{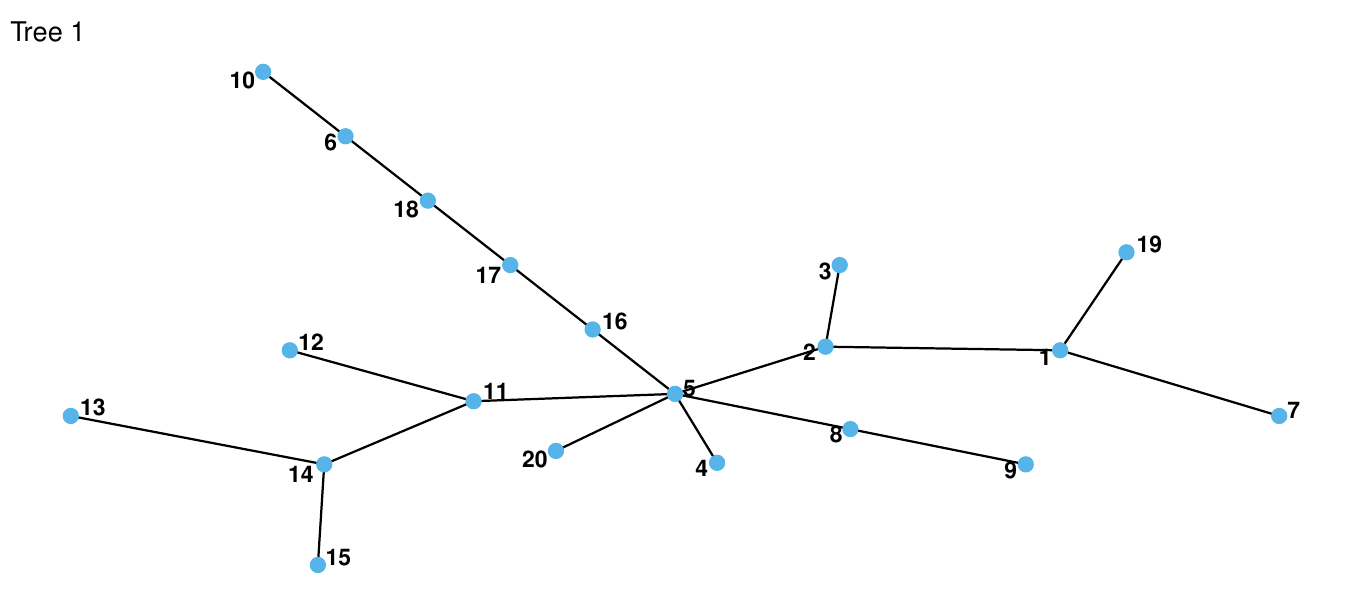}
	\caption{First tree of an S-vine fitted on the whole data set. Trees across time-steps are connected at $(i_1, j_1) = (15, 6)$ (with $\wh \tau \approx 0.16)$.}
	\label{figRTVine}
\end{figure}

The corresponding tree of the S-vine can be seen in \Cref{figRTVine}. The cross-sectional connection is described by a regular vine. We can identify some  clusters of industry branches: IT (variables 11--15), insurance (1--5), and oil and gas (16--18). Interestingly, regional factors seem to be more important than the branch for aerospace and automotive stocks, however. The European manufacturers BMW (7) and Airbus (19) are attached to the European insurance cluster (1--3). American counterparts General Motors (8) and Boeing (20) are linked to the American insurances MetLife and Prudential (4, 5). Some of these links can also be identified from the M-/D-vine structure in \Cref{figMDVine}, but not as prominently. This plus in interpretability is one of the big advantages of using general R-vines as the cross-sectional structure.

The inter-serial connection of the S-vine is made at $(i_1, j_2) = (15, 6)$ (Alibaba$\rightarrow$Ping An) with an empirical Kendall's $\tau$ of 0.16. The dependence here is much stronger than for the serial connections of the M- and D-vine models. This reflects the greater flexibility of the S-vine model. Recall that compatibility does not restrict the connection in the first tree. We are thus free to choose from all possible in-/out- pairs. The linking edge is interesting in itself. First, it links to different companies across subsequent time points. Hence, this dependence must be stronger than any inter-serial dependence of a single stock. Second, it links Alibaba, a Chinese IT/Consumer goods company, to Ping An, a Chinese insurance company, which makes sense economically. Further,  this link did not appear in the cross-sectional parts of either of the vine models. So while the cross-sectional dependence between the companies is comparably weak, their inter-temporal dependence is still quite strong.

\begin{table}[!t]
\centering
			\begin{tabular}{r|r|r|r|r|r}
				 S-vine & M-vine & D-vine    & VAR & GARCH-vine & DCC-GARCH \\ \hline
				-163\,371 & -163\,257 & -163\,258   & -156\,360  & -162\,279 & -159\,857
			\end{tabular}
	\caption{Aikaike's information criterion  for the three vine copula time series models.}\label{tab-AIC}
\end{table}

The fit of the models is compared by AIC in \Cref{tab-AIC}. We only consider  parametric vine models, but also include three popular competitor models: 
\begin{itemize}
    \item  \emph{VAR}: A \emph{vector autoregressive} model of order 1 \citep[using the \texttt{vars} R package][]{vars}.
    \item \emph{GARCH-vine}:  A combination of ARMA-GARCH marginal models (with skew-t residuals) and a vine copula for the residuals \citep[using \texttt{rugarch} and \texttt{rvinecopulib},][]{rugarch, rvinecopulib}. The ARMA-GARCH orders are selected for each marginal series individually by AIC. This model is an instance of the general residuals method proposed by \citet{nasri2019a}. Other choices for marginal models would also be possible.
    \item \emph{DCC-GARCH}: the DCC-GARCH model of \citet{engle2001theoretical} based on a multivariate t distribution \citep[using \texttt{rmgarch},][]{rmgarch}. 
\end{itemize}

The VAR model clearly performs worst, since it cannot account for heteroscedasticity. We further see that the vine models outperform the GARCH-vine and DCC-GARCH models. The S-vine provides the best fit.

\subsection{Out-of-sample predictions} \label{sec:oos}

We now compare the forecasting abilities by a backtest. We fit all models on three years' data (one year has 252 trading days). On each of the following days, we make predictions for the cumulative portfolio return over the next day or week and compare them to the observed data. Every half year the models are fitted again on three years' data.

\begin{figure}
	\centering
	\includegraphics[width=0.9\textwidth]{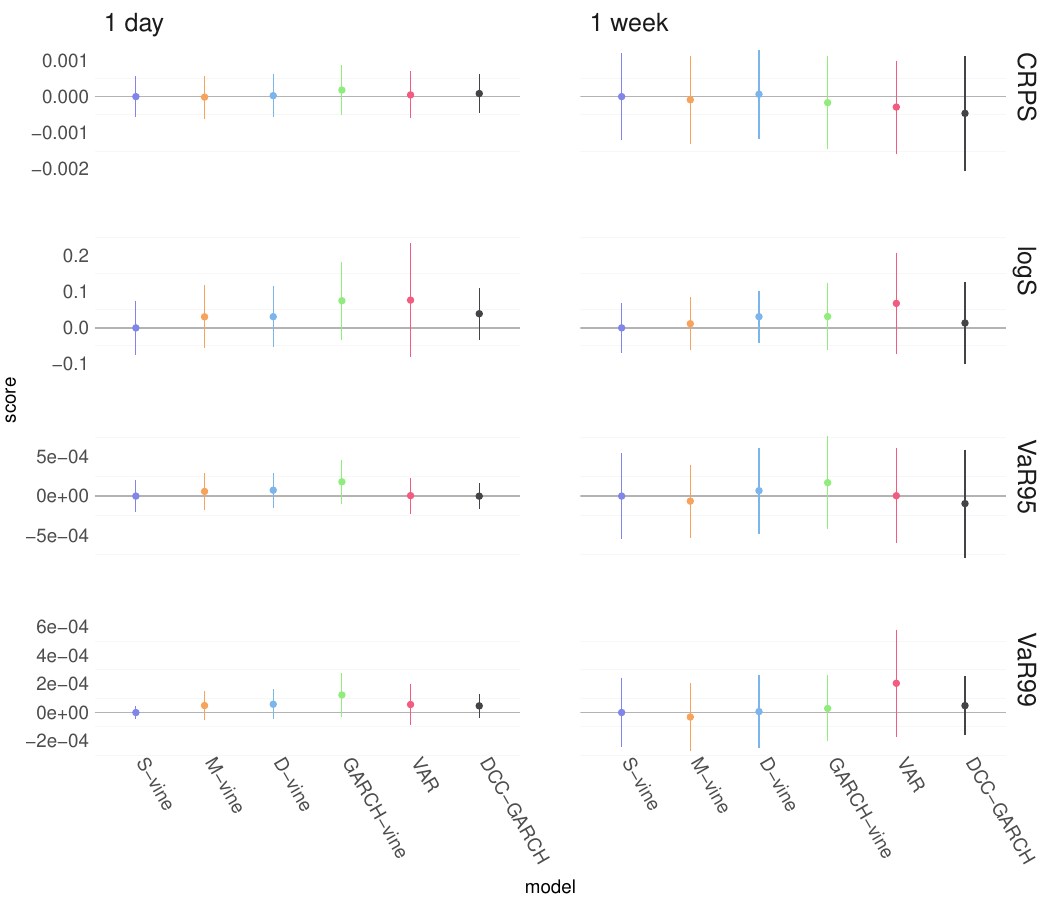}
	\caption{Forecast performance of various time series models. Dots are mean performance, error bars indicate 90\%-confidence intervals (accounting for 30 lags of autocorrelation). The left panel corresponds to 1-day-ahead, the right to 1-week-ahead forecasts. Scores are centered such that S-vine has score 0.}
	\label{fig:perf} 
\end{figure}

Our predictions take the form of a Monte-Carlo sample drawn from the predictive distribution. They are evaluated with three types of measures:
\begin{itemize}
	\item \texttt{CRPS}: The \emph{continuous ranked probability score} of \citet{gneiting2007}.
	\item \texttt{logS}: The negative predictive log-likelihood.
	\item \texttt{VaR95}, \texttt{VaR99}: The \emph{check-loss} known from quantile-regression \citep[e.g.,][]{koenker2002inference} computed for predicted  quantiles at levels 0.05 and 0.01. Such quantiles are popular risk measures in banking and insurance, where they are called \emph{Value-at-Risk (VaR)}.
\end{itemize}
\texttt{CRPS} and \texttt{logS} are computed with the \texttt{scoringRules} R package \citep{scoring}, \text{VaR}s as empirical quantiles of the Monte-Carlo sample. The measures are averaged across 1000 randomly sampled portfolios. The first 19 weights are drawn uniformly from a $\mathrm{Uniform}(-0.15, 0.25)$ distribution and the 20th set such that weights sum up to one.

The forecast performance is shown in \Cref{fig:perf}. The dots are the average measure over the full period, the error bars indicate 90\%-confidence intervals (adjusted for serial dependence). The left panel corresponds to 1-day-ahead, the right to 1-week-ahead forecasts. Scores are centered such that S-vine has score 0. Some observations:
\begin{itemize}
	\item A general observation is that uncertainty (as indicated by the confidence intervals) is rather larger compared to the differences between models. Everything that follows should therefore be taken with a grain of salt.
	\item Overall all methods seem to provide reasonable predictions, including the classical residuals method (GARCH-vine).
	\item The three stationary vine models perform similarly in all scenarios. The  S-vine and M-vine tend to perform slightly better than the long D-vine. The S-vine is uniformly best for 1-day-ahead forecasts, but slightly outperformed by the M-vine for 1-week-ahead forecasts (except for \texttt{logS}). After all, the vine structures are found by heuristics and there is no guarantee that the best is found. 
	\item For 1-day-ahead forecasts, the S-vine performs best for all measures, especially for extreme quantiles and the predictive log-likelihood (\texttt{logS}). For 1-week-ahead forecasts the stationary vine models are outperformed by the competitors models for \texttt{CRPS} and for \texttt{VaR95} by the DCC-GARCH. It compares favorably for the other measures.
\end{itemize}

We conclude that stationary vine models provide good forecasts for financial time series. This is somewhat remarkable since, in contrast to the vine models, the GARCH-vine and DCC-GARCH models were specifically designed for such data. Recently, some copula families have been specifically designed for modeling serial dependence in economic time series \citep[e.g.,][]{bladt2020time, loaiza2018time}, but were not used in this article. We expect that incorporating such families will lead to a further increase in performance. 

\section{Discussion} \label{sec:conclusion}
This work deals with vine copula models for the joint distribution of a stationary time series. We derived the maximal class of vine structures that guarantee stationarity under practicable conditions. The underlying principle is intuitive: we start with a vine model for the dependence at a specific time point and connect copies of this model serially in a way that preserves time ordering. This class includes previously proposed models of \citet{BeareSeo2015} and \citet{smith2015copula} as special cases. The COPAR model of \citet{brechmann2015copar} was shown to be inadequate in this sense because it fails to guarantee stationarity under simple conditions. The simulations and application suggest that the added flexibility leads to improvements over the previous models. Another benefit is the greater interpretability of the model structure. But more importantly, our contribution gives a final answer in the search for vine copula models suitable for stationary time series.

We developed methods for parameter estimation, model selection, simulation, prediction, and uncertainty quantification in such models. All methods are designed with computational efficiency in mind, such that the full modeling pipeline runs in no more than a few minutes on a customary laptop. The proposed bootstrap procedure avoids refitting the models through a one-step approximation. The method appears to be new and may  prove useful beyond the current scope.  The bootstrap technique also does not require explicit estimation of the rather complicated limiting variances in our theorems. It might be possible to achieve this even more efficiently using a blocking technique similar to \citet{ibragimov2010t}.

We further provide theoretical justifications  in the form of asymptotic results. To the best of our knowledge, these are the first results applicable to vine copula models under serial dependence. Even when specialized to the \emph{iid} case, they extend the existing literature in several ways. In particular, they provide post-hoc justification for what is already practiced widely: step-wise estimation and simulation-based inference in fully parametric, but usually misspecified R-vine models. 
Our main results are empowered by more abstract theorems given in \Cref{sec:gen}.
They deal with general semiparametric method-of-moment type estimators with potentially  non-negligible nuisance parameter.
As this is a common setup, especially in copula models, these abstract results shall prove powerful beyond the present paper.
For example, generalizations of the results in \citet{tsukahara2005semiparametric}  are obtained as easy corollaries.
The results shall also help in other interesting extensions of our model, for example accounting for long-memory dependence or non-stationarity \citep[see, e.g.,][]{ibragimov2017,chen2020}.

Despite confirmatory numerical experiments, a limitation of the results is an assumption on the decay of mixing coefficients (required only for the asymptotic distribution). Judging from earlier work in a narrower context, we do not believe this poses a serious issue. However, we do not yet know any easily verifiable sufficient conditions. Investigating the mixing properties of stationary vine copulas --- and multivariate copula models more generally --- is therefore an urging problem for future research.

\section*{Acknowledgements}

We thank the Editor and three anonymous referees for many helpful suggestions that substantially improved our paper.

%---------------------------------------------------------------%

\appendix

\section{General results for semiparametric method-of-moment estimation} \label{sec:gen}

\numberwithin{theorem}{section}
\counterwithin*{theorem}{section} 
\setcounter{theorem}{0}
\numberwithin{lemma}{section}
\counterwithin*{lemma}{section} 
\setcounter{lemma}{0}

The proofs for the parametric and semiparametric cases are largely similar. To avoid duplication, we first establish general results that cover both cases. The statements and proofs make extensive use of empirical process techniques \citep[e.g.,][]{van1996weak, dehling2002empirical} and the associated short notation $\PT g = \frac 1 T \sum_{t = 1}^T g(\bm X_t)$ and $P g = \E\{g(\bm X_t)\}$
for the empirical measure and expectation over an arbitrary function $g$.

Suppose we want to estimate a Euclidean parameter $\bm \alpha^* \in {\mathcal A} \subseteq \R^p$ in the presence of a nuisance parameter $\bm \nu^* \in  \mathfrak N$, possibly infinite-dimensional. Let $\bm \phi_{\bm \alpha, \bm \nu} = (\phi_{\bm \alpha, \bm  \nu, 1}, \dots, \phi_{\bm\alpha, \bm  \nu, r})$ be a map $\R^{s} \to \R^r$ such that $P \bm \phi_{\bm \alpha^*, \bm  \nu^*} = 0$.  Given an estimator $\wb \nu$ of $\bm \nu^*$ define $\wb\alpha$ as the solution to $\PT \bm \phi_{ \bm \alpha, \wb \nu} = 0$. We shall assume that $\mathfrak N$ is a subset of a Banach space and define 
\begin{align*}
	{\mathcal A}(\delta) = \{\bm \alpha \in {\mathcal A}\colon \|\bm \alpha - \bm \alpha^*\| \le \delta\}, \qquad {\mathfrak N}(\delta) = \{\bm \nu \in {\mathfrak N}\colon \|\bm \nu - \bm \nu^*\| \le \delta\}.
\end{align*}

We impose the following general conditions:

\begin{enumerate}[label = (C\arabic*)]
	\item \label{cond:gen_mixing} 
	The series $(\bm X_t)_{t \in \mathds Z}$ is strictly stationary and absolutely regular.

	\item \label{cond:beta_consistency} 
	For every $\delta > 0$, $\Pr(\| \wb  \nu - \bm  \nu^*\| \le \delta) \to 1$ as $T \to \infty$. 

	\item \label{cond:gen_identify} 
	For every $\epsilon > 0$, it holds  $\inf_{\|\bm \alpha - \bm \alpha^* \| > \epsilon} \|P\bm \phi_{\bm \alpha, \bm  \nu^*}\| > 0.$ 
		
	\item \label{cond:lipschitz_class} 
	For every $K > 0$, it holds $P \sup_{\bm \alpha \in \mathcal A(K)}\|\bm \phi_{\bm \alpha, \bm  \nu^*}\|  < \infty$ and there is a $\delta > 0$ such that
	\begin{align*}
		P\left\{\sup_{\bm \alpha_1, \bm \alpha_2 \in \mathcal A(K)}\sup_{\bm \nu_1, \bm \nu_2 \in \mathfrak N(\delta)} \frac{\bigl\| \bm \phi_{ \bm \alpha_1, \bm  \nu_1} - \bm \phi_{ \bm \alpha_2,  \bm  \nu_2} \bigr \|}{\|\bm \alpha_1 - \bm \alpha_2\| + \|\bm \nu_1 - \bm \nu_2\|} \right\} < \infty.
   \end{align*}

	\item \label{cond:mixed_lipschitz_class} 
	It holds   and there is $\delta > 0$ such that 
	\begin{align*}
		P\left\{\sup_{\bm \alpha_1, \bm \alpha_2 \in \mathcal A(\delta)}\sup_{\bm \nu_1, \bm \nu_2 \in \mathfrak N(\delta)} \frac{\bigl\| (\bm \phi_{ \bm \alpha_1, \bm  \nu_1} - \bm \phi_{ \bm \alpha_2, \bm  \nu_1}) - (\bm \phi_{ \bm \alpha_1,  \bm  \nu_2} - \bm \phi_{ \bm \alpha_2,  \bm  \nu_2}) \bigr \|}{\|\bm \alpha_1 - \bm \alpha_2\| \|\bm \nu_1 - \bm \nu_2\|} \right\} < \infty.
   \end{align*}

   \item \label{cond:beta_weakly}
   $T^{1/2}(\wb \nu - \bm \nu^*)$ converges weakly to a tight, centered Gaussian limit $\bm N$.
	
   \item \label{cond:hadamard} 
   The map $(\bm \alpha, \bm \nu) \mapsto P \bm\phi_{ \bm \alpha,  \bm \nu}$ from $\mathcal A \times \mathfrak N$ to $\R^r$ is Fr\'echet differentiable  at  $(\bm \alpha^*, \bm \nu^*)$  with derivative $(\bm a, \bm b) \mapsto \Phi_{\bm \alpha^*, \bm \nu^*, 1}(\bm a) + \Phi_{\bm \alpha^*, \bm \nu^*, 2}(\bm b)$. That is, $\Phi_{\bm \alpha^*, \bm \nu^*, 1}, \Phi_{\bm \alpha^*, \bm \nu^*, 2}$ are continuous, linear maps such that for every $\|\bm a \| \to 0$, $\|\bm b \| \to 0$,
   \begin{align*}
	   \bigl\|P \bm\phi_{ \bm \alpha^* + \bm a,  \bm \nu^* + \bm b} - P \bm\phi_{ \bm \alpha^*,  \bm \nu^*} - \Phi_{\bm \alpha^*, \bm \nu^*, 1}(\bm a) - \Phi_{\bm \alpha^*, \bm \nu^*, 2}(\bm b) \bigr\| = o(\|\bm a\| + \| \bm b \|).
   \end{align*}
	 Further assume that $\Phi_{\bm \alpha^*, \bm \nu^*, 1}$ is invertible.

   \item \label{cond:gen_mixing2} 
   The $\beta$-mixing coefficients of $(X_t)_{t \in \mathds Z}$ satisfy $\sum_{t = 0}^\infty \beta(t) < \infty$ and 
   $\sum_{t = 0}^\infty \int_0^{\beta(t)} Q^2(u) du < \infty,$
   where $Q$ is the inverse survival function of $\|\bm \phi_{\bm \alpha^*, \bm \nu^*}\|$.
\end{enumerate}

Conditions \ref{cond:beta_consistency} and \ref{cond:beta_weakly} make 
convergence of $\wb \nu$, the estimator of the nuisance parameter, a prerequisite. 
The other conditions concern the regularity of the time series and identifying functions $\bm \phi_{\bm \alpha, \bm \nu}$.
The following theorems establish consistency and asymptotic normality of $\wb \alpha$, our estimator for the parameter of interest. 
%The results are similarly broad in scope as those by \citet{andrews1994} and, in the \emph{iid} setting, by  \citet{ma2005}.
%However, both restrict to settings where the effect of estimating the nuisance parameter is negligible, which is rare in copula models.
%Our results below alleviate this constraint. 
%For asymptotic normality, this comes at the cost of requiring weak $\sqrt{T}$-convergence of $\wb \nu$.
%This is more restrictive than in previous results, but still interesting: complying examples for $\wb \nu$ are empirical distribution functions, functional principal components, or certain shape-constrained estimators.

\begin{theorem} \label{thm:aux_consistency}
	If \ref{cond:gen_mixing}--\ref{cond:lipschitz_class} hold, then $\| \wb \alpha - \bm \alpha^* \| = o_p(1)$.
\end{theorem}

\begin{theorem} \label{thm:aux_an}
		Under conditions \ref{cond:gen_mixing}--\ref{cond:gen_mixing2}, it holds
		\begin{align*}
		\wb \alpha - \bm \alpha^* &= -\Phi_{\bm \alpha^*, \bm \nu^*, 1}^{-1} \bigl\{\PT \bm\phi_{ \bm \alpha^*,  \bm \nu^*} + \Phi_{\bm \alpha^*, \bm \nu^*, 2}(\wb \nu - \bm \nu^*) \bigr\} + o_p(T^{-1/2}),
		\end{align*}
		and 
		\begin{align*}
			T^{1/2}(\wb \alpha - \bm \alpha^*)			&\to_d  \mathcal N\{0, \Phi_{\bm \alpha^*, \bm \nu^*, 1}^{-1} \bm \Sigma_{\bm \alpha^*, \bm \beta^*} (\Phi_{\bm \alpha^*, \bm \nu^*, 1}^{-1})^{\prime}\},
			\end{align*}
		where $\bm \Sigma_{\bm \alpha^*, \bm \beta^*}$ is the limiting covariance of $T^{1/2}\{\PT \bm\phi_{ \bm \alpha^*,  \bm \nu^*} + \Phi_{\bm \alpha^*, \bm \nu^*, 2}(\wb \nu - \bm \nu^*)\}$.
\end{theorem}

Now let $F_{\bm \alpha, \bm \nu}$ be a cumulative distribution function parametrized by $(\bm \alpha, \bm \nu)$. 
Suppose the parameter of interest is defined as $\mu^* = \psi(F_{\bm \alpha^*, \bm \nu^*})$ for some functional $\psi$. 
Denote $F_{N, \bm \alpha, \bm \nu}$ the empirical measure over $N$ \emph{iid} realizations from $F_{\bm \alpha, \bm \nu}$. 
For estimators $\wb \alpha, \wb \nu$, define $\wh \mu = \psi(F_{N, \wb \alpha, \wb \nu})$.

\begin{theorem} \label{thm:aux_mu}
	Suppose that the maps $F \mapsto \psi(F)$ and $(\bm \alpha, \bm \nu) \to \psi(F_{\bm \alpha, \bm \nu})$ are Frechet differentiable at $F_{\bm \alpha^*, \bm \nu^*}$ and $(\bm \alpha^*, \bm \nu^*)$ respectively.
	\begin{enumerate}
		\item If $T \to \infty$ and  $(\wb \alpha, \wb \nu) \to_p (\bm \alpha^*, \bm \nu^*)$, then $\wh \mu \to_p \mu^*$.
		\item If $T = o(N)$ and $T^{1/2}\{(\wb \alpha, \wb \nu) - (\bm \alpha^*, \bm \nu^*)\}$ converges weakly to a tight process $(\bm A, \bm N)$, then 
		\begin{align*}
			T^{1/2}(\wh \mu - \mu^*) \to_d \Psi_{(\bm \alpha^*)}(\bm A) +  \Psi_{(\bm \nu^*)}(\bm N),
		\end{align*}
		where $(\bm a, \bm b) \mapsto \Psi_{(\bm \alpha^*)}(\bm a) +  \Psi_{(\bm \nu^*)}(\bm b)$ is the Frechet derivative of the map $(\bm a, \bm b) \mapsto \psi(F_{\bm a, \bm b}) $ at $(\bm \alpha^*, \bm \nu^*)$.
	\end{enumerate}
\end{theorem}

In the context of our paper, $\wb \nu$ is a vector of empirical distribution functions. 
Lemma 4.1 of \citet{chen2006estimation} establishes \ref{cond:beta_consistency} and \ref{cond:beta_weakly}, but under conditions slightly stronger than our \ref{cond:gen_mixing} and \ref{cond:gen_mixing2}.
The following lemma  improves their result accordingly. For sake of completeness, we give a detailed proof in Section S6.6 of the supplementary material. 

\begin{lemma} \label{lem:weighted}
	Let $Z_1, \dots, Z_T \in \R$ be a a stationary time series with $\beta$-mixing coefficients $\beta(t), t\ge 0$. 
Define $F_T(z) = (T + 1)^{-1} \sum_{t = 1}^T \ind(Z_t \le z)$ and $W_T = (F_T - F_Z)(z) / w\{F_Z(z)\}$, where  $w(u) = u^\gamma(1 - u)^\gamma$, $\gamma \in [0, 1)$.
	\begin{enumerate}
		\item If $\beta(t) \to 0$, $\sup_{z \in \R} |W_T(z)|\to 0$ almost surely.
		\item If $\gamma \in [0, 1/2)$ and $\beta(t) = O(t^{-a})$ with $ a > 1 / (1 - 2\gamma)$, the process $W_T$ converges weakly in $\ell^\infty(\R)$ to	a tight Gaussian limit $W$ with mean zero and covariance
		\begin{align*}
			E\{W(z_1) W(z_2)\} = \frac{\var\{\ind(Z_1 \le z_1), \ind(Z_1 \le z_2)\} + 2\sum_{t = 2}^\infty \cov\{\ind(Z_1 \le z_1), \ind(Z_t \le z_2)\}}{w(z_1) w(z_2)}.
		\end{align*}
	\end{enumerate}
\end{lemma}

\bibliographystyle{chicago}

\bibliography{literature}

\clearpage

\newpage



\section*{Supplementary material}

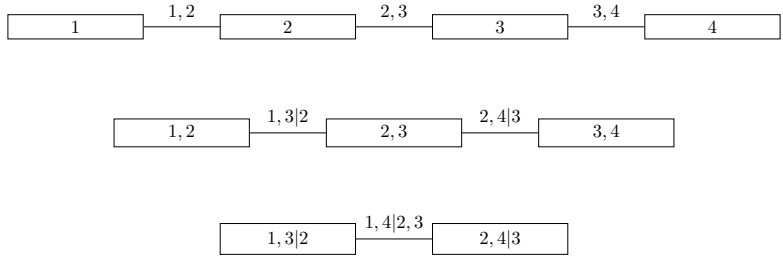

Figure S1: A four-dimensional D-vine.

# S1 Graph theoretical concepts: additional illustrations

In this section we provide additional illustrations for graph theoretical concepts. For convenience, we repeat the definitions from the main manuscript (with numbering preserved).

## S1.1 Regular vines

**Definition 1.** *A collection of trees $\mathcal{V} = (V_k, E_k)_{k=1}^{d-1}$ on a set $V_1$ with $d$ elements is called R-vine if*

(i) *$T_1$ is a tree with vertices $V_1$ and edges $E_1$,*

(ii) *for $k = 2, \ldots, d-1$, $T_k$ is a tree with vertices $V_k = E_{k-1}$,*

(iii) *(proximity condition) for $k = 2, \ldots, d-1$: if vertices $a, b \in V_k$ are connected by an edge $e \in E_k$, then the corresponding edges $a = \{a_1, a_2\}$, $b = \{b_1, b_2\} \in E_{k-1}$, must share a common vertex: $|a \cap b| = 1$.*

A graphical example of a regular vine on four variables is shown in [Figure S1]. Let us discuss the three conditions of the definition in the context of this example:

(i) The tree $T_1$ consists of vertices $V_1 = \{1, 2, 3, 4\}$ and edges $E_1 = \{(1, 2), (2, 3), (3, 4)\}$. We see that the graph is connected and contains no cycles, so it is indeed a tree.

(ii) The edges in $E_1$ become the vertices $V_2$ of the second tree. That is, $V_2 = E_1 = \{(1, 2), (2, 3), (3, 4)\}$. The vertices in $V_2$ are connected by some edges $E_2$. The edges $E_2$ then become the vertices in $T_3$, i.e., $V_3 = E_2$. The labeling of edges in the second and third tree of the figure is explained below.

(iii) The proximity condition only concerns tree levels $k \geq 2$. Recall that the second tree has vertices $V_2 = \{(1, 2), (2, 3), (3, 4)\}$. We connect the vertex $(1, 2)$ to the vertex $(2, 3)$. The proximity condition permits this, because the edges $(1, 2) \in E_1$ and $(2, 3) \in E_1$ share a common vertex, namely $2 \in V_1$. Such a common vertex does not exist for the edges $(1, 2) \in E_1$ and $(3, 4) \in E_1$. The proximity condition therefore forbids to connect the corresponding vertices $(1, 2) \in V_2$ and $(3, 4) \in V_2$.

## S1.2 Labeling of edges

The edge labels in Figure S1 follow the convention dictated by Definitions 2 and 3.

**Definition 2.** *The complete union of an edge $e \in E_k$ is given by*

$$\mathcal{U}_e = \{i \in V_1 | \ i \in e_1 \in e_2 \in \cdots \in e \ \text{for some} \ (e_1, \ldots, e_{k-1}) \in E_1 \times \cdots \times E_{k-1}\}$$

*and for a singleton $i \in V_1$ it is given by the singleton, i.e. $\mathcal{U}_i = \{i\}$.*

**Definition 3.**

(i) *The* conditioning set *of an edge $e$ connecting $v_1$ with $v_2$ is $D_e = \mathcal{U}_{v_1} \cap \mathcal{U}_{v_2}$.*

(ii) *The* conditioned set *of an edge $e$ connecting $v_1$ with $v_2$ is defined as $(a_e, b_e)$, where $a_e = \mathcal{U}_{v_1} \setminus D_e$ and $b_k = \mathcal{U}_{v_2} \setminus D_e$.*

*We will then label an edge by $e = (a_e, b_e | D_e)$.*

As an example, consider the edge $e \in E_2$ connecting $v_1 = (1, 2) \in V_2$ to $v_2 = (2, 3) \in V_2$ in Figure S1. The complete union of an edge is obtained by collecting all variables indices appearing in the edges. In particular, we have $\mathcal{U}_{v_1} = \mathcal{U}_{(1,2)} = \{1, 2\}$, $\mathcal{U}_{v_2} = \mathcal{U}_{(2,3)} = \{2, 3\}$. Note also that the complete union can always be determined recursively via $\mathcal{U}_e = \mathcal{U}_{v_1} \cup \mathcal{U}_{v_2}$. Hence, we have $\mathcal{U}_e = \{1, 2, 3\}$. We can now compute the conditioning set of $e$ as $D_e = \mathcal{U}_{v_1} \cap \mathcal{U}_{v_2} = \{2\}$ and the conditioned set as $a_e = \mathcal{U}_{v_1} \setminus D_2 = \{1\}$ and $b_e = \mathcal{U}_{v_2} \setminus D_2 = \{3\}$. We therefore label this edge by $(a_e, b_e | D_e) = (1, 3|2)$. The same principal applies to edges in later trees, where it is convenient to use the fact that $\mathcal{U}_e = a_e \cup b_e \cup D_e$.

## S1.3 D- and C-vines

Figure S1 displays a four-dimensional D-vine. A D-vine is a vine where each tree is a path, i.e., each vertex is connected to at most two other vertices. The proximity condition then already fixes all the following trees. Edge labels indicate the pair-copulas involved in the four-dimensional D-vine copula:

$$c = c_{12} \cdot c_{23} \cdot c_{34} \cdot c_{13|2} \cdot c_{24|3} \cdot c_{14|23},$$

where we omit the arguments of copula densities for simplicity.

Figure S2 displays a four-dimensional C-vine. A C-vine is a vine where each tree is a star, i.e., there is one vertex connected to all the others. In the first tree, vertex 1 is connected with other three vertices. One may check that the proximity condition does not impose any constraints on the second tree. Each vertex in the second tree can be connected with any of the remaining ones, as long as no cycle occurs (otherwise the graph would not be a tree). In the second tree of Figure S2, vertices $(1, 2)$ and $(1, 3)$ as well as vertices $(1, 2)$ and $(1, 4)$ are connected. Because each star on three variables also forms a path, the resulting third tree is then fixed.

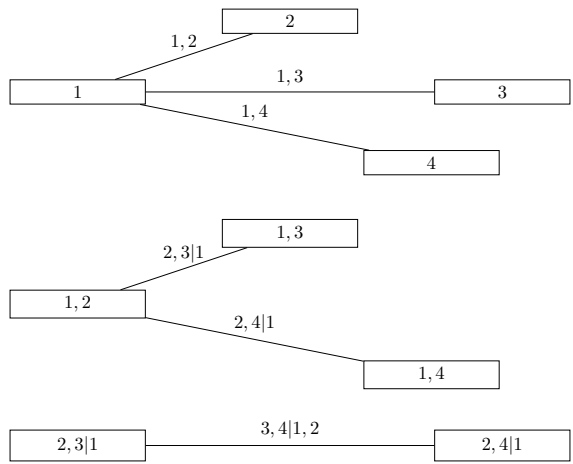

Figure S2: A four-dimensional C-vine.

## S1.4 R-vine matrices

Any R-vine can be represented compactly in an upper-left triangular matrix

$$M = (m_{i,j})_{i,j=1,\ldots,d}.$$

In particular, all edge labels can be read out of this matrix. Using notations from the main manuscript, the $j$th edge $e = (a_e, b_e|D_e)$ in the $k$th tree is encoded by the entries $(m_{d+1-j,j}, m_{k,j}|m_{k-1,j}, \ldots, m_{1,j})$. Less formally, the diagonal and $k$th element of the $j$th column form the conditioned set, the remaining entries above the $k$th element form the conditioning set. For example, the R-vine matrix of a four-dimensional D-vine is

$$M_D = \begin{bmatrix} 3 & 2 & 1 & 1 \\ 2 & 1 & 2 & 0 \\ 1 & 3 & 0 & 0 \\ 4 & 0 & 0 & 0 \end{bmatrix}$$

and its graphical representation is shown in Figure S1. From $M$, we are able to reconstruct the trees of the regular vine and vice versa. From the first column of the matrix $M_D$, we read out edge labels $(3,4), (2,4|3)$ and $(1,4|2,3)$. Similarly, the second column contains information on edge labels $(2,3)$ and $(1,3|2)$. The third column indicates the edge label $(1,2)$.

The R-vine matrix of a four-dimensional C-vine is

$$M_C = \begin{bmatrix} 1 & 1 & 1 & 1 \\ 2 & 2 & 2 & 0 \\ 3 & 3 & 0 & 0 \\ 4 & 0 & 0 & 0 \end{bmatrix}$$

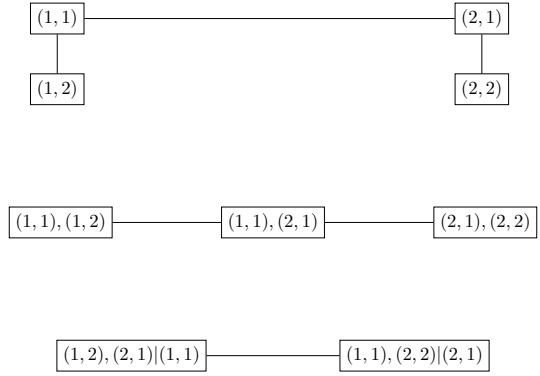

Figure S3: A vine on two variables and three time points.

and its graphical representation is shown in Figure S2. From $M_C$, we are able to reconstruct the trees of the regular vine and vice versa. Later, edge labels also indicate all pair-copulas of a vine-copula.

The matrix representation is unique as soon as the diagonal entries are fixed. For details, we refer to Dissmann et al. (2013) or Czado (2019), who use lower-left or upper-right triangular matrices, respectively.

## S1.5 Translation invariance

We now move on to the time series context. Here the vertices of the first vine tree are identified by tuple $(t, j)$, where the index $t$ indicates the time point and the index $j$ the variable. Figure S3 shows such a vine for two variables and two time points.

**Definition 4** (Translation invariance). *A vine copula model $(\mathcal{V}, \mathcal{C}(\mathcal{V}))$ on the set $V_1 = \{1, \ldots, T\} \times \{1, \ldots, d\}$ is called* translation invariant *if $c_{a_e, b_e | D_e} = c_{a_{e'}, b_{e'} | D_{e'}}$ holds for all edges $e, e' \in \bigcup_{k=1}^{T d - 1} E_k$ for which there is $\tau \in \mathbb{Z}$ such that*

$$a_e = a_{e'} + (\tau, 0), \quad b_e = b_{e'} + (\tau, 0), \quad D_e = D_{e'} + (\tau, 0),$$

*where the last equality is short for $D_e = \{v + (\tau, 0) \colon v \in D_{e'}\}$.*

Translation invariance is a condition on the pair-copulas associated with edges of a vine. For every edge, we check if the same edge appears shifted in time. If so, the copula associated with the two edges must be the same. For example, consider the edges $e_1 = \big((1,1), (1,2)\big)$ and $e_2 = \big((2,1), (2,2)\big)$ in Figure S3. Note that the edge $e_2$ is obtained by adding 1 to the time index of both vertices of $e_1$, i.e., $e_2$ is the same as $e_1$ but shifted by one time point. Therefore, translation invariance requires $c_{(1,1),(1,2)} = c_{(2,1),(2,2)}$. One can further check that this is the only invariance constraint in this example.

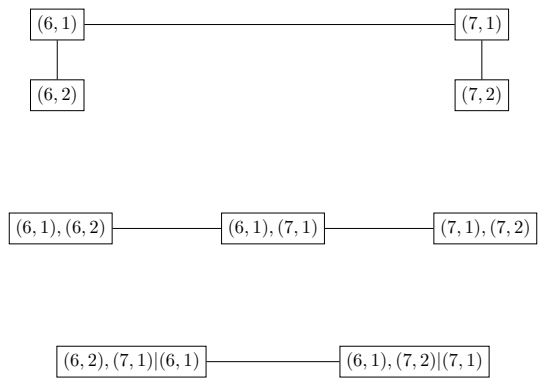

Figure S4: A translation of the vine in [Figure S3](#). All vertices and edges are shifted by 5 time steps.

## S1.6 Restriction of vines

**Definition 5** (Restriction of vines). *Let $\mathcal{V} = (V_k, E_k)_{k=1}^{Td-1}$ be a vine on $\{1 \dots, T\} \times \{1, \dots, d\}$ and $V_1' = \{t, \dots, t+m\} \times \{1, \dots, d\}$ for some $t, m$ with $1 \leq t \leq T$, $0 \leq m \leq T - t$. For all $k \geq 1$, define $E_k' = E_k \cap \binom{V_k'}{2}$ and $V_{k+1}' = E_k'$. Then the sequence of graphs $\mathcal{V}_{t,t+m} = (V_k', E_k')_{k=1}^{(m+1)d-1}$ is called* restriction of $\mathcal{V}$ *on the time points* $t, \dots, t+m$.

The graphical interpretation of this definition is straightforward. In the first tree of the regular vine $\mathcal{V}$, one deletes all vertices not in $V_1'$ along with corresponding edges. In the remaining trees, all edges and vertices of the tree $(V_k, E_k)$ affected by the deletion in the previous tree $(V_{k-1}, E_{k-1})$ are discarded. Or even more simply: to restrict a vine on time points $t$ to $t+m$, we delete all edges and vertices where time indices outside the range $[t, t+m]$ appear in the labels. The restriction $\mathcal{V}_{1,1}$ of the vine shown in [Figure S3](#) to time point $t = 1$, we therefore delete all vertices and edges where a time index $t = 2$ appears. This leaves us with only the vertices $V_1' = \{(1,1),(1,2)\}$ and the edge $E_1' = \{((1,1),(1,2))\}$.

## S1.7 Translation of vines

**Definition 6.** *Let $m \geq 0$, and $\mathcal{V} = (V_k, E_k)_{k=1}^{(m+1)d-1}$ be vine on $\{t, \dots, t+m\} \times \{1, \dots, d\}$ and $\mathcal{V}' = (V_k', E_k')_{k=1}^{(m+1)d-1}$ be a vine on $\{s, \dots, s+m\} \times \{1, \dots, d\}$. We say that $\mathcal{V}$ is a* translation *of $\mathcal{V}'$ (denoted by $\mathcal{V} \sim \mathcal{V}'$) if for all $k = 1, \dots, d-1$ and edges $e \in E_k$, there is an edge $e' \in E_k'$ such that $e = e' + (t - s, 0)$ (and vice versa).*

The translation of a vine $\mathcal{V}$ on $\{t, \dots, t+m\} \times \{1, \dots, d\}$ is obtained by shifting all vertices and edges in time by the same amount. An exemplary translation of the vine shown in [Figure S3](#) is given in [Figure S4](#). All vertices and edges are shifted by five time steps.

## S1.8 Compatible permutations

**Definition 8** (Compatible permutations). *We call a permutation* $(i_1, \ldots, i_d)$ *of* $(1, \ldots, d)$ compatible *with a vine* $\mathcal{V}$ *on* $\{1, \ldots, d\}$ *if for all* $k = 2, \ldots, d$, *there is an edge* $e \in E_{k-1}$ *with conditioned set* $\{i_k, i_r\}$ *and conditioning set* $\{i_1, \ldots, i_{k-1}\} \setminus i_r$ *for some* $r \in \{1, \ldots, k-1\}$.

he first index of the permutation $(i_1)$ is not constrained by compatibility, but the remaining ones are. A permutation is only compatible if the vine contains the edge $\{i_2, i_1\}$ in the first tree. Further, the vine must contain an edge with either a) conditioned set $\{i_3, i_1\}$ and conditioning set $\{i_2\}$, or b) an edge with conditioned set $\{i_3, i_2\}$ and conditioning set $\{i_1\}$, etc.

As an example, consider the vine in Figure S1. The first index $i_1$ is not constrained, so we may make the arbitrary choice $i_1 = 4$. Then we require, an edge with conditioned set $\{i_2, 4\}$ and empty conditioning set. Since there is only one such edge, it must hold $i_2 = 3$. Two examples of compatible conditions for this vine are $(1, \ldots, 4)$ as used by the D-vine model of Smith (2015) and $(4, \ldots, 1)$ as in the M-vine model of Beare and Seo (2015).

For tree $k$, the entries $i_k$ and $j_k$ of the compatible permutations $(i_1, \ldots, i_d)$ and $(j_1, \ldots, j_d)$, respectively, indicate how cross-sectional trees are connected. A cross-sectional vertex at time $t$ with $(t, i_k)$ in the conditioned set is connected to a vertex at time $t + 1$ with $(t + 1, j_1)$ in the conditioned set. The conditioning set of the connecting edge label consists of $(t, i_{k-1}), \ldots, (t, i_1)$. Conversely, the index $j_k$ indicates that a cross-sectional vertex at time $t + 1$ with $(t, j_k)$ in the conditioned set is connected to a vertex at time $t$ with $(t, i_1)$ in the conditioned set. The conditioning set of the connecting edge label consists of $(t, j_{k-1}), \ldots, (t, j_1)$. For $\mathcal{V}^{(0)}$, the compatible permutation $(i_1, \ldots, i_d)$ determines *out-vertices* of its cross-sectional trees and the compatible permutation $(j_1, \ldots, j_d)$ determines *in-vertices* of its cross-sectional trees.

## S2 Vine copula models for multivariate time series: additional illustrations

Suppose $(\boldsymbol{X_t})_{t=1,\ldots,T} = (X_{t,1},\ldots,X_{t,d})'_{t=1,\ldots,T}$ is a strictly stationary time series, whose cross-sectional and temporal dependence we model by a vine copula. By fixing the stationary marginal distributions $F_1,\ldots,F_d$, it suffices to consider time series of marginally standard uniform variables $(\boldsymbol{U_t})_{t=1,\ldots,T} = (U_{t,1},\ldots,U_{t,d})'_{t=1,\ldots,T}$, where $U_{t,1} = F_1(X_{t,1}),\ldots,U_{t,d} = F_d(X_{t,d})$.

The existing models make specific choices for the cross-sectional structure and connecting edge. To trace the marginal variables and the time point, we use variable sub-indices with two numbers $t$ and $i$ separated by a comma, i.e. $U_{t,i}$. The first sub-index $t$ indicates the time point and the second sub-index $i$ determines the marginal variable. In the time series context, each vertex of a vine's first tree is identified with a tuple $(t,i)$, where $t$ is the time index and $i$ is the variable index. The vertex $(t,i)$ corresponds to the random variable $U_{t,i}$. In particular, the components of edge labels $e = (a_e, b_e | D_e)$ (i.e., $a_e, b_e$ as well as elements of $D_e$) are tuples.

### S2.1 D-vine of Smith (2015)

In the *D-vine* of Smith (2015), the cross-sectional structure is a D-vine and two cross-sectional D-vines at time points $t$ and $t+1$ are connected at the two distinct variables (vertices) that lie at their opposite borders. Without loss of generality, we can assume that the last variable at time $t$ is connected with the first variable at time $t+1$, i.e. an edge is added for $(U_{t,d}, U_{t+1,1})$. With these choices, there is only one global vine model satisfying proximity condition, which is a long D-vine spanning all variables at all time points. Now, a technical advantage of using a D-vine for cross-sectional dependence becomes clear, since all trees of the D-vine are uniquely specified. Moreover, the cross-sectional D-vines are combined in one global D-vine. Therefore, there is no need in any further rules for tree constructions.

Figure S5 illustrates the first five trees of a four-dimensional D-vine on three time points. For a better visibility, we drop edge labels since they are vertex labels in subsequent figures. Variables $U_{t,i}$ for $t = 1, 2, 3$ and $i = 1, 2, 3, 4$ are expressed through their sub-indices in parentheses as $(t, i)$. Figure S6 shows the R-vine matrix of this long D-vine. The red entries of this matrix are related to the cross-sectional dependence at time points $t = 1, 2, 3$. If we ignore the first variable $t$ in $(t, i)$ then these entries are identical to the entries of the R-vine matrix for a four-dimensional D-vine. The black entries of this matrix, which are not on the diagonal, capture the cross-temporal dependence between $\boldsymbol{U_1}$ and $\boldsymbol{U_2}$ as well as between $\boldsymbol{U_2}$ and $\boldsymbol{U_3}$. Similarly, the blue entries of this matrix describe the cross-temporal dependence between $\boldsymbol{U_1}$ and $\boldsymbol{U_3}$. If the multivariate time series is a Markov process of order 1 then the pair copulas corresponding to the blue entries of this matrix are just independence copulas. If $\boldsymbol{U_1}, \boldsymbol{U_2}, \boldsymbol{U_3}$ are independent then all black and blue entries correspond to the independence copula.

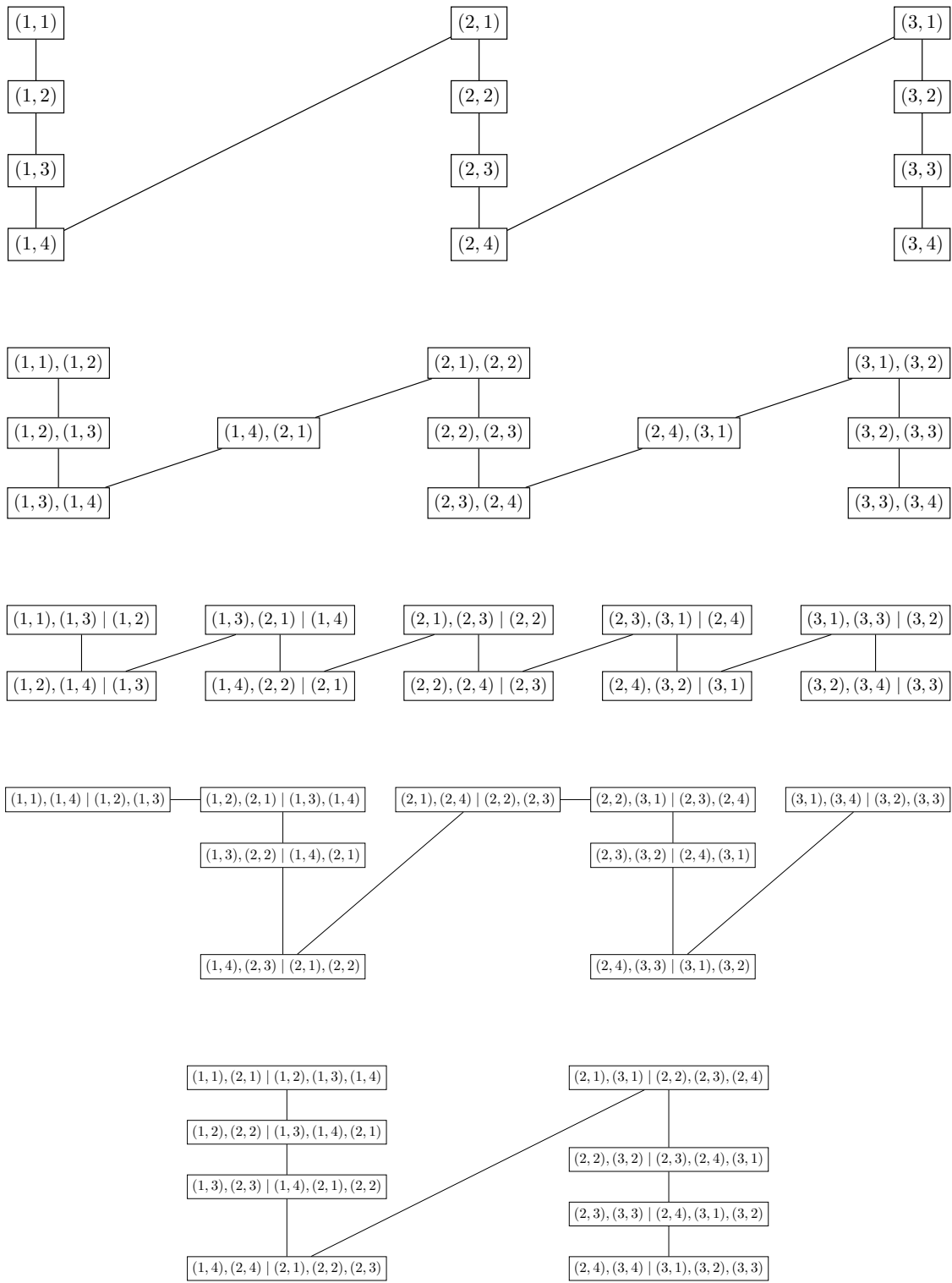

Figure S5: The first five trees of a four-dimensional D-vine on three time points.

$$\begin{pmatrix}
(3,3) & (3,2) & (3,1) & (2,4) & (2,3) & (2,2) & (2,1) & (1,4) & (1,3) & (1,2) & (1,1) & (1,1) \\
(3,2) & (3,1) & (2,4) & (2,3) & (2,2) & (2,1) & (1,4) & (1,3) & (1,2) & (1,1) & (1,2) & \\
(3,1) & (2,4) & (2,3) & (2,2) & (2,1) & (1,4) & (1,3) & (1,2) & (1,1) & (1,3) & & \\
(2,4) & (2,3) & (2,2) & (2,1) & (1,4) & (1,3) & (1,2) & (1,1) & (1,4) & & & \\
(2,3) & (2,2) & (2,1) & (1,4) & (1,3) & (1,2) & (1,1) & (2,1) & & & & \\
(2,2) & (2,1) & (1,4) & (1,3) & (1,2) & (1,1) & (2,2) & & & & & \\
(2,1) & (1,4) & (1,3) & (1,2) & (1,1) & (2,3) & & & & & & \\
(1,4) & (1,3) & (1,2) & (1,1) & (2,4) & & & & & & & \\
(1,3) & (1,2) & (1,1) & (3,1) & & & & & & & & \\
(1,2) & (1,1) & (3,2) & & & & & & & & & \\
(1,1) & (3,3) & & & & & & & & & & \\
(3,4) & & & & & & & & & & & 
\end{pmatrix}$$

Figure S6: Matrix representation of long D-vine (Smith) with $d = 4$ and $T = 3$.

## S2.2 M-vine of Beare and Seo (2015)

In the *M-vine* of Beare and Seo (2015), the cross-sectional structure is a D-vine and two crossectional D-vines at time points $t$ and $t + 1$ are connected at one variable that lies at the border of the D-vines. Without loss of generality, we can assume that the first variable at time $t$ is connected with the first variable at time $t + 1$, i.e. an edge is added for $(U_{t,1}, U_{t+1,1})$. With the additional restriction that vertices of adjacent time points are connected first, this also fixes all further trees of the vine (see also, Begin et al., 2020, for their connection to vector autoregressive models).

Figure S7 illustrates the first five trees of a four-dimensional M-vine on three time points. Again, variables are expressed through their sub-indices in parentheses. Figure S8 shows the corresponding R-vine matrix. The red entries of this matrix are related to the cross-sectional dependence at time points $t = 1, 2, 3$ and coincide with red entries from Figure S6. Similarly, black and blue entries of the matrix are related to the cross-temporal dependence for lag 1 and 2, correspondingly. A comparison of the R-vine matrices from Figures S6 and S8 shows that they have the same elements and the difference is only at their position/placement resulting in different vine-copulas. Further, red entries are identical due to the cross-sectional D-vine.

## S2.3 COPAR of Brechmann and Czado (2015)

In the *COPAR* of Brechmann and Czado (2015), the cross-sectional structure is a C-vine and two C-vines at time points $t$ and $t + 1$ are connected at the root vertex of the C-vine. Without loss of generality, we can assume that the root variable is the first variable, i.e. an edge is added for $(U_{t,1}, U_{t+1,1})$. This leaves a lot of flexibility for higher trees and the authors settled on a specific set of rules.

In particular, the model contains all edges of a D-vine on the variables $U_{1,1}, U_{2,1}, \ldots, U_{T,1}$. Further serial dependencies are modelled by iteratively conditioned D-vine copulas and appropriate between-series copulas. To specify the COPAR model, let us introduce the vector $\boldsymbol{U}_{s:t,i:j}$

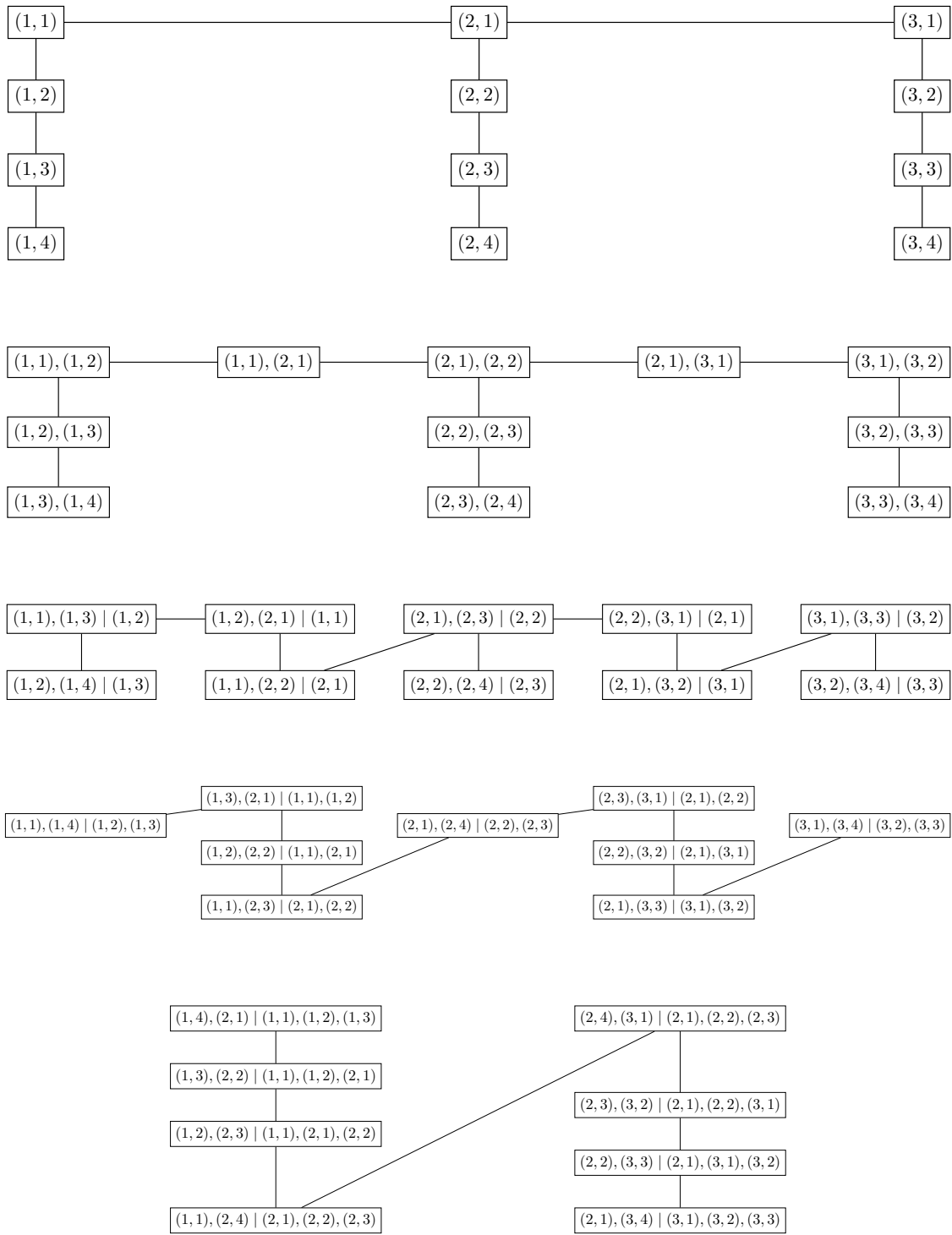

Figure S7: The first five trees of a four-dimensional M-vine on three time points.

$$\begin{pmatrix}
(3,3) & (3,2) & (3,1) & (2,1) & (2,3) & (2,2) & (2,1) & (1,1) & (1,3) & (1,2) & (1,1) & (1,1) \\
(3,2) & (3,1) & (2,1) & (2,2) & (2,2) & (2,1) & (1,1) & (1,2) & (1,2) & (1,1) & (1,2) & \\
(3,1) & (2,1) & (2,2) & (2,3) & (2,1) & (1,1) & (1,2) & (1,3) & (1,1) & (1,3) & & \\
(2,1) & (2,2) & (2,3) & (2,4) & (1,1) & (1,2) & (1,3) & (1,4) & (1,4) & & & \\
(2,2) & (2,3) & (2,4) & (1,1) & (1,2) & (1,3) & (1,4) & (2,1) & & & & \\
(2,3) & (2,4) & (1,1) & (1,2) & (1,3) & (1,4) & (2,2) & & & & & \\
(2,4) & (1,1) & (1,2) & (1,3) & (1,4) & (2,3) & & & & & & \\
(1,1) & (1,2) & (1,3) & (1,4) & (2,4) & & & & & & & \\
(1,2) & (1,3) & (1,4) & (3,1) & & & & & & & & \\
(1,3) & (1,4) & (3,2) & & & & & & & & & \\
(1,4) & (3,3) & & & & & & & & & & \\
(3,4) & & & & & & & & & & &
\end{pmatrix}$$

Figure S8: Matrix representation of M-vine with $d = 4$ and $T = 3$ .

given by

$$\boldsymbol{U}_{s:t,i:j} = (U_{s,i}, \ldots, U_{t,i}, \ldots, U_{s,j}, \ldots, U_{t,j})',$$

where $1 \leq s \leq t \leq T$ and $1 \leq i \leq j \leq d$. For $i > j$ or $s > t$, the vector $\boldsymbol{U}_{s:t,i:j}$ is just the empty set. The serial dependence of the first variable is modelled for $U_{s,1}$ and $U_{t,1}$ given $\boldsymbol{U}_{(s+1):(t-1),1}$ for $1 \leq s < t \leq T$. Between series dependence of variable $i$ and variable $j$ for $1 \leq i < j \leq d$ is modelled for $U_{s,i}$ and $U_{t,j}$ given $\boldsymbol{U}_{1:t,1:(i-1)}$ and $\boldsymbol{U}_{(s+1):t,i}$ for $1 \leq s \leq t \leq T$ as well as for $U_{t,i}$ and $U_{s,j}$ given $\boldsymbol{U}_{1:t,1:(i-1)}$, $\boldsymbol{U}_{1:(t-1),i:(j-1)}$ and $\boldsymbol{U}_{(s+1):(t-1),j}$ for $1 \leq s < t \leq T$. Finally, conditional serial dependence of variable $i$ with $2 \leq i \leq d$ is modelled for $U_{s,i}$ and $U_{t,i}$ given $\boldsymbol{U}_{1:t,1:(i-1)}$ and $\boldsymbol{U}_{(s+1):(t-1),i}$ for $1 \leq s < t \leq T$.

Figure S9 illustrates the first five trees of a four-dimensional COPAR vine on three time points. Again, variable sub-indices are displayed in parentheses. Figure S10 shows the corresponding R-vine matrix. The red entries of this matrix are related to the cross-sectional dependence at time points $t = 1, 2, 3$. Unfortunately, black entries cannot be related to to the cross-temporal dependence for lag 1 since it contains variables for $t = 1, 2, 3$. In general, the same conclusion holds also for the blue entries and we could observe it for $T = 4$.

## S2.4 Difference between M-vine and COPAR for bivariate time series

To illustrate the difference between the M-vine and COPAR, consider a two-dimensional time series with three times points, i.e. $d = 2$ and $T = 3$. Figure S12 shows the corresponding M-vine. Since adjacent time points are connected, vertex $\big((1,1),(2,1)\big)$ in the second tree is connected with vertices $\big((1,1),(1,2)\big)$ and $\big((2,1),(2,2)\big)$. Similarly, vertex $\big((2,1),(3,1)\big)$ in the second tree is connected with vertices $\big((2,2),(1,2)\big)$ and $\big((3,1),(3,2)\big)$.

Figure S11 shows the corresponding COPAR, whose first tree is identical to the first tree of the M-vine. The first difference appears in the second tree since the serial dependence in the COPAR is first modelled for the key variable, i.e. variable 1. Therefore, vertices $\big((1,1),(2,1)\big)$ and $\big((2,1),(3,1)\big)$ in the second tree are connected and the corresponding edge is in red. With

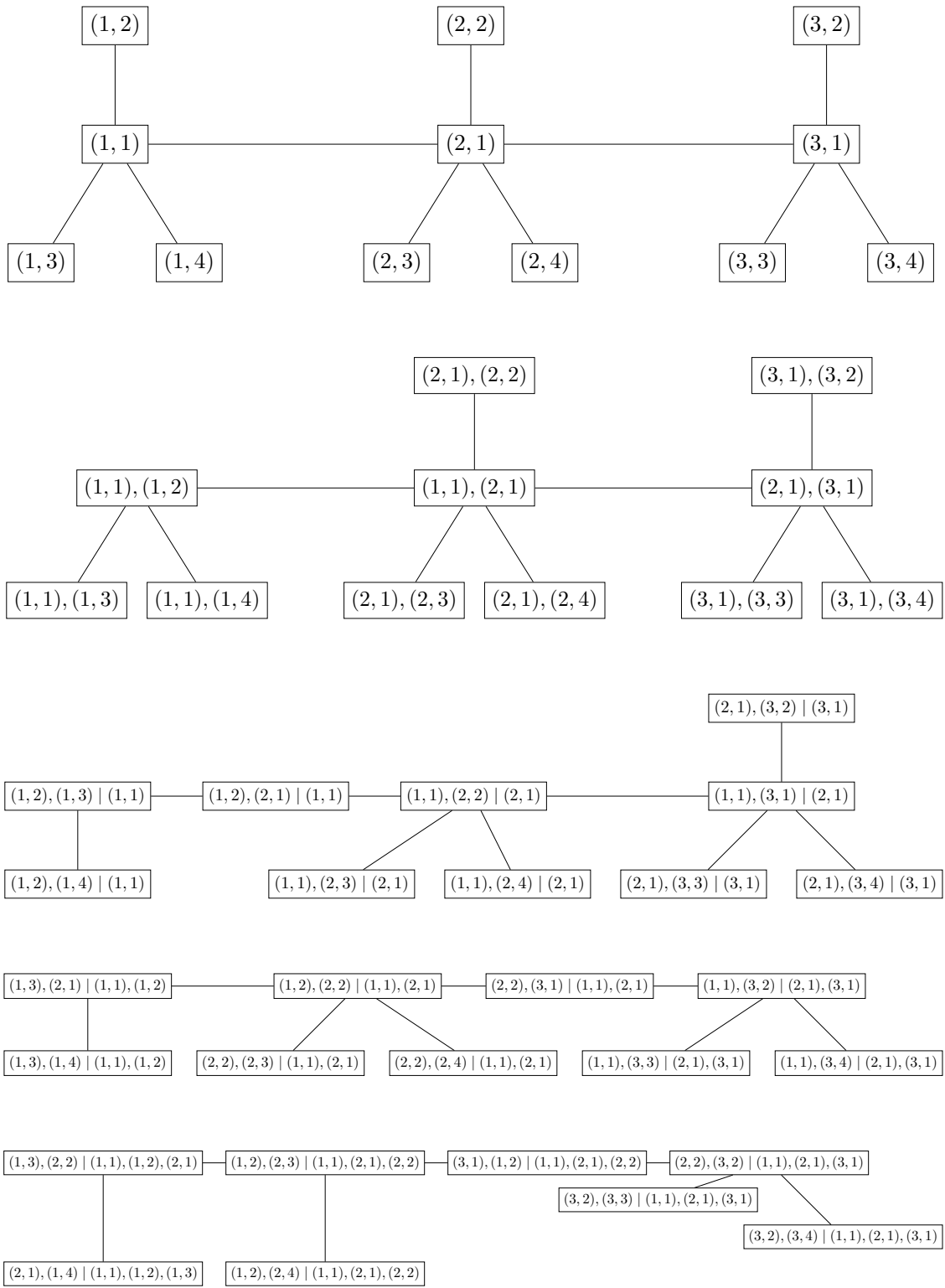

Figure S9: First five trees level of a four-dimensional COPAR on three time points.

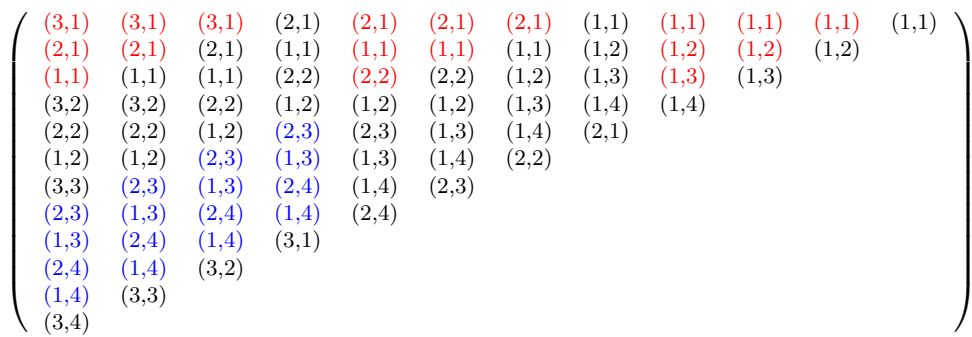

$$
\begin{pmatrix}
(3,1) & (3,1) & (3,1) & (2,1) & (2,1) & (2,1) & (2,1) & (1,1) & (1,1) & (1,1) & (1,1) & (1,1) \\
(2,1) & (2,1) & (2,1) & (1,1) & (1,1) & (1,1) & (1,1) & (1,2) & (1,2) & (1,2) & (1,2) \\
(1,1) & (1,1) & (1,1) & (2,2) & (2,2) & (2,2) & (1,2) & (1,3) & (1,3) & (1,3) \\
(3,2) & (3,2) & (2,2) & (1,2) & (1,2) & (1,2) & (1,3) & (1,4) & (1,4) \\
(2,2) & (2,2) & (1,2) & (2,3) & (2,3) & (1,3) & (1,4) & (2,1) \\
(1,2) & (1,2) & (2,3) & (1,3) & (1,3) & (1,4) & (2,2) \\
(3,3) & (2,3) & (1,3) & (2,4) & (1,4) & (2,3) \\
(2,3) & (1,3) & (2,4) & (1,4) & (2,4) \\
(1,3) & (2,4) & (1,4) & (3,1) \\
(2,4) & (1,4) & (3,2) \\
(1,4) & (3,3) \\
(3,4)
\end{pmatrix}
$$

Figure S10: Matrix representation of COPAR with $d = 4$ and $T = 3$.

respect to the M-vine, this results in a different vertex $\big((1,1),(3,1)|(2,1)\big)$ in the third tree. In the corresponding M-vine, vertex $\big((2,1),(3,1)\big)$ should be connected with vertex $\big((2,1),(2,2)\big)$ from the adjacent cross sectional D-vine. To illustrate the difference to the M-vine, we have displayed all different edges of the COPAR in red. These red edges will correspond to different pair-copula in the COPAR vine representation with respect to the M-vine.

As it is mentioned in the main manuscript, COPAR are not stationary in general. The problem already occurs in its second tree. The edge connecting $\big(1,1),(1,2)\big)$ with $\big(1,1),(2,1)\big)$ characterizes the conditional dependence $\big((1,2),(2,1) \mid (1,1)\big)$. However, it does not appear again as a translation. Translation invariance is then insufficient for ensuring that the shifted conditional dependence $\big((2,2),(3,1) \mid (2,1)\big)$ is identical.

## S2.5 S-vine

There is obvious potential for generalization of M-vines. First, we would like to allow for arbitrary R-vines in the cross-sectional structure. Second, we would like to connect two cross-sectional trees at arbitrary variables. Specific verisons of such models were constructed in preliminary work by Krüger (2018) (called *temporal vine (T-vine)*) and in unpublished work by Harry Joe. But where should we stop? In principle, we could take any $(T \times d)$-dimensional vine as a model for the vector $(\boldsymbol{U}_1, \ldots, \boldsymbol{U}_T)$.

Figure S13 illustrates the first five trees of a five-dimensional S-vine on three time points. Again, variables are expressed through their sub-indices in parentheses. We consider a five-dimensional time series since four-dimensional vines are either D-vines or C-vines. More general R-vine structures appear starting from dimension 5.

For S-vine in Figure S13, two compatible permutations from Theorem 2 of the main manuscript are given by out-vertices $(i_1, i_2, i_3, i_4, i_5) = (2, 1, 3, 4, 5)$ and in-vertices $(j_1, j_2, j_3, j_4, j_5) = (3, 4, 5, 2, 1)$. Let us describe the connection of the vertices for the first $d = 5$ trees, since the remaining trees are D-vines, which are uniquely determined. In tree $k$ for $k = 1, \ldots, 5$, the cross-sectional out-vertex with $(t, i_k)$ in the conditioned set is connected to vertex with $(t+1, j_1)$

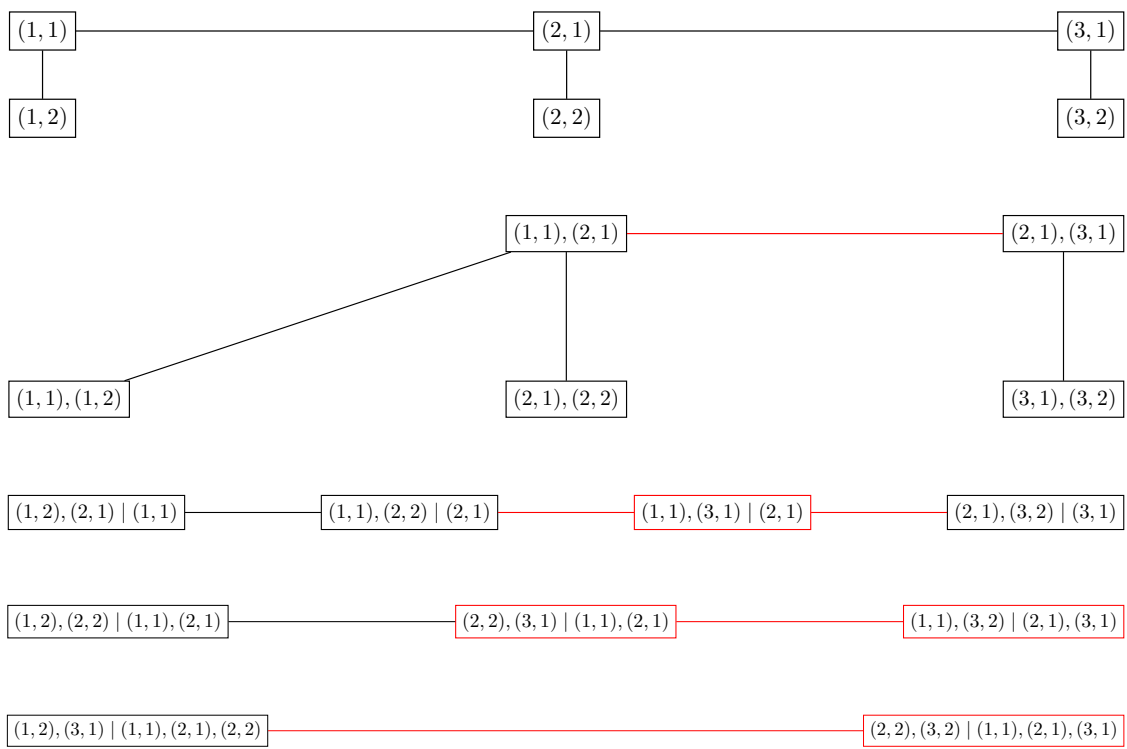

Figure S11: Five trees of a two-dimensional COPAR on three time points.

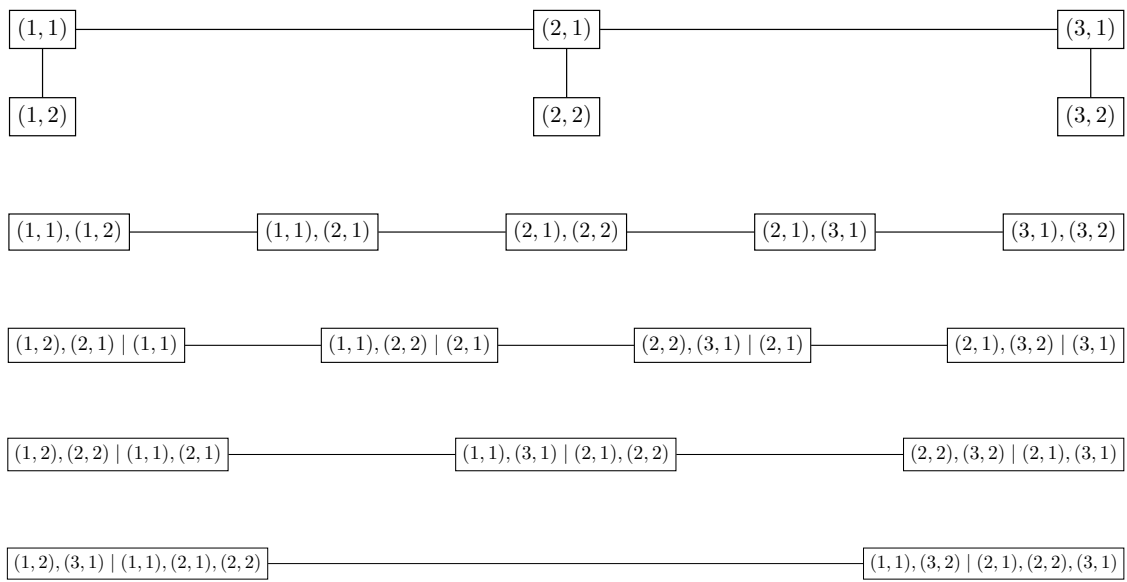

Figure S12: Five trees of a two-dimensional M-vine on three time points.

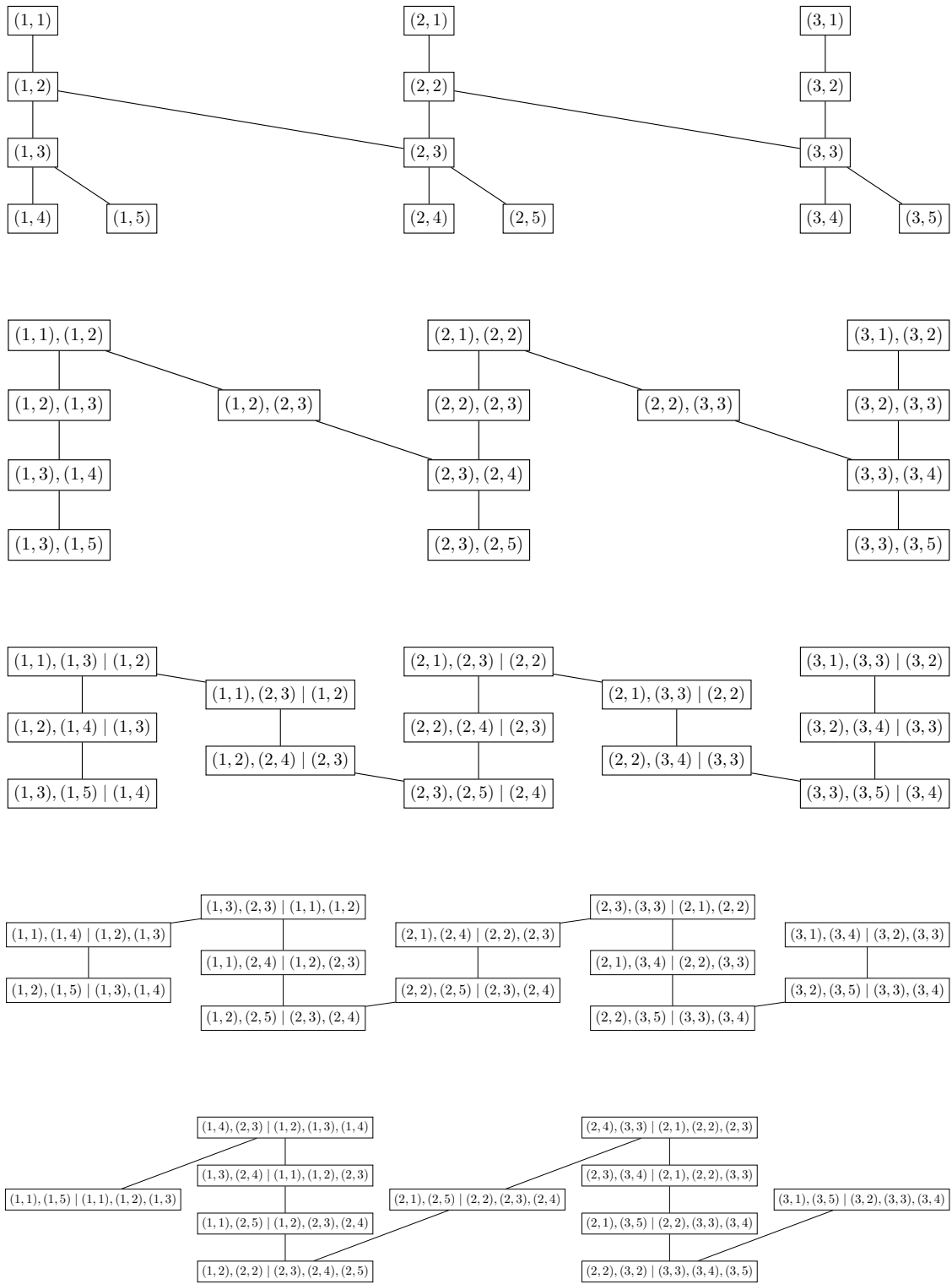

Figure S13: First five trees level of a five-dimensional S-vine on three time points.

$$\begin{pmatrix}
* & * & * & \iddots & \iddots & (1,i_1) & * & * & * & (1,j_1) \\
* & * & \iddots & \iddots & \iddots & \vdots & * & * & \iddots \\
* & \iddots & \iddots & \iddots & \iddots & (1,i_{d-1}) & * & (1,j_{d-1}) \\
(T-1,i_1) & \iddots & \iddots & \iddots & \iddots & (1,i_d) & (1,j_d) \\
\vdots & \iddots & \iddots & \iddots & \iddots & (2,j_1) \\
\vdots & \iddots & \iddots & \iddots & \iddots \\
(1,i_1) & \iddots & \iddots & \iddots \\
\vdots & \iddots & \iddots \\
(1,i_d) & (n,j_{d-1}) \\
(T,j_d)
\end{pmatrix}$$

Figure S14: Matrix representation of stationary vines.

in the conditioned set for $t = 1, 2$. Therefore, the corresponding edge label is given by $(t, i_k)$ and $(t + 1, j_1)$ conditioned on $(t, i_{k-1}), \ldots, (t, i_1)$. This procedure is similar for the in-vertices. In tree $k$ for $k = 1, \ldots, 5$, the cross-sectional in-vertex with $(t + 1, j_k)$ in the conditioned set is connected to vertex with $(t, i_1)$ in the conditioned set for $t = 1, 2$. Therefore, the corresponding edge label is given by $(t, j_k)$ and $(t + 1, i_1)$ conditioned on $(t, j_{k-1}), \ldots, (t, j_1)$.

Let us describe the out-vertices at time point $t = 1$ and the in-vertices at time point $t = 2$ for the first three trees. In tree 1, the cross-sectional out-vertex is $(1, 2)$ and the respective in-vertex is $(2, 3)$ corresponding to $i_1 = 2$ and $j_1 = 3$, respectively. In tree 2, the cross-sectional out-vertex is $\big((1,1),(1,2)\big)$ and the respective in-vertex is $\big((2,3),(2,4)\big)$ corresponding to $i_2 = 1$ and $j_2 = 4$, respectively. In tree 3, the cross-sectional out-vertex is $\big((1,1),(1,3)|(1,2)\big)$ and the respective in-vertex is $\big((2,3),(2,5)|(2,4)\big)$ corresponding to $i_3 = 3$ and $j_3 = 5$, respectively.

Theorem 2 implies that all stationary vines can be represented by a matrix of the form shown in Figure S14, where the stars correspond to the cross-sectional structure.

The class of S-vines also allows to construct a stationary version of the COPAR model. More specifically, we keep the first tree of the COPAR, but construct later trees in a way that preserves stationarity. Five trees of the corresponding vine are illustrated in Figure S15. Starting from the fifth tree, its trees coincide with those of the M-vine.

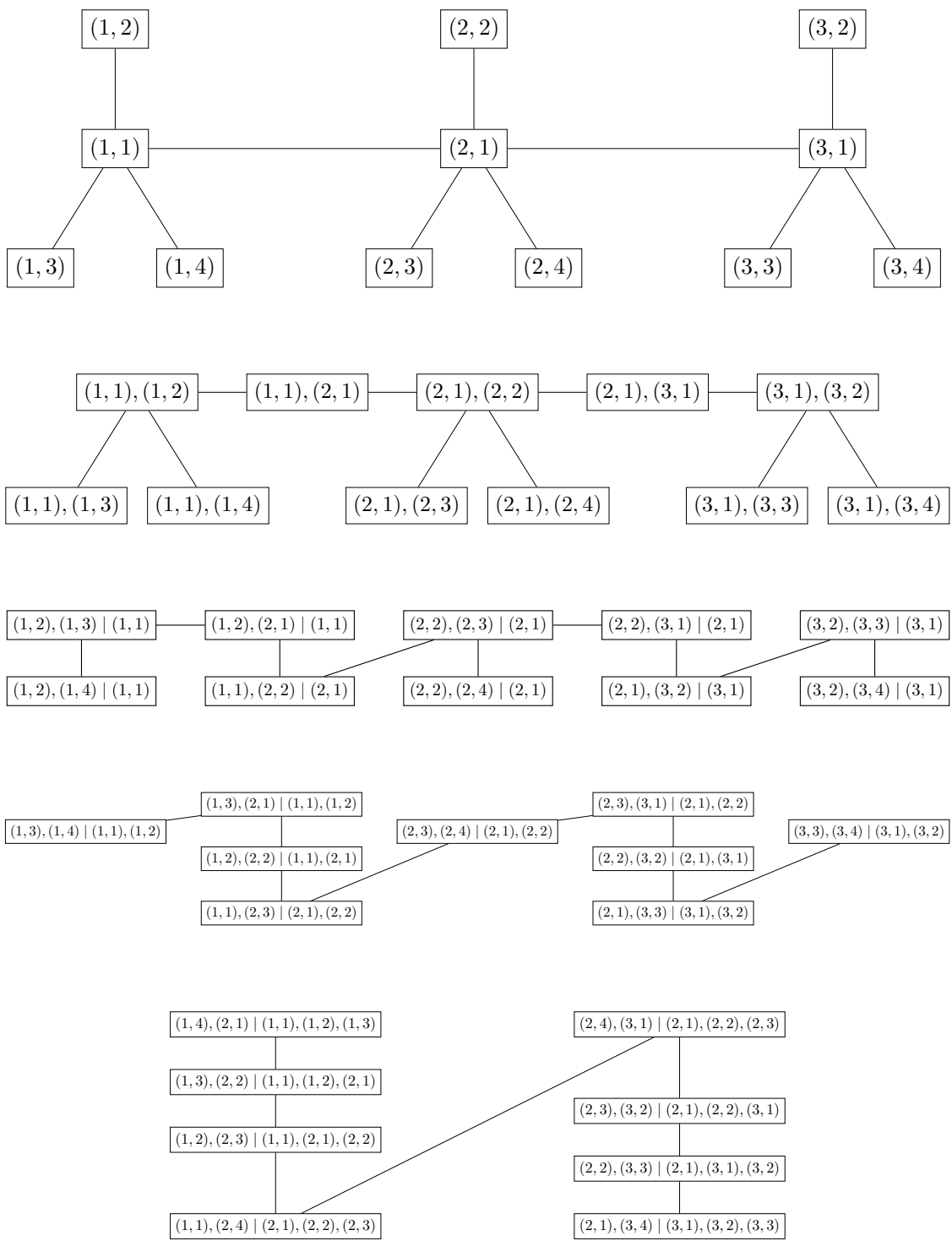

Figure S15: First five trees level of a four-dimensional S-vine on three time points with a cross-section C-vine.

# S3 Implementation details

## S3.1 Model selection

Parametric models are easy to select in a fully step-wise procedure. The most common selection criteria are AIC or BIC. For the margins, these criteria are computed from the maximal log-likelihood

$$\max_{\boldsymbol{\eta}_j} \sum_{t=1}^{n} \ln f_j(X_{t,j}; \boldsymbol{\eta}_j), \quad j = 1, \ldots, d.$$

Then we continue step-wise with the pair-copulas, where AIC and BIC are computed from the maximal pair-copula log-likelihood

$$\max_{\boldsymbol{\theta}_{[e']}} \sum_{e \sim e'} \ln c_{[e]} \big\{ C_{a_{[e]}|D_{[e]}}(\widehat{U}_{a_e} \mid \widehat{\boldsymbol{U}}_{D_e}; \widehat{\boldsymbol{\theta}}_{S_a([e])}), C_{b_{[e]}|D_{[e]}}(\widehat{U}_{b_e} \mid \widehat{\boldsymbol{U}}_{D_e}; \widehat{\boldsymbol{\theta}}_{S_b([e])}); \boldsymbol{\theta}_{[e']} \big\}.$$

For the vine structure of stationary time series, we propose a similar heuristic extending the popular algorithm of Dissmann et al. (2013). Its idea is to capture the strongest dependencies as early as possible in the tree structure. In addition to the cross-sectional structure $\mathcal{V}^{(0)}$, we also need to select the in- and out-vertices $(j_1, \ldots, j_d)$ and $(i_1, \ldots, i_d)$.

We first focus on the cross-sectional structure. We compute the (absolute) empirical Kendall's $\tau$ between all pairs of variables and find the maximum spanning tree. Then we find the optimal in-/out-vertices by computing all pair-wise empirical Kendall's $\tau$ between the original time series and a lagged version and choose the edge with maximal $|\tau|$. Now the first tree is completely specified. Then we estimate all parameters in this tree and generate pseudo-observations for the next. We again build a maximum spanning tree for the cross-sectional part with the proximity condition as side constraints. Then we find the compatible in-/out-vertices by maximizing absolute Kendall's $\tau$ for the corresponding edges in Theorem 2 (ii). We continue this way until the first $d$ trees are selected, which completely determines the remaining structure.

The procedure can be simplified for M- and D-vines by imposing appropriate constraints on the cross-sectional structure (which becomes a shortest path problem) and the in-/out-vertices (where only $i_1$ is a free parameter). It must be emphasized that all these methods are heuristic and give no guarantees to find the optimal model in their class.

Finally, the Markov order is selected by adding lags to model as long as the overall AIC improves.

**Algorithm 1** Simulating $K$ time steps from a $p$-Markovian vine copula model conditional on the past.

---

**Input**: A $p$-Markovian, stationary vine copula model $(\mathcal{V}, \mathcal{C}(\mathcal{V}))$, stationary marginal distributions $F_1, \ldots, F_d$, and past observations $\boldsymbol{X}_{t-p+1}, \ldots, \boldsymbol{X}_t$.

**Output**: Samples $\boldsymbol{X}_{t+1}, \ldots, \boldsymbol{X}_{t+K}$ from the conditional distribution given $\boldsymbol{X}_{t-p+1}, \ldots, \boldsymbol{X}_t$.

---

(i) Set $U_{t-p+k,j} = F_j(X_{t-p+k,j})$ for $k = 1, \ldots, p$, $j = 1, \ldots, d$.

(ii) **for** $k = 0, \ldots, K-1$:

    a) Set $(\boldsymbol{W}_1, \ldots, \boldsymbol{W}_p) = R_{\mathcal{C}(\mathcal{V}_{t-p+1,t})}(\boldsymbol{U}_{t-p+1+k}, \ldots, \boldsymbol{U}_{t+k})$.

    b) Simulate $\boldsymbol{W}_{p+1} \sim \mathrm{Uniform}([0,1]^d)$.

    c) Set $(\boldsymbol{U}_{t-p+1+k}, \ldots, \boldsymbol{U}_{t+k+1}) = R^{-1}_{\mathcal{C}(\mathcal{V}_{t-p+1,t+1})}(\boldsymbol{W}_1, \ldots, \boldsymbol{W}_{p+1})$.

    **end for**

(iii) Set $X_{t+k,j} = F_j^{-1}(U_{t+k,j})$ for $k = 1, \ldots, K$, $j = 1, \ldots, d$.

## S3.2 Simulation algorithm

The simulation algorithm is based on the *Rosenblatt transform*, whose inverse also appears in the standard algorithm. The Rosenblatt transform applies certain conditional distribution functions to a random vector to turn it into independent uniforms. Conversely, the inverse Rosenblatt transform turns independent uniforms into a vector with arbitrary joint distribution. Let $M = (m_{i,j})_{i,j=1}^d$ be the R-vine matrix corresponding to a $d$-dimensional vine copula model $(\mathcal{V}, \mathcal{C}(\mathcal{V}))$. Our version of the Rosenblatt transform is defined as $R_{\mathcal{C}(\mathcal{V})} \colon [0,1]^d \mapsto [0,1]^d$ with

$$\left(R_{\mathcal{C}(\mathcal{V})}(\boldsymbol{u})\right)_{m_{d+1-j,j}} = C_{m_{d+1-j,j}|m_{d-j,j},\ldots,m_{1,j}}(u_{m_{d+1-j,j}} \mid \boldsymbol{u}_{m_{d-j,j},\ldots,m_{1,j}}), \quad j = 1, \ldots, d.$$

The conditional distributions in this formula can be computed recursively from pair-copulas in $\mathcal{C}(\mathcal{V})$ as explained in Section [Section 2.3](). The inverse transformation is

$$\left(R^{-1}_{\mathcal{C}(\mathcal{V})}(\boldsymbol{u})\right)_{m_{d+1-j,j}} = C^{-1}_{m_{d+1-j,j}|m_{d-j,j},\ldots,m_{1,j}}(u_{m_{d+1-j,j}} \mid \boldsymbol{u}_{m_{d-j,j},\ldots,m_{1,j}}), \quad j = 1, \ldots, d,$$

which can also be computed recursively from pair-copulas in $\mathcal{C}(\mathcal{V})$, see [Czado](https://) ([2019]()).

[Algorithm 1]() can also be used to simulate unconditionally by giving it an unconditional draw $(\boldsymbol{U}_{t-p+1}, \ldots, \boldsymbol{U}_t)$ from the vine copula model $(\mathcal{V}_{t-p+1,t}, \mathcal{C}(\mathcal{V}_{t-p+1,t}))$ as an input. Note that in step (ii) c), only $U_{t+k+1}$ changes; all other entries on the left remain unchanged by the definition of the (inverse) Rosenblatt transform.

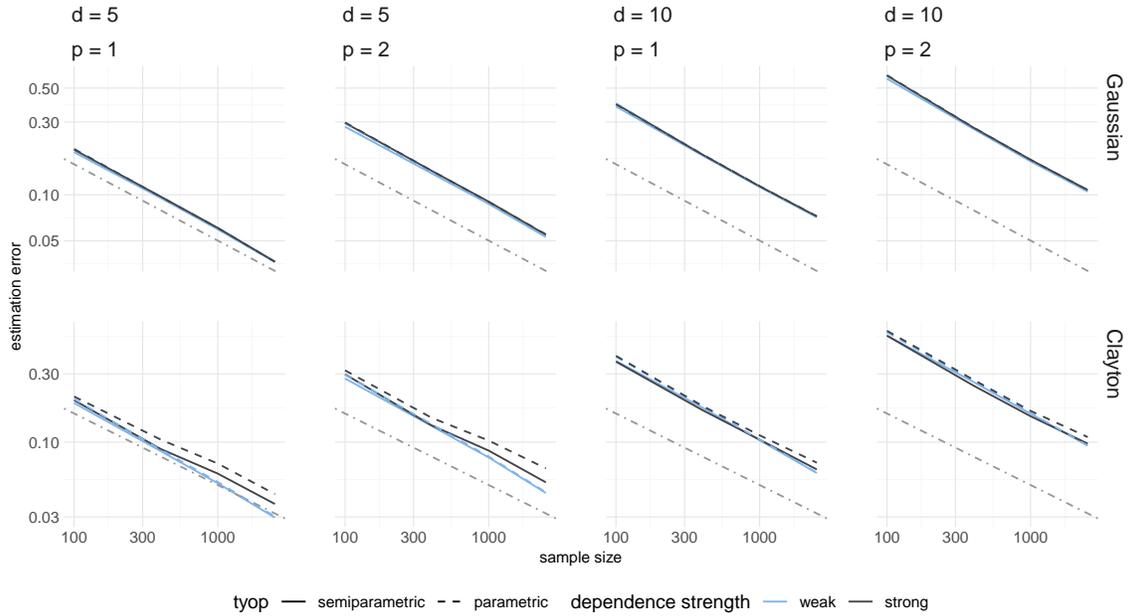

Figure S16: Estimation error $\sum_k \sum_{e \in E_k} |k\widehat{\tau}_e - \tau_0| / \sum_k |E_k|$ of the step-wise estimator in an S-vine model with Gaussian margins. Both axes have logarithm scale. The errors decay with rate $T^{-1/2}$ if they are parallel to the dash-dotted line.

## S4 Numerical experiments

To validate the methodology proposed in the paper, we work with the following default setup: We repeatedly simulate time series of length $T = 100, 400, 1\,000, 2\,500$ from models of varying size (cross-sectional dimension $d = 5, 10$, Markov order $p = 1, 2$), containing either only Gaussian or only Clayton pair-copulas. The pair-copula parameters are set such that the Kendall's $\tau$ in tree $k$ equals $\tau_0/k$ with either $\tau_0 = 0.2$ (weak dependence) or $\tau_0 = 0.7$ (strong dependence). This decay in dependence strength serves two purposes. First, it provides control over the strength of dependence for implicit pairs. Our specific choice produces models where the implicit pairs have a Kendall's $\tau$ of approximately the same magnitude than the explicit ones. Second, it reflects the common practice to work with vine structure that capture strongest dependencies early (see Section S3.1). The marginal distributions are set to a heavy tailed Student t distribution with mean zero, unit scale, and three degrees of freedom.

### S4.1 Parameter estimation

The step-wise estimation procedure of Section 4 is validated in Figure S16. The x-axis shows the sample size and the y-axis the average estimation error for a rescaled version of Kendall's $\tau$: $\sum_k \sum_{e \in E_k} |k\widehat{\tau}_e - \tau_0| / \sum_k |E_k|$. The rescaling of $\widehat{\tau}_e$ by $k$ ensures that errors have comparable scale across trees. Both axes are on a logarithmic scale, so that $\sqrt{T}$-convergence would show as a straight line with slope $-1/2$. As a guide, a dash-dotted line with slope $-1/2$ is added to each

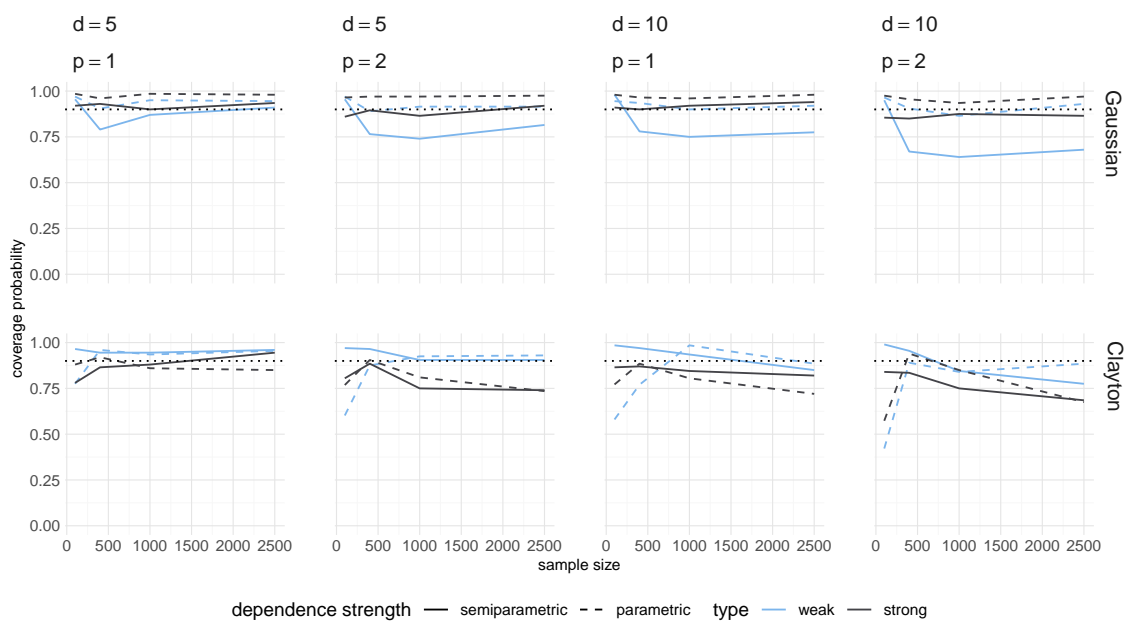

Figure S17: Coverage probabilities of bootstrapped confidence intervals for estimating the 90%-quantile of $(X_{t,1} + \cdots + X_{t,d})/d$. The dotted line indicates the nominal level of 90%.

plot. We see that $\sqrt{T}$-convergence is achieved in all scenarios. However, the vertical upwards shift from the left to right panels indicates that the estimation error increases with model size. The difference between parametric and nonparametric margins is generally negligible.

## S4.2 Uncertainty quantification

To validate the bootstrap method proposed in Section 6, we simulate from the same models as before, but now the goal is to estimate the conditional 90%-quantile of $\mu_t = (X_{t,1} + \cdots + X_{t,d})/d$ given the past. We fit the models, estimate the quantile of $\mu_t$, and construct 90%-confidence intervals using the bootstrap method. In all scenarios, we use the rule of thumb $\ell_T = 8T^{1/5}$, which is known to be the optimal rate (see, Bühlmann, 1993, Bücher and Kojadinovic, 2016).

Figure S17 shows the coverage probability on the y-axis with the target level of 0.9 indicated by the dotted line. In most scenarios the coverage is approximately right. Exception are the larger semiparametric Gaussian models and large models with the Clayton copula. The latter might be explained by the fact that the simulated time series mix fairly slowly, see also Ibragimov (2009). The coverage can further be improved by tuning the block length parameter $\ell_T$, but our rule of thumb appears to give acceptable results in a wide range of scenarios.

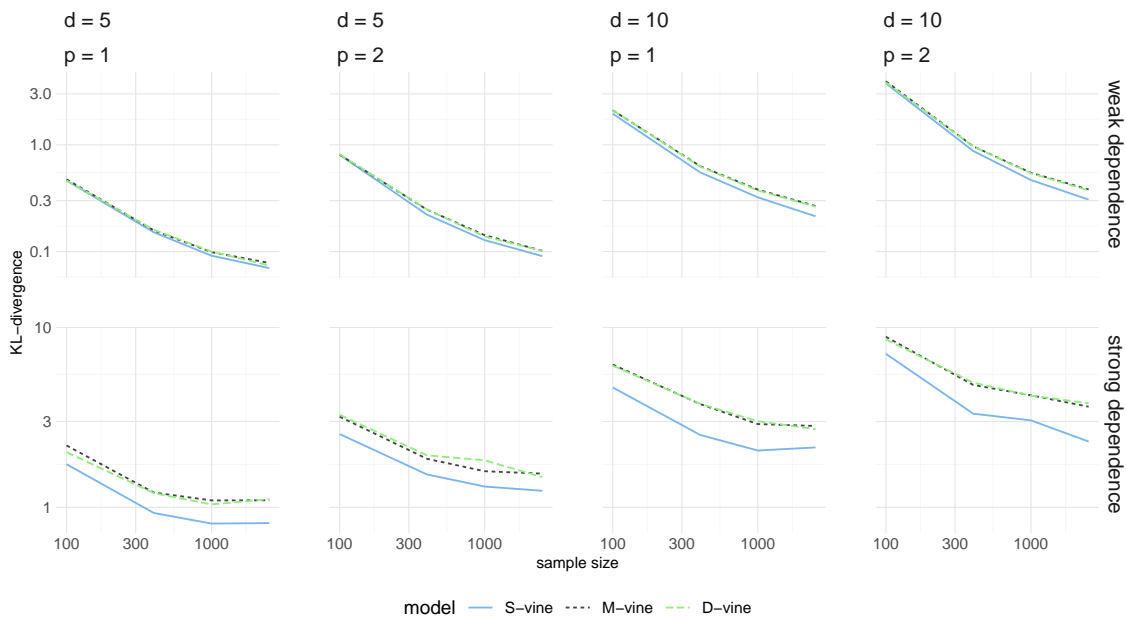

Figure S18: Averaged empirical KL-divergence for S-, M- and D-vine models

### S4.3  Model selection

We now assess the model selection heuristic proposed in Section S3.1 by comparing selected S-vine to M- and D-vine models. We shall use a randomized version of the above setup, to avoid a ground truth model that coincidentally favors one of the methods. Randomizing the vine structure and pair-copulas allows us to see how the structure and family selection heuristics perform over a wider range of data sets.

In each iteration, we draw a random S-vine structure. For the pair-copulas, we draw with equal probability from the Gaussian and Clayton copulas and rotate them randomly at 0, 90, 180 and 270 with equal probability. The parameters of pair copulas are set according to a randomly drawn Kendall's $\tau$s from a Beta distribution with parameters $\beta = 5\tau_0/(k - \tau_0)$ and $\beta = 5$ with $\tau_0 = 0.2$ (weak dependence) or $\tau_0 = 0.7$ (strong dependence). The expected value of Kendall's $\tau$ in tree $k$ is then $\tau_0/k$ as before. We generate a time series from the model and run the selection heuristics outlined above, allowing for all parametric families implemented in `rvinecopulib` (Nagler and Vatter, 2020). The model fit is then measured in terms of the fitted models' KL-divergence from the true model (computed via Monte-Carlo integration). A small divergence indicates better fit.

The x-axis in Figure S18 shows the sample size and the y-axis the KL-divergence between the true and estimated models. Again both axes are on a logarithmic scale and we would expect to see straight lines in a correctly specified model. However, none of the lines are straight, indicating that the heuristics rarely identify the structure that generated the data. This underlines how important it is to base inference on results and tools that allow for misspecification. The M- and

D-vine models are hardly distinguishable, but the S-vine model performs best in all scenarios. The gap increases with increasing dependence strength and sample size. This suggests that the added flexibility of the general stationary model can make a difference in applications.

## S5 Proofs of graph theoretic results

*Remark:* Referenced equations or results not prepended by an 'S' refer to the main manuscript.

### S5.1 Proof of Theorem 1

It is easy to see that (ii) together with translation invariance implies that the model is stationary. Now consider the reverse implication. Condition (2) is obviously true for $m = T - 1$ and $t = T - m = 1$. Now take $m = T - 2$.

We first show that $\mathcal{V}_{1,T-1}$ must be a vine, i.e., a sequence of trees satisfying the proximity condition. The proximity condition cannot be violated because $\mathcal{V}$ is a vine. We thus need to show that the restriction $\mathcal{V}_{1,T-1} = (V_{1,T-1,k}, E_{1,T-1,k})_{k=1,\ldots,(T-1)d-1}$ is a sequence of trees, which are connected, acyclic graphs. The restriction $\mathcal{V}_{1,T-1}$ cannot contain cycles, because it is constructed from deleting vertices and edges out of the sequence of trees $\mathcal{V}$. It thus remains to show that the graphs $(V_{1,T-1,k}, E_{1,T-1,k})_{k=1,\ldots,(T-1)d-1}$ are connected. We will prove this by contradiction.

Let $k \geq 1$ be the smallest level, where the graph $(V_{1,T-1,k}, E_{1,T-1,k})$ is not connected. Denote by $E_k$ the corresponding edge set of the complete vine $\mathcal{V}$. Recall that the $k$th graph $(V_k, E_k)$ of the whole vine $\mathcal{V}$ is a tree and therefore connected. For the restricted graph to be disconnected, $(V_k, E_k)$ must contain a path $P = (e_1, \ldots, e_{\ell+1})$, $1 \leq \ell \leq d$, connecting vertices $(v, v_1, \ldots, v_\ell, v')$ with $v, v' \in V_{1,T-1,k}$ and $v_1, \ldots, v_\ell \notin V_{1,T-1,k}$. Suppose that all pair-copulas in trees below and above the $k$th level are independence copulas. Then the joint density of $(U_v, U_{v'})$ can be written as

$$\int_{[0,1]^\ell} c_{e_1}(u_v, u_{v_1}) \times \cdots \times c_{e_{\ell+1}}(u_{v_\ell}, u_{v'}) du_{v_1} \cdots du_{v_\ell}. \tag{S1}$$

Now consider the density of $(U_w, U_{w'}) = (U_{v+(1,0)}, U_{v'+(1,0)})$. Since $(V_k, E_k)$ is connected, there must be another path $P' = (e'_1, \ldots, e'_{L+1})$, $L \geq 0$, connecting vertices $(w, w_1, \ldots, w_L, w')$. Hence, the joint density of $(U_{v+(1,0)}, U_{v'+(1,0)})$ is

$$\int_{[0,1]^L} c_{e'_1}(u_w, u_{w_1}) \times \cdots \times c_{e'_{L+1}}(u_{w_L}, u_{w'}) du_{w_1} \cdots du_{w_L}. \tag{S2}$$

For the model to be stationary, (S1) and (S2) must be equal for all $(u_v, u_{v'}) = (u_w, u_{w'}) \in (0,1)^2$. The edges $e_1$ and $e_{\ell+1}$ connect vertices from $V_{1,T-1,k}$ and $V \setminus V_{1,T-1,k}$. Because there is no time point $T + 1$, their translations $e'_1 + (1,0)$ and $e'_{\ell+1} + (1,0)$ cannot be in $E_k$. Hence, the edges $e'_1$

and $e'_{L+1}$ cannot be translations of $e_1$ and $e_{\ell+1}$. Therefore, translation invariance of $\mathcal{C}(\mathcal{V})$ does not imply equality of (S1) and (S2), which contradicts our premise. Hence, $\mathcal{V}_{1,T-1}$ must be a vine. That $\mathcal{V}_{2,T}$ is also a vine follows from symmetric arguments.

It remains to show that $\mathcal{V}_{1,T-1} \sim \mathcal{V}_{2,T}$. If $\mathcal{V}_{1,T-1} \nsim \mathcal{V}_{2,T}$, then there is an edge $e_1 \in E_{k,1,T-1}$ for which there is no translation in $E_{k,2,T}$. Similarly, there must be an edge $e_2 \in E_{k,2,T}$ for which there is no translation in $E_{k,1,T-1}$. Suppose all edges in $\mathcal{V}$ except the translations of $e_1$ and $e_2$ are independence copulas. Then the joint densities of $(\boldsymbol{U}_1, \dots, \boldsymbol{U}_{T-1})$ and $(\boldsymbol{U}_2, \dots, \boldsymbol{U}_T)$ are

$$c_{1,n-1}(\boldsymbol{u}) = \prod_{e \sim e_1} c_e(u_{a_e}, u_{b_e}), \quad \text{and} \quad c_{2,n}(\boldsymbol{u}) = \prod_{e \sim e_2} c_e(u_{a_e}, u_{b_e}).$$

For all $\boldsymbol{u} \in [0,1]^{d(T-1)}$, it must hold $c_{1,T-1}(\boldsymbol{u}) = c_{2,T}(\boldsymbol{u})$, but because $e_1 \nsim e_2$, this is not ensured by translation invariance, which is a contradiction. Hence, $\mathcal{V}_{1,T-1} \sim \mathcal{V}_{2,T}$.

We have shown that (ii) holds for $m = T-2$. This and (i) also imply that (i) holds for vine copula models $(V_{1,T-1}, \mathcal{C}(V_{1,T-1}))$ and $(\mathcal{V}_{2,T}, \mathcal{C}(V_{2,T}))$. We now define $T' = T-1$ and use the same arguments, to show that $(V_{1,T'-1}, \mathcal{C}(V_{1,T'-1}))$ and $(V_{2,T'}, \mathcal{C}(V_{2,T'}))$, satisfy (i) and (ii), and so on. This way, we can inductively show that (ii) must hold for all $m = T-3, \dots, 1$. $\qquad\square$

## S5.2 Proof of Theorem 2

It is easy to check that a vine satisfying (i)–(ii) is stationary. It is sufficient to specify the edges of the first $d$ trees, since the $d$th tree is a path. This fact and the proximity condition fix all edges in later trees.

Now consider the reverse implication. Let $\mathcal{V}$ be a stationary vine. In the first tree of $\mathcal{V}$ there can only be one edge between cross-sectional trees at adjacent time points, since $E_1$ must not contain cycles. We denote this edge by $\{(t, i_1), (t+1, j_1)\}$. Now suppose that there is some $k \le d-1$ such that (ii) holds for all $k'$ with $1 \le k' \le k$. We will show that it must also hold for $k+1$.

First observe that the edges

$$\bigcup_{r=1}^{k} \left\{ e \colon a_e = (t, i_{k+1-r}), b_e = (t+1, j_r), D_e = \bigcup_{s=1}^{k-r}\{(t, i_s)\} \cup \bigcup_{s=1}^{r-1}\{(t+1, j_s)\} \right\} \subset E_k$$

form a path. This together with the proximity condition fixes edges $\{e_{r,t} \colon r = 2, \dots, k, t = 1, \dots, T-1\} \subset E_{k+1}$ with

$$a_{e_{r,t}} = (t, i_{k+2-r}), \quad b_{e_{r,t}} = (t+1, j_r), \quad D_{e_{r,t}} = \bigcup_{s=1}^{k+1-r}\{(t, i_s)\} \cup \bigcup_{s=1}^{r-1}\{(t+1, j_s)\}.$$

For $E_{k+1}$ to form a tree, there must be two more edges: one connects the path to an edge in $E_{k+1}^{(0)} + (t, 0)$, the other connects it to an edge in $E_{k+1}^{(0)} + (t+1, 0)$. By the proximity condition,

these edges must have the form

$$a_e = (t, i_{k+1}), \quad b_e = (t+1, j_1), \quad D_e = \bigcup_{s=1}^{k} \{(t, i_s)\}$$

and

$$a_{e'} = (t, i_1), \quad b_{e'} = (t+1, j_{k+1}), \quad D_{e'} = \bigcup_{s=1}^{k} \{(t+1, j_s)\},$$

where $i_{k+1} \notin \bigcup_{s=1}^{k} \{i_s\}$ and $j_{k+1} \notin \bigcup_{s=1}^{k} \{j_s\}$. Furthermore, these edges are only permitted by the proximity condition if $E_k^{(0)}$ contains edges $e, e'$ with

$$a_e = i_{k+1}, \quad b_e = i_r, \quad D_e = \{i_1, \dots, i_k\} \setminus i_r,$$
$$a_{e'} = j_{k+1}, \quad b_{e'} = j_{r'}, \quad D_{e'} = \{j_1, \dots, j_k\} \setminus j_{r'}.$$

for some $r, r' \in \{1, \dots, k\}$. Hence (ii) holds for all $1 \le k \le d$ and the two permutations must be compatible with $\mathcal{V}^{(0)}$. □

### S5.3 Proof of Lemma 1

Let $i_1 \in \{1, \dots, d\}$ be arbitrary. Because $\mathcal{V}$ is a vine, its first tree clearly contains an edge with conditioned set $\{i_1, i_2\}$ for some $i_2 \ne i_1$. Now suppose we have found $2 \le k < d - 1$ indices $i_1, \dots, i_k$ indices that do not violate the condition in Definition 8. In particular, there is $e \in E_{k-1}$ with conditioned set $\{i_k, i_r\}$ and conditioning set $\{i_1, \dots, i_{k-1}\} \setminus i_r$ for some $r \in \{1, \dots, k-1\}$. In the $(k+1)$th level, this edge becomes a vertex. Because $\mathcal{V}$ is a vine, there must be an edge leaving this vertex. This edge must have conditioned set $\{i, j\}$ and conditioning set $\{i_1, \dots, i_{k-1}\} \setminus i$ for some $j \notin \{i_1, \dots, i_k\}$ and $i \in \{i_k, i_r\}$. Setting $i_{k+1} = j$, we see that the condition is also satisfied for $k + 1$. □

## S6 Proofs of asymptotic results

*Remark:* Referenced equations or results not prepended by an 'S' refer to the main manuscript.

### S6.1 Proofs of Theorems 4 and 5

Set $\mathcal{A} = \mathcal{H} \times \Theta$, $\mathfrak{N} = \emptyset$, and $\phi_{\boldsymbol{\alpha}, \boldsymbol{\nu}} = \phi_{\boldsymbol{\alpha}} = \phi_{\boldsymbol{\eta}, \boldsymbol{\theta}}^{(P)}$. It suffices to check the conditions of Theorem A.1. (C2) is void because $\mathfrak{N} = \emptyset$. (P1) is equivalent to (C3) and (P2) implies (C4). This proves Theorem 4.

To prove Theorem 5, we check the additional conditions of Theorem A.2. Conditions (C5)–(C6) are void because of $\mathfrak{N} = \emptyset$. Further, it follows from (P2) that (C7) holds with $\Phi_{\boldsymbol{\alpha}^*, 2} \equiv 0$

and $\boldsymbol{a} \mapsto \Phi_{\boldsymbol{\alpha}^*,1}(\boldsymbol{a}) = P(\nabla'_{\boldsymbol{\alpha}}\boldsymbol{\phi}_{\boldsymbol{\alpha}^*})\boldsymbol{a}$, which is invertible by (P3). To see this, let $\|\boldsymbol{a}\| \to 0$ be arbitrary. By the fundamental theorem of calculus and Fubini's theorem,

$$\big\|P\boldsymbol{\phi}_{\boldsymbol{\alpha}^*+\boldsymbol{a}} - P\boldsymbol{\phi}_{\boldsymbol{\alpha}^*} - P(\nabla'_{\boldsymbol{\alpha}}\boldsymbol{\phi}_{\boldsymbol{\alpha}^*})\boldsymbol{a}\big\| = \bigg\|\int_0^1 P(\nabla'_{\boldsymbol{\alpha}}\boldsymbol{\phi}_{\boldsymbol{\alpha}^*+t\boldsymbol{a}}\boldsymbol{a} - \nabla'_{\boldsymbol{\alpha}}\boldsymbol{\phi}_{\boldsymbol{\alpha}^*}\boldsymbol{a})dt\bigg\|$$
$$\leq \int_0^1 P\big\|\nabla'_{\boldsymbol{\alpha}}\boldsymbol{\phi}_{\boldsymbol{\alpha}^*+t\boldsymbol{a}} - \nabla'_{\boldsymbol{\alpha}}\boldsymbol{\phi}_{\boldsymbol{\alpha}^*}\big\|dt \times \|\boldsymbol{a}\|.$$

By (P2), the expectation in the integral on the right is bounded and, because the derivative $\nabla'_{\boldsymbol{\alpha}}\boldsymbol{\phi}_{\boldsymbol{\alpha}}$ is continuous, converges to zero by dominated convergence. Finally, (P4) implies (C8) and the claim follows. $\square$

## S6.2 Proofs of Theorems 6 and 7

We again check the conditions of Theorem A.1 and Theorem A.2. Set $\mathcal{A} = \Theta$ and $\mathfrak{N} = \{\nu\colon \mathbb{R} \to \mathbb{R}, \sup_x |\nu(x)| < \infty\}^d$. Further, let $\boldsymbol{\nu}^* \equiv \boldsymbol{0}$ and $\widehat{\nu}_j(x) = \{\widehat{F}_j(x) - F_j(x)\}/w\{F_j(x)\}$. Define $F_{\nu_j}(x) = F_j(x) + w\{F_j(x)\}\nu_j(x)$, $\boldsymbol{F_\nu(x)} = \big(F_{\nu_1}(x_1), \ldots, F_{\nu_d}(x_d)\big)$, and

$$\boldsymbol{\phi}_{\boldsymbol{\alpha},\boldsymbol{\nu}}(\boldsymbol{X}_t, \ldots, \boldsymbol{X}_{t+p}) = \boldsymbol{\phi}_{\boldsymbol{\alpha}}^{(SP)}\{\boldsymbol{F_\nu}(\boldsymbol{X}_t), \ldots, \boldsymbol{F_\nu}(\boldsymbol{X}_{t+p})\}.$$

(C2) follows from Lemma A.1, (SP1) is equivalent to (C3), and (SP2) implies (C4). This proves Theorem 6.

To prove Theorem 7, we check the additional conditions of Theorem A.2. Condition (SP3) implies (C5) and Lemma A.1 with (SP5) implies (C6). Further, it follows from (SP2) and an argument similar to the parametric case that (C7) holds with $\boldsymbol{a} \mapsto \Phi_{\boldsymbol{\alpha}^*,\boldsymbol{\nu}^*,1}(\boldsymbol{a}) = P(\nabla_{\boldsymbol{\alpha}}\boldsymbol{\phi}_{\boldsymbol{\alpha}^*,\boldsymbol{\nu}^*})\boldsymbol{a}$ and $\boldsymbol{b} \mapsto \Phi_{\boldsymbol{\alpha}^*,\boldsymbol{\nu}^*,2}(\boldsymbol{b}) = \sum_{j=1}^d P(\boldsymbol{\phi}_{\boldsymbol{\alpha}^*,\boldsymbol{\nu}^*}^{(j)}b_j)$, where

$$\boldsymbol{\phi}_{\boldsymbol{\alpha}^*,\boldsymbol{\nu}^*}^{(j)}(\boldsymbol{x}_1, \ldots, \boldsymbol{x}_{1+p}) = \sum_{t=1}^{1+p}\frac{\partial}{\partial\{\nu_j(x_{t,j})\}}\boldsymbol{\phi}_{\boldsymbol{\alpha}^*,\boldsymbol{\nu}^*}(\boldsymbol{x}_1, \ldots, \boldsymbol{x}_{1+p})$$
$$= \sum_{t=1}^{1+p}w\{F_j(x_{t,j})\}\frac{\partial}{\partial F_j(x_{t,j})}\boldsymbol{\phi}_{\boldsymbol{\alpha}}^{(SP)}\{\boldsymbol{F}(\boldsymbol{x}_1), \ldots, \boldsymbol{F}(\boldsymbol{x}_{1+p})\},$$

which gives the asymptotic variance in Theorem 7. Finally, (SP5) implies (C8) and the claim follows. $\square$

## S6.3 Proof of Theorem A.1

We first show that for all $K > 0$,

$$\sup_{\boldsymbol{\alpha}\in\mathcal{A}(K)} \|\mathbb{P}_T\boldsymbol{\phi}_{\boldsymbol{\alpha},\widehat{\boldsymbol{\nu}}} - P\boldsymbol{\phi}_{\boldsymbol{\alpha},\boldsymbol{\nu}^*}\| = o_p(1). \tag{S3}$$

Condition (C2) implies that there is a sequence $\delta_T \to 0$ such that $\|\widehat{\boldsymbol{\nu}} - \boldsymbol{\nu}\| \le \delta_T$ with probability tending to 1. On events where $\|\widehat{\boldsymbol{\nu}} - \boldsymbol{\nu}\| \le \delta_T$, the triangle inequality implies

$$\sup_{\boldsymbol{\alpha} \in \mathcal{A}(K)} \|\mathbb{P}_T \boldsymbol{\phi}_{\boldsymbol{\alpha}, \widehat{\boldsymbol{\nu}}} - P\boldsymbol{\phi}_{\boldsymbol{\alpha}, \boldsymbol{\nu}^*}\|$$

$$\le \sup_{\boldsymbol{\alpha} \in \mathcal{A}(K)} \sup_{\|\boldsymbol{\nu} - \boldsymbol{\nu}^*\| \le \delta_T} \|\mathbb{P}_T \boldsymbol{\phi}_{\boldsymbol{\alpha}, \boldsymbol{\nu}} - \mathbb{P}_T \boldsymbol{\phi}_{\boldsymbol{\alpha}, \boldsymbol{\nu}^*}\| + \sup_{\boldsymbol{\alpha} \in \mathcal{A}(K)} \|\mathbb{P}_T \boldsymbol{\phi}_{\boldsymbol{\alpha}, \boldsymbol{\nu}^*} - P\boldsymbol{\phi}_{\boldsymbol{\alpha}, \boldsymbol{\nu}^*}\|$$

$$\le \delta_T \mathbb{P}_T \left\{ \sup_{\boldsymbol{\alpha} \in \mathcal{A}(K)} \sup_{\|\boldsymbol{\nu} - \boldsymbol{\nu}^*\| \le \delta_T} \frac{\|\boldsymbol{\phi}_{\boldsymbol{\alpha}, \boldsymbol{\nu}} - \boldsymbol{\phi}_{\boldsymbol{\alpha}, \boldsymbol{\nu}^*}\|}{\|\boldsymbol{\nu} - \boldsymbol{\nu}^*\|} \right\} + \sup_{\boldsymbol{\alpha} \in \mathcal{A}(K)} \|\mathbb{P}_T \boldsymbol{\phi}_{\boldsymbol{\alpha}, \boldsymbol{\nu}^*} - P\boldsymbol{\phi}_{\boldsymbol{\alpha}, \boldsymbol{\nu}^*}\|.$$

By (C4) and the ergodic theorem, the first term on the right is of order $O_p(\delta_T) = o_p(1)$. Furthermore, (C4) implies that the function classes

$$\mathcal{F}_j = \{\phi_{\boldsymbol{\alpha}, \boldsymbol{\nu}^*, j} \colon \boldsymbol{\alpha} \in \mathcal{A}(K)\}, \quad j = 1, \dots, r,$$

have finite $L_1(P)$-bracketing numbers (see, van der Vaart and Wellner, 1996, Theorem 2.7.11). Then (C1) and the Glivenko-Cantelli theorem for absolutely regular sequences (Nobel and Dembo, 1993, Theorem 1),

$$\sup_{\boldsymbol{\alpha} \in \mathcal{A}(K)} \|\mathbb{P}_T \boldsymbol{\phi}_{\boldsymbol{\alpha}, \boldsymbol{\nu}^*} - P\boldsymbol{\phi}_{\boldsymbol{\alpha}, \boldsymbol{\nu}^*}\| = o_p(1),$$

which proves (S3). Because the norm is a continuous, (S3) and the continuous mapping theorem further imply

$$\sup_{\boldsymbol{\alpha} \in \mathcal{A}(K)} \left| \|\mathbb{P}_T \boldsymbol{\phi}_{\boldsymbol{\alpha}, \widehat{\boldsymbol{\nu}}}\| - \|P\boldsymbol{\phi}_{\boldsymbol{\alpha}, \boldsymbol{\nu}^*}\| \right| = o_p(1).$$

Now observe that $\widehat{\boldsymbol{\alpha}}$ maximizes $\boldsymbol{\alpha} \mapsto -\|\mathbb{P}_T \boldsymbol{\phi}_{\boldsymbol{\alpha}, \widehat{\boldsymbol{\nu}}}\|$ and, by (C3)–(C4), $\boldsymbol{\alpha}^*$ is the unique maximizer of the continuous map $\boldsymbol{\alpha} \mapsto -\|P\boldsymbol{\phi}_{\boldsymbol{\alpha}, \boldsymbol{\nu}^*}\|$. Now the claim follows from the Argmax theorem (van der Vaart and Wellner, 1996, Theorem 3.2.2). $\square$

## S6.4 Proof of Theorem A.2

Expand

$$\mathbb{P}_T \boldsymbol{\phi}_{\widehat{\boldsymbol{\alpha}}, \widehat{\boldsymbol{\nu}}} = (\mathbb{P}_T - P)(\boldsymbol{\phi}_{\widehat{\boldsymbol{\alpha}}, \widehat{\boldsymbol{\nu}}} - \boldsymbol{\phi}_{\boldsymbol{\alpha}^*, \boldsymbol{\nu}^*}) + P(\boldsymbol{\phi}_{\widehat{\boldsymbol{\alpha}}, \widehat{\boldsymbol{\nu}}} - \boldsymbol{\phi}_{\boldsymbol{\alpha}^*, \boldsymbol{\nu}^*}) + \mathbb{P}_T \boldsymbol{\phi}_{\boldsymbol{\alpha}^*, \boldsymbol{\nu}^*}$$
$$= (\mathbb{P}_T - P)(\boldsymbol{\phi}_{\widehat{\boldsymbol{\alpha}}, \widehat{\boldsymbol{\nu}}} - \boldsymbol{\phi}_{\boldsymbol{\alpha}^*, \boldsymbol{\nu}^*}) + \Phi_{\boldsymbol{\alpha}^*, \boldsymbol{\nu}^*, 1}(\widehat{\boldsymbol{\alpha}} - \boldsymbol{\alpha}^*) + \Phi_{\boldsymbol{\alpha}^*, \boldsymbol{\nu}^*, 2}(\widehat{\boldsymbol{\nu}} - \boldsymbol{\nu}^*) + \mathbb{P}_T \boldsymbol{\phi}_{\boldsymbol{\alpha}^*, \boldsymbol{\nu}^*}$$
$$+ o_p(\|\widehat{\boldsymbol{\alpha}} - \boldsymbol{\alpha}^*\| + T^{-1/2}),$$

where the last equality follows from (C7). Since $\mathbb{P}_T \boldsymbol{\phi}_{\widehat{\boldsymbol{\alpha}}, \widehat{\boldsymbol{\nu}}} = 0$ by the definition of $(\widehat{\boldsymbol{\alpha}}, \widehat{\boldsymbol{\nu}})$, and $\Phi_{\boldsymbol{\alpha}^*, \boldsymbol{\nu}^*, 1}$ has linear inverse by (C7), the last display can be rearranged as

$$\widehat{\boldsymbol{\alpha}} - \boldsymbol{\alpha}^* = -\Phi_{\boldsymbol{\alpha}^*, \boldsymbol{\nu}^*, 1}^{-1} \{ \mathbb{P}_T \boldsymbol{\phi}_{\boldsymbol{\alpha}^*, \boldsymbol{\nu}^*} + \Phi_{\boldsymbol{\alpha}^*, \boldsymbol{\nu}^*, 2} (\widehat{\boldsymbol{\nu}} - \boldsymbol{\nu}^*) + (\mathbb{P}_T - P)(\boldsymbol{\phi}_{\widehat{\boldsymbol{\alpha}}, \widehat{\boldsymbol{\nu}}} - \boldsymbol{\phi}_{\boldsymbol{\alpha}^*, \boldsymbol{\nu}^*}) \}$$
$$+ o_p(\| \widehat{\boldsymbol{\alpha}} - \boldsymbol{\alpha}^* \| + T^{-1/2}),$$

It holds $\Phi_{\boldsymbol{\alpha}^*, \boldsymbol{\nu}^*, 2}(\widehat{\boldsymbol{\nu}} - \boldsymbol{\nu}^*) = O_p(T^{-1/2})$ by (C6) and linearity of $\Phi_{\boldsymbol{\alpha}^*, \boldsymbol{\nu}^*, 2}$. Furthermore, Corollary 1.2 of Rio (2017) implies $\mathbb{E}\{(\mathbb{P}_T \boldsymbol{\phi}_{\boldsymbol{\alpha}^*, \boldsymbol{\nu}^*})^2\} \leq 4T^{-1} \sum_{t=0}^{\infty} \int_0^{\beta(t)} Q^2(u) du$, which is of order $O(T^{-1})$ by (C8). Hence, the first display of theorem is proven if we can show that $(\mathbb{P}_T - P)(\boldsymbol{\phi}_{\widehat{\boldsymbol{\alpha}}, \widehat{\boldsymbol{\nu}}} - \boldsymbol{\phi}_{\boldsymbol{\alpha}^*, \boldsymbol{\nu}^*}) = o_p(T^{-1/2})$. Theorem A.1 implies that there exists a sequence $\epsilon_T \to 0$, such that $\| \widehat{\boldsymbol{\alpha}} - \boldsymbol{\alpha}^* \| \leq \epsilon_T$ with probability tending to one. Hence, it suffices to show that,

$$\sup_{\boldsymbol{\alpha} \in \mathcal{A}(\epsilon_T)} \left| (\mathbb{P}_T - P)(\boldsymbol{\phi}_{\boldsymbol{\alpha}, \widehat{\boldsymbol{\nu}}} - \boldsymbol{\phi}_{\boldsymbol{\alpha}, \boldsymbol{\nu}^*}) \right| = o_p(T^{-1/2}), \tag{S4}$$

for every $\epsilon_T \to 0$.

To prove (S4), let $\phi_{\boldsymbol{\alpha}, \boldsymbol{\nu}, j}$ denote the $j$th component of the vector $\boldsymbol{\phi}_{\boldsymbol{\alpha}, \boldsymbol{\nu}}$ and define the function classes $\mathfrak{N}_\infty = \{ \boldsymbol{\nu} \in \mathfrak{N} \colon \sup_{\boldsymbol{x}} \| \boldsymbol{\nu}(\boldsymbol{x}) \| < \infty \}$ and

$$\mathcal{G}_{T, \boldsymbol{\alpha}, \boldsymbol{b}_0, \delta, j} = \left\{ \phi_{\boldsymbol{\alpha}, \boldsymbol{\nu}^* + T^{-1/2} \boldsymbol{b}, j} - \phi_{\boldsymbol{\alpha}, \boldsymbol{\nu}^* + T^{-1/2} \boldsymbol{b}_0, j} \colon \boldsymbol{b} \in \mathfrak{N}_\infty, \| \boldsymbol{b} - \boldsymbol{b}_0 \| < \delta \right\}, \quad j = 1, \ldots, r, \quad \boldsymbol{b}_0 \in \mathfrak{N}_\infty.$$

Let $G_{T, \boldsymbol{\alpha}, \boldsymbol{b}_0, j, \delta}$ be an arbitrary envelope to this class, i.e., a function satisfying $\sup_{g \in \mathcal{G}_{T, \boldsymbol{\alpha}, \boldsymbol{b}_0, j}} |g(\boldsymbol{x})| \leq G_{T, \boldsymbol{\alpha}, \boldsymbol{b}_0, j, \delta}(\boldsymbol{x})$. Following Theorem 2.3 in Wellner and van der Vaart (2007), equation (S4) holds if for all $j = 1, \ldots, r$:

(i) $\sup_{\boldsymbol{\alpha} \in \mathcal{A}_{\epsilon_T}} | T^{1/2}(\mathbb{P}_T - P)(\phi_{\boldsymbol{\alpha}, \boldsymbol{\nu}^* + T^{-1/2} \boldsymbol{b}, j} - \phi_{\boldsymbol{\alpha}, \boldsymbol{\nu}^*, j}) | = o_p(1)$, for every $\boldsymbol{b} \in \mathfrak{N}_\infty$,

(ii) $\sup_{\boldsymbol{\alpha} \in \mathcal{A}_{\epsilon_T}} | T^{1/2}(\mathbb{P}_T - P) G_{T, \boldsymbol{\alpha}, \boldsymbol{b}, j, \delta} | = o_p(1)$, for every small $\delta > 0$ and $\boldsymbol{b} \in \mathfrak{N}_\infty$,

(iii) $\sup_{\boldsymbol{\alpha} \in \mathcal{A}_{\epsilon_T}} \sup_{\boldsymbol{b} \in \mathfrak{N}_\infty} T^{1/2} P G_{T, \boldsymbol{\alpha}, \boldsymbol{b}, j, \delta_T} = o(1)$, for every $\delta_T \to 0$.

We first check conditions (ii) and (iii) by constructing an appropriate envelope. For some $\epsilon > 0$ small enough, define

$$G(\boldsymbol{x}) = \max_{1 \leq j \leq r} \sup_{\boldsymbol{\alpha} \in \mathcal{A}(\epsilon)} \sup_{\boldsymbol{\nu}_1, \boldsymbol{\nu}_2 \in \mathfrak{N}(\epsilon)} \frac{|\phi_{\boldsymbol{\alpha}, \boldsymbol{\nu}_1, j}(\boldsymbol{x}) - \phi_{\boldsymbol{\alpha}, \boldsymbol{\nu}_2, j}(\boldsymbol{x})|}{\| \boldsymbol{\nu}_1 - \boldsymbol{\nu}_2 \|}$$

and note that $PG < \infty$ by (C4). For every $\delta \leq \epsilon$, we can now set $G_{T, \boldsymbol{\alpha}, \boldsymbol{b}, j, \delta} = G T^{-1/2} \delta$, for which conditions (ii) and (iii) are immediate. For (i), the triangle inequality gives

$$\sup_{\boldsymbol{\alpha} \in \mathcal{A}_{\epsilon_T}} \left| T^{1/2}(\mathbb{P}_T - P)(\phi_{\boldsymbol{\alpha}, \boldsymbol{\nu}^* + T^{-1/2} \boldsymbol{b}, j} - \phi_{\boldsymbol{\alpha}, \boldsymbol{\nu}^*, j}) \right|$$
$$\leq \sup_{\boldsymbol{\alpha} \in \mathcal{A}_{\epsilon_T}} \left| T^{1/2}(\mathbb{P}_T - P)\{ (\phi_{\boldsymbol{\alpha}, \boldsymbol{\nu}^* + T^{-1/2} \boldsymbol{b}, j} - \phi_{\boldsymbol{\alpha}, \boldsymbol{\nu}^*, j}) - (\phi_{\boldsymbol{\alpha}^*, \boldsymbol{\nu}^* + T^{-1/2} \boldsymbol{b}, j} - \phi_{\boldsymbol{\alpha}^*, \boldsymbol{\nu}^*, j}) \} \right|$$

$$+ \big|T^{1/2}(\mathbb{P}_T - P)(\phi_{\boldsymbol{\alpha}^*, \boldsymbol{\nu}^* + T^{-1/2}\boldsymbol{b}, j} - \phi_{\boldsymbol{\alpha}^*, \boldsymbol{\nu}^*, j})\big|.$$

The expectation of the first term on the right is bounded by

$$2T^{1/2}P \sup_{\boldsymbol{\alpha} \in \mathcal{A}_{\epsilon_T}} \big|(\phi_{\boldsymbol{\alpha}, \boldsymbol{\nu}^* + T^{-1/2}\boldsymbol{b}, j} - \phi_{\boldsymbol{\alpha}, \boldsymbol{\nu}^*, j}) - (\phi_{\boldsymbol{\alpha}^*, \boldsymbol{\nu}^* + T^{-1/2}\boldsymbol{b}, j} - \phi_{\boldsymbol{\alpha}^*, \boldsymbol{\nu}^*, j})\big|,$$

which is of order $O(\epsilon_T) = o(1)$ by (C5). For the second term, note that the sequence $T^{1/2}|\phi_{\boldsymbol{\alpha}^*, \boldsymbol{\nu}^* + T^{-1/2}\boldsymbol{b}, j} - \phi_{\boldsymbol{\alpha}^*, \boldsymbol{\nu}^*, j}|$ is dominated by $G\|\boldsymbol{b}\|$ and, thus, uniformly integrable. Then the weak law of large numbers for $\alpha$-mixing triangular arrays (Kanaya, 2017, Theorem 1 and Remark 2) establishes condition (i). Therefore, (S4) holds which establishes the first display of the theorem.

For the second part of the theorem, we have $P\big(\mathbb{P}_T \boldsymbol{\phi}_{\boldsymbol{\alpha}^*, \boldsymbol{\nu}^*}\big)^2 = O(T^{-1})$ by (C8) and (Rio, 2017, Corollary 1.2). Further, because $\Phi_{\boldsymbol{\alpha}^*, \boldsymbol{\theta}^*, 2}$ is linear and continuous by (C7) and, hence, bounded, it holds $T^{1/2}\Phi_{\boldsymbol{\alpha}^*, \boldsymbol{\theta}^*, 2}(\widehat{\boldsymbol{\nu}} - \boldsymbol{\nu}^*) = O_p(1)$ by (C6) and the continuous mapping theorem. Therefore, $\|\widehat{\boldsymbol{\alpha}} - \boldsymbol{\alpha}^*\| = O_p(T^{-1/2})$. Convergence in distribution follows from the central limit theorem for strongly mixing variables (Rio, 2017, Theorem 4.2) and the continuous mapping theorem. $\square$

## S6.5 Proof of Theorem A.3

Because the set of indicators $\{\mathbb{1}(\cdot \leq \boldsymbol{z} \in \mathbb{R}^d) : \boldsymbol{z} \in \mathbb{R}\}$ forms a VC class, Theorem 2.8.3 of van der Vaart and Wellner (1996) implies

$$\lim_{K \to \infty} \sup_{\|(\boldsymbol{\alpha}, \boldsymbol{\nu}) - (\boldsymbol{\alpha}^*, \boldsymbol{\nu}^*)\| \leq \delta} \mathrm{P}\left\{N^{1/2} \sup_{\boldsymbol{z} \in \mathbb{R}^d} \big|F_{N, \boldsymbol{\alpha}, \boldsymbol{\nu}}(\boldsymbol{z}) - F_{\boldsymbol{\alpha}, \boldsymbol{\nu}}(\boldsymbol{z})\big| > K\right\} \to 0.$$

Since Frechet differentiability implies continuity, part $(i)$ of the theorem follows from $(\widehat{\boldsymbol{\alpha}}, \widehat{\boldsymbol{\nu}}) \to_p (\boldsymbol{\alpha}^*, \boldsymbol{\nu}^*)$. For part (ii), we Frechet differentiability of $\psi$ implies

$$\psi(F_{N, \widehat{\boldsymbol{\alpha}}, \widehat{\boldsymbol{\nu}}}) - \psi(F_{\widehat{\boldsymbol{\alpha}}, \widehat{\boldsymbol{\nu}}}) = O(N^{-1/2}).$$

and

$$\psi(F_{\widehat{\boldsymbol{\alpha}}, \widehat{\boldsymbol{\nu}}}) - \psi(F_{\boldsymbol{\alpha}^*, \boldsymbol{\nu}^*}) = \Psi_{(\boldsymbol{\alpha}^*)}(\boldsymbol{\alpha}^* - \widehat{\boldsymbol{\alpha}}) + \Psi_{\boldsymbol{\nu}^*}(\boldsymbol{\nu}^* - \widehat{\boldsymbol{\nu}}) + o_p(T^{-1/2}).$$

This yields

$$T^{1/2}(\widehat{\mu} - \mu^*) = T^{1/2}\{\psi(F_{N, \widehat{\boldsymbol{\alpha}}, \widehat{\boldsymbol{\nu}}}) - \psi(F_{\widehat{\boldsymbol{\alpha}}, \widehat{\boldsymbol{\nu}}})\} + T^{1/2}\{\psi(F_{\widehat{\boldsymbol{\alpha}}, \widehat{\boldsymbol{\nu}}}) - \psi(F_{\boldsymbol{\alpha}^*, \boldsymbol{\nu}^*})\}$$
$$= T^{1/2}\Psi_{(\boldsymbol{\alpha}^*)}(\boldsymbol{\alpha}^* - \widehat{\boldsymbol{\alpha}}) + T^{1/2}\Psi_{\boldsymbol{\nu}^*}(\boldsymbol{\nu}^* - \widehat{\boldsymbol{\nu}}) + o_p(1).$$

Part (ii) then follows from the continuous mapping theorem and Sltusky's lemma. $\square$

## S6.6 Proof of Lemma A.1

We follow the strategy of Chen and Fan (2006, Lemma 4.1). The process $W_T$ is an empirical process indexed by the function class $\mathcal{G} = \{\mathbb{1}(\cdot \leq z)/w\{F_Z(z)\} \colon z \in \mathbb{R}\}$. It shall be convenient that the function $z \mapsto w\{F_Z(z)\}$ is monotone on either $\{z \leq F_Z^{-1}(1/2)\}$ or $\{z > F_Z^{-1}(1/2)\}$. We thus split the process into two parts, $\mathcal{G}_1 = \{\mathbb{1}(\cdot \leq z)/w\{F_Z(z)\} \colon z \leq F_Z^{-1}(1/2)\}$ and $\mathcal{G}_2 = \{\mathbb{1}(\cdot \leq z)/w\{F_Z(z)\} \colon z > F_Z^{-1}(1/2)\}$, and establish (i) and (ii) separately. Since the proofs are symmetric for $\mathcal{G}_1$ and $\mathcal{G}_2$, we only give arguments for the first.

The proof relies on a bracketing argument, which is constructed as follows. Let $0 = u_0 < u_1 < \cdots < u_K = 1/2$ and $z_k = F_Z^{-1}(u_k), k = 0, \ldots, K$. Let $l_1(z) = 0$ and $h_1(z) = \mathbb{1}(z \leq z_1)/w\{F_Z(z)\}$. Because $w$ is in increasing, we get $l_1(z) \leq \mathbb{1}(z \leq z')/w\{F_Z(z')\} \leq h_1(z)$ for all $z' \leq z_1$ and $z \in \mathbb{R}$. The pair $l_1, h_1$ is therefore called a bracket. For $k \geq 2$, define brackets $l_k(z) = \mathbb{1}(z \leq z_{k-1})/w\{F_Z(z_k)\}$, $h_k(z) = \mathbb{1}(z \leq z_k)/w\{F_Z(z_{k-1})\}$. Then one can easily check that for every $z' \leq F_Z^{-1}(1/2)$, there is a $k \in \{1, \ldots, K\}$ such that $l_k(z) \leq \mathbb{1}(z \leq z')/w\{F_Z(z')\} \leq h_k(z)$. Both parts of the theorem are proved by controlling the size of these brackets, albeit in different norms.

In the following bounds, we make repeated use of the fact that $u_k \in [0, 1/2]$ and $\gamma \in [0, 1)$. We compute

$$\mathbb{E}\{|h_1(Z) - l_1(Z)|\} = \int_0^{u_1} 1/w(u)du \leq 2^\gamma \int_0^{u_1} u^{-\gamma}du \leq 2u_1^{1-\gamma}.$$

Hence, we have $\mathbb{E}\{|h_1(Z) - l_1(Z)|\} \leq \epsilon$ whenever $u_1 \leq (\epsilon/2)^{1/(1-\gamma)}$. For $k \geq 2$, we use the that

$$|(1/w(u))'| = [u(1-u)]^{-2\gamma}\gamma[u(1-u)]^{\gamma-1}(1-2u) \leq [u(1-u)]^{-2} \leq 4u^{-2},$$

which is decreasing in $u$. Thus,

$$\begin{aligned}
\mathbb{E}\{|h_k(Z) - l_k(Z)|\} &= \mathbb{E}\{\mathbb{1}(Z \leq z_k)\}/w(u_{k-1}) - E\{\mathbb{1}(Z \leq z_{k-1})\}/w(u_k) \\
&= \mathbb{E}\{\mathbb{1}(Z \leq z_{k-1})\}\{1/w(u_{k-1}) - 1/w(u_k)\} + \mathbb{E}\{\mathbb{1}(z_{k-1} \leq Z \leq z_k)\}/w(u_{k-1}) \\
&= u_{k-1}\{1/w(u_{k-1}) - 1/w(u_k)\} + (u_k - u_{k-1})/w(u_{k-1}) \\
&\leq 4u_{k-1}^{-1}(u_k - u_{k-1}) + 2u_{k-1}^{-1}(u_k - u_{k-1}).
\end{aligned}$$

Solving for $\mathbb{E}\{|h_k(Z) - l_k(Z)|\} \leq \epsilon$ gives the progression $u_k = u_{k-1} + \epsilon u_{k-1}/6$. The brackets cover the entire set $\mathcal{G}_1$ when $u_K \geq 1/2$. With $u_1 = (\epsilon/2)^{1/(1-\gamma)}$ from before, we get

$$u_K = u_1 + \sum_{k=2}^K (u_k - u_{k-1}) = u_1 + \sum_{k=2}^K \epsilon u_{k-1}/6 \geq (K-1)\epsilon u_1/6 \geq (K-1)\epsilon(\epsilon/2)^{1/(1-\gamma)}/6,$$

and thus need $K = O(\epsilon^{-1-1/(1-\gamma)})$ $\epsilon$-brackets to cover $\mathcal{G}_1$. Furthermore, the class $\mathcal{G}_1$ admits an envelope $G(z) = 1/w\{F_Z(z)\}$ with $\mathbb{E}\{|G(Z)|\} = \int_0^1 1/w(u)du < \infty$. Part (i) of the theorem

then follows from the bracketing Glivenko-Cantelli theorem (van der Vaart and Wellner, 1996, 2.4.1 Theorem), and Theorem 1 of Nobel and Dembo (1993) for the equivalence of the $\beta$-mixing and *iid* cases.

For part (ii), we need to control the size of brackets in a stronger norm, see Doukhan et al. (1995). Let $\|X\|_\beta^2 = \sum_{t=0}^\infty \int_0^{\beta(t)} Q_X(u)^2 du$, where $Q_X$ is the inverse survival function of $|X|$. We first note that the envelope $G$ mentioned above satisfies

$$\|G(Z)\|_\beta^2 = \sum_{t=0}^\infty \int_0^{\beta(t)} w(u)^{-2} du \le 4 \sum_{t=0}^\infty \beta(t)^{1-2\gamma} < \infty,$$

since $\beta(t) = O(t^{-a})$ with $a > 1/(1-2\gamma)$ by assumption. Now we compute the bracket sizes under the $\|\cdot\|_\beta$-norm. The inverse survival function of $|h_1(Z) - l_1(Z)|$ is given by

$$Q_1(u) = \mathbb{1}(u \le u_1)/w(u) \le 2\mathbb{1}(u \le u_1)u^{-\gamma},$$

and hence

$$\|h_1(Z) - l_1(Z)\|_\beta^2 \le 4 \sum_{t=0}^\infty \int_0^{\beta(t)\wedge u_1} u^{-2\gamma} du \le C_\gamma \sum_{t=0}^\infty \{\beta(t) \wedge u_1\}^{1-2\gamma}$$
$$= C_\gamma t^* u_1^{1-2\gamma} + C_\gamma \sum_{t=t^*+1}^\infty \beta(t)^{1-2\gamma},$$

with $C_\gamma = 4/(1-2\gamma)$. The right hand side is increasing in $u_1$ and finite by our assumption on the decay of $\beta(t)$: because $\beta(t)^{1-2\gamma} = O(t^{-a(1-2\gamma)})$, the series $\sum_{t=t^*+1}^\infty \beta(t)^{1-2\gamma}$ decays as $(t^*)^{1-a(1-2\gamma)}$ provided $a > 1/(1-2\gamma)$. Hence, we can pick $t^*$ and $u_1$ such that $\|h_1(Z) - l_1(Z)\|_\beta \le \epsilon$. We choose $t^* = C_*\epsilon^b$ for $b = 1/\{1-a(1-2\gamma)\} < 0$ and some constant $C_* < \infty$, such that $C_\gamma \sum_{t=t^*+1}^\infty \beta(t)^{1-2\gamma} \le \epsilon/2$. Next, define $u_1 = \{\epsilon/(2C_\gamma t^*)\}^{1/(1-2\gamma)}$ such that also $C_\gamma t^* u_1^{1-2\gamma} \le \epsilon/2$. We therefore have $\|h_1(Z) - l_1(Z)\|_\beta \le \epsilon$ with $u_1 = C_1\epsilon^{(1-b)/(1-2\gamma)}$ for some constant $C_1 \in (0, \infty)$.

The brackets with $k \ge 2$ are bounded, so we can use more lenient bounds. Define $B^* = \exp(2/q)\{\sum_{t=0}^\infty (t+1)^{2/(q-2)}\beta(t)\}^{1-2/q}$ for some $q > 0$. Due to our assumption on the decay of $\beta(t)$, $B^*$ is finite for large enough $q$ (see, Rio, 2017, eq. 1.25b). As in Rio (2017, Section 1.4), we can now use Hölder's inequality to obtain the bound

$$\|h_k(Z) - l_k(Z)\|_\beta^2 = \sum_{t=0}^\infty \int_0^{\beta(t)} Q_k(u)^2 du \le B^* \mathbb{E}\{|h_k(Z) - l_k(Z)|^q\}^{2/q}.$$

Further,

$$\mathbb{E}\{|h_k(Z) - l_k(Z)|^q\} \le \mathbb{E}\{\mathbb{1}(Z \le z_{k-1})\}\{1/w(u_{k-1}) - 1/w(u_k)\}^q + \mathbb{E}\{\mathbb{1}(z_{k-1} \le Z \le z_k)\}/w(u_{k-1})^q$$
$$\le (4u_{k-1}^{-(\gamma+1)})^q (u_k - u_{k-1})^q + (2u_{k-1}^{-\gamma})^q (u_k - u_{k-1})$$

$$\leq (4^q + 2^q) u_{k-1}^{-2q} (u_k - u_{k-1}).$$

Hence,

$$\|h_k(Z) - l_k(Z)\|_\beta \leq 6\sqrt{B^*} u_{k-1}^{-2} (u_k - u_{k-1})^{1/q}.$$

Solving for $\|h_k(Z) - l_k(Z)\|_\beta \leq \epsilon$ yields $u_k = u_{k-1} + \epsilon^q u_{k-1}^{2q} (B^*)^{-q/2} 6^{-q}$. Using $u_1 = C_1 = \epsilon^{(1-b)/(1-2\gamma)}$, we now have

$$u_K = u_1 + \sum_{k=2}^{K} (u_k - u_{k-1}) \geq (K-1)\epsilon^q u_1^{2q} (B^*)^{-q/2} 6^{-q} \geq (K-1)\epsilon^{q+2q(1-b)/(1-2\gamma)} C_1^{2q} (B^*)^{-q/2} 6^{-q}.$$

Hence, we can cover $\mathcal{G}_1$ with number of brackets $K$ growing polynomially in $\epsilon^{-1}$. Part (ii) of the theorem is now a consequence of Theorem 1 by Doukhan et al. (1995). $\square$



\end{document}